\documentclass[12pt]{article}
\pdfoutput=1

\usepackage{color}
\usepackage{amssymb,amsmath,bm}
\usepackage{epsf}
\usepackage{epsfig}
\usepackage{afterpage}
\usepackage{longtable}
\usepackage[dvipsnames]{xcolor}
\usepackage[linktoc=page,bookmarks=false,colorlinks=false,linkbordercolor=RoyalBlue,citebordercolor=ForestGreen,urlbordercolor=CornflowerBlue]{hyperref}
\usepackage{latexsym,mathrsfs}
\usepackage[normalem]{ulem} 
\usepackage[compress]{cite}
\usepackage{graphicx}
\usepackage{url}
\usepackage{paralist}
\usepackage{slashed}
\usepackage{multirow}

\usepackage[hypcap]{caption, subcaption}

\usepackage{booktabs}
\usepackage{listings}

\usepackage{textcomp}

\usepackage[titles]{tocloft}
\allowdisplaybreaks[1]

\numberwithin{equation}{section}

\def \nnb{\nonumber}
\def \dps{\displaystyle}
\def \ts{\textstyle}
\def\ub{\bar{u}}

\def\slash2#1{#1 \hskip-0.45em /}

\def\be{\begin{equation}}
\def\ee{\end{equation}}
\def\bea{\begin{eqnarray}}
\def\eea{\end{eqnarray}}

\definecolor{darkgreen}{rgb}{0.0,0.6,0.0}

\newcounter{MBQ}

\newcounter{GBQ}

\newcounter{THQ}

\newcounter{XQLQ}

\newcommand{\calO}{\mathcal{O}}
\newcommand{\calH}{\mathcal{H}}

\newcommand{\as}{\alpha_s}

\newcommand{\asfourpi}{\frac{\alpha_s}{4\pi}}
\newcommand{\braket}[1]{\langle #1 \rangle}

\newcommand{\ESGamma}{S_{\Gamma}}

\newcommand{\vs}[1]{\vspace*{#1 pt}}

\newcommand{\eps}{\epsilon}
\newcommand{\ep}{\epsilon}

\newcommand{\nms}{\slashed{n}_-}
\newcommand{\nps}{\slashed{n}_+}

\begin{document}

\allowdisplaybreaks

\begin{titlepage}

\begin{flushright}
{\small
TUM-HEP-1250/20 \\
SI-HEP-2019-17 \\
QFET-2019-12\\
February 7, 2020
}
\end{flushright}

\vspace{0.9cm}
\begin{center}
{\Large\bf\boldmath
Two-loop non-leptonic penguin amplitude  \\[0.2cm] 
in QCD factorization
}
\\[12mm]
{\sc
Guido~Bell$^{a}$,
Martin~Beneke$^{b}$,
Tobias~Huber$^{a}$
and
Xin-Qiang~Li$^{c}$
}
\\[1.2cm]

$^{a}${\it Theoretische Physik 1, Naturwissenschaftlich-Technische Fakult\"at,\\
Universit\"at Siegen, Walter-Flex-Strasse 3, D-57068 Siegen, Germany\\[0.2cm]}
$^{b}${\it Physik Department T31, James-Franck-Stra\ss e~1, \\
Technische Universit\"at M\"unchen, D--85748 Garching, Germany\\[0.2cm]}
$^{c}${\it Institute of Particle Physics and \\
Key Laboratory of Quark and Lepton Physics~(MOE),\\
Central China Normal University, Wuhan, Hubei 430079, P.\ R.\ China\\[0.2cm]}
\end{center}

\vspace{0.9cm}
\begin{abstract}
\vskip0.2cm\noindent
We complete the calculation of the QCD penguin amplitude at 
next-to-next-to-leading order in the QCD factorization approach 
to non-leptonic $B$-meson decays. This provides the last 
missing piece in the computation of the QCD correction to direct 
CP asymmetries at leading power in the heavy-quark expansion.

\end{abstract}
\end{titlepage}







%
%
%

\section{Introduction}
\label{sec:intro}

Direct CP violation arises from the interference of amplitudes 
with different CP-violating (CKM)  and rescattering phases. 
The values of the CKM parameters in the Standard Model (SM) 
imply that direct CP violating asymmetries are either very small or 
require the measurement of rare decays. In $B$-meson physics, 
rare decays to charmless final states provide the best opportunity 
to study such CP violation, when there is an interference of 
an amplitude generated primarily by a tree-level $W$-boson 
mediated process and a loop-induced $b\to D g^* \,(\to q\bar{q})$ 
($D=d,s$) amplitude, called the QCD penguin amplitude. 

The theoretical calculation of direct CP asymmetries 
is very challenging, since it is usually not 
possible to calculate the rescattering phases in a process 
involving hadrons. The best prospects are offered by 
charmless decays to two (pseudoscalar or vector) mesons, 
in which case the QCD factorization 
approach~\cite{Beneke:1999br,Beneke:2000ry,Beneke:2001ev}
provides a rigorous and systematic approximation to the 
non-leptonic decay amplitudes for the leading term in the 
heavy quark expansion.\footnote{
Larger direct CP asymmetries have been observed in three-body 
decays \cite{Aaij:2014iva}. However, these are caused by 
interference with strong phases which are generated by 
long-distance, hadronic resonance physics.} The matrix element of 
the effective Hamiltonian operators $Q_i$ (to be specified below) 
responsible for the 
$B\to M_1 M_2$ decay can be expressed as\footnote{
The overall sign refers to the case of two pseudoscalar 
final-state mesons. The factor 
of 1/4 arises in relation to the corresponding equation in 
\cite{Beneke:2009ek}, since the operators in the effective 
Hamiltonian (\ref{eq:Heff}) below are now defined with an 
overall factor $4 G_F/\sqrt{2}$ rather than $G_F/\sqrt{2}$. 
Note that this factor was missing in the corresponding 
equation in \cite{Bell:2015koa}.}
\begin{eqnarray}
\label{factformula}
\langle M_1 M_2 | Q_i | \bar{B} \rangle & = &
i \,\frac{m_B^2}{4} \,\bigg\{F^{BM_1}(0)
\int_0^1 \! du \; T_{i}^I(u) \, f_{M_2}\phi_{M_2}(u)
+ (M_1\leftrightarrow M_2) \nnb \\
&& \hspace*{-1.5cm}
+ \,\int_0^\infty \! d\omega \int _0^1 \! du dv \;
T_{i}^{II}(\omega,v,u) \, f_B \phi_B(\omega)  \; 
f_{M_1}\phi_{M_1}(v) \;
f_{M_2} \phi_{M_2}(u) \bigg\}
\end{eqnarray}
in terms of non-perturbative $B\to M$ form factors $F^{BM}(0)$, 
light-cone distribution amplitudes (LCDAs) 
$f_{M}\phi_{M}(u)$, and perturbatively calculable 
hard-scattering kernels $T_{i}^I(u)$, $T_{i}^{II}(\omega,v,u)$. 
Importantly, the rescattering phases are present only in 
the latter, hence direct CP violation can be calculated once 
the form factors and LCDAs are known. It also follows that 
direct CP asymmetries are either of ${\cal O}(\alpha_s)$, 
since phases arise from loop contributions to the kernels 
above, or of next-to-leading power $\mathcal{O}(\Lambda/m_b)$, 
where $\Lambda \ll m_b$ denotes the strong interaction scale, 
in the heavy-quark expansion. 
Since both parameters are $\mathcal{O}(1/10)$ it is a priori 
unclear whether the direct CP asymmetries in charmless decays 
are short- or long-distance dominated.

The calculation of the short-distance direct CP asymmetry has 
therefore been of long-standing interest. The first non-vanishing 
${\cal O}(\alpha_s)$ contribution has been known  
for a long time from the first QCD factorization 
calculations \cite{Beneke:1999br,Beneke:2001ev,Beneke:2003zv}. 
As usual, the next term in the $\alpha_s$ expansion is 
needed to check the reliability of the expansion and to reduce 
theoretical scale uncertainties. In the present case of 
direct CP asymmetries, this implies the next-to-next-to-leading 
order (NNLO) $\mathcal{O}(\alpha_s^2)$ correction to both interfering 
amplitudes. The  $\mathcal{O}(\alpha_s^2)$ correction to the 
spectator-scattering kernels $T_{i}^{II}(\omega,v,u)$ is 
somewhat simpler to compute, since it is a one-loop effect, 
and has been obtained in~\cite{Beneke:2005vv,Kivel:2006xc,Pilipp:2007mg} and~\cite{Beneke:2006mk} for the 
tree and the QCD and electroweak penguin amplitudes, respectively. 
The NNLO calculation of the kernel $T_{i}^I(u)$ of the 
form-factor term in (\ref{factformula}) has been performed about 
ten years ago~\cite{Bell:2007tv,Bell:2009nk,Beneke:2009ek} for 
the tree-induced amplitudes $T$, $C$, but only a partial result 
is currently available for the QCD penguin amplitude, $P$, from 
1) the one-loop matrix element of the chromomagnetic dipole 
operator $Q_{8g}$~\cite{Kim:2011jm}, 
and 2) the two-loop matrix elements 
of the current-current operators $Q_{1,2}^p$ \cite{Bell:2015koa}. 

In the present paper we present the calculation of the last 
missing, and most difficult piece of the NNLO computation, 
the matrix elements of the penguin operators  $Q_{3-6}$ 
in the effective weak Hamiltonian. Two-loop vertex 
and two-loop penguin diagrams contribute to these matrix 
elements, which we compute by extending previous results 
from \cite{Bell:2009nk,Beneke:2009ek,Bell:2014zya,Bell:2015koa}. 
The NNLO computation of the penguin amplitudes provides the 
basis for a comprehensive reanalysis of the phenomenology 
of all 130 final states of two pseudoscalar and/or vector mesons 
from the ground-state nonet \cite{Beneke:2003zv,Beneke:2006hg}.
This analysis is deferred to a separate study, while here 
we present only a numerical result of the penguin amplitude, 
that puts the new NNLO contribution into perspective.

The paper is organized as follows. In section~\ref{sec:theory} 
we lay out the theoretical framework and formulate the 
problem as a matching calculation of the matrix elements of 
operators from the effective weak Hamiltonian 
to a certain four-quark operator in soft-collinear effective 
theory (SCET), which factorizes into the product of the $B\to M$ 
form factor and the LCDA of the light meson at the 
matrix-element level. 
Section~\ref{sec:calculation} provides some technical details 
of the two-loop computations in QCD and SCET, and the structure 
of the result for the hard-scattering kernel before and 
after the convolution with the Gegenbauer expansion of the 
light-meson LCDA. The numerical size of the NNLO correction 
and the residual renormalization scale dependence of the 
full QCD penguin amplitude is evaluated in 
section~\ref{sec:amplitudes}. We conclude in 
section~\ref{sec:conclusion}. Two appendices summarize the 
renormalization constants for the operators from the effective 
weak Hamiltonian required for this work, 
and list numerical tables from which the convoluted kernels 
used in section~\ref{sec:amplitudes} can be reconstructed, 
including their dependence on the internal charm-quark mass. 

%
%
%

\section{Theoretical framework}
\label{sec:theory}


\subsection{Penguin amplitude in QCD factorization}
\label{sec:pengampsQCDF}

On the fundamental level, what is commonly called {\em the} QCD 
penguin amplitude in non-leptonic 
decays is generated by the loop-induced weak-interaction process 
$b\to D g^*$, where $D$ refers to a down or strange quark, followed 
by $g^*\to \sum_{q=u,d,s} q\bar{q}$, where the $q$ and $\bar{q}$ 
end up in different mesons in the final state. This basic 
process defined by the quark flavour configuration can be dressed 
by quark and gluon loops. 

Following the notation introduced in \cite{Beneke:2003zv}, the 
charmless two-body decay matrix element of the effective weak  
Hamiltonian can be decomposed in terms of CKM structures  
$\lambda_p^{(D)} = V_{pD}^* V_{pb}$ and flavour operators 
as 
\begin{equation}
\label{eq:toperators}
   \langle M_1'M_2'|{\cal H}_{\rm eff}|\bar B\rangle
   = \sum_{p=u,c} \lambda_p^{(D)}\,
   \langle M_1' M_2'|{\cal T}_A^p + {\cal T}_B^p|\bar B\rangle \,.
\end{equation}
The term ${\cal T}_A^p$ accounts for the flavour topologies of the 
form-factor and spectator-scattering terms in (\ref{factformula}), 
and ${\cal T}_B^p$ is reserved for the $1/m_b$ suppressed weak 
annihilation amplitudes.
The QCD penguin amplitude corresponds to 
\begin{eqnarray}
\label{alphaidef}
{\cal T}_A^p
   \supset \alpha_4^p(M_1 M_2)\,\sum_{q=u,d,s} 
A([\bar q_s q][\bar q D])\,, 
\end{eqnarray}
where $\bar q_s$ denotes the spectator anti-quark in the $\bar{B}$ 
meson. The coefficient $\alpha_4^p(M_1 M_2)$ contains the dynamical 
information, while the arguments of $A$ encode the flavour 
composition of the final state $M_1 M_2$ and hence determine the 
final states $M_1'M_2'$ to which the QCD penguin amplitude can 
contribute. 

The matrix element of ${\cal T}_A^p$ contains kinematical factors 
as well as the form factors and decay constants that appear in 
(\ref{factformula}), as defined in 
\cite{Beneke:2003zv,Beneke:2006hg}, such that $\alpha_4^p(M_1 M_2)$ 
is a dimensionless number composed of the Wilson coefficients 
$C_i$ of the operators $Q_i$ and the convolutions of the 
hard-scattering kernels with the meson LCDAs. 
In the QCD factorization approach $\alpha_4^p(M_1 M_2)$ 
is further divided into the quantities  $a_4^p$, $a_6^p$, 
\begin{equation}
\alpha_4^p(M_1 M_2) = a_4^p(M_1 M_2) \pm 
r_\chi^{M_2} a_6^p(M_1 M_2)
\label{eq:a4hat}
\end{equation}
where the plus (minus) sign applies to the decays where $M_1$ is
a pseudoscalar (vector) meson. We focus on 
$a_4^p(M_1 M_2)$ in this paper, which is
the only leading-power contribution in the heavy-quark expansion. 
The normalization of $a_4^p(M_1 M_2)$ is such that
\begin{equation}
a_4^p = \frac{1}{3} C_3 + \frac{4}{9} C_4+\frac{16}{3} C_5
+\frac{64}{9} C_6+\mathcal{O}(\alpha_s)\,,\qquad p=u,c
\end{equation}
in terms of the Wilson coefficients of the effective weak  
Hamiltonian given in the following subsection. We note that 
at tree-level $a_4^p(M_1 M_2)$ is independent of the final 
state and real. However, in higher orders in $\alpha_s$, 
the QCD penguin coefficient 
in QCD factorization depends on the identity of the final-state 
mesons through their LCDAs and acquires an imaginary part from 
loop corrections to the hard-scattering kernels. In the following 
we describe the calculation of and provide results for the 
$\mathcal{O}(\alpha_s^2)$ correction to  $a_4^p(M_1 M_2)$.


\subsection{Operator bases}
\label{sec:opbases}

The calculation is done in the framework of the effective weak 
Hamiltonian for $b\to D$ transitions, which is given by
\begin{align}
\calH_\text{eff} =
    \frac{4G_F}{\sqrt{2}} \; \sum_{p=u,c} V_{pD}^* V_{pb}
    \,\bigg( C_1 Q_1^p + C_2 Q_2^p
	+ \sum_{i=3}^{10} C_i Q_i
	+ C_{7\gamma} Q_{7\gamma}
    + C_{8g} Q_{8g} \bigg)
    + \text{h.c.}
\label{eq:Heff}
\end{align}
We adopt the CMM operator basis~\cite{Chetyrkin:1997gb}, where the 
current-current and QCD penguin operators are defined as
\begin{align}
Q_1^p &=
  (\bar p_L \gamma^\mu T^A b_L )\;
  (\bar D_L \gamma_\mu T^A p_L),\nnb\\
Q_2^p &=
  (\bar p_L \gamma^\mu b_L)\;
  (\bar D_L \gamma_\mu p_L),\nnb\\
Q_3 &=
  (\bar D_L \gamma^\mu b_L) \;
  {\ts \sum_q}\; (\bar q \gamma_\mu q),\nnb\\
Q_4 &=
  (\bar D_L \gamma^\mu T^A b_L)\;
  {\ts \sum_q}\; (\bar q \gamma_\mu T^A q),\nnb\\
Q_5 &=
  (\bar D_L \gamma^\mu\gamma^\nu\gamma^\rho b_L) \;
  {\ts \sum_q}\; (\bar q \gamma_\mu\gamma_\nu\gamma_\rho q),\nnb\\
Q_6 &=
  (\bar D_L \gamma^\mu\gamma^\nu\gamma^\rho T^A b_L) \;
  {\ts \sum_q}\; (\bar q \gamma_\mu\gamma_\nu\gamma_\rho T^A q) \, .
\label{eq:cmmbasis}
\end{align}
Here the sums run over the five quark flavours $q = u$, $d$, $s$, 
$c$, $b$. The electroweak penguin operators $Q_{7-10}$ and the 
electromagnetic dipole operator $Q_{7\gamma}$ are irrelevant for the 
QCD penguin amplitude. Our 
definition of the chromo\-magnetic dipole operator, 
\begin{align}
Q_{8g} &= \frac{-g_s}{32\pi^2} \,\overline{m}_b \; 
\bar D \sigma_{\mu\nu} (1+\gamma_5) G^{\mu\nu} b,
\label{eq:Q8gdef}
\end{align}
corresponds to the sign convention $iD_\mu = i\partial_\mu + 
g_s A^A_\mu T^A$, and $\overline{m}_b$ denotes the bottom quark 
mass in the $\overline{\rm MS}$ scheme at the scale $\mu$.

In dimensional regularization the operator basis has to be supplemented by a set of evanescent operators. In the CMM basis the one-loop evanescent operators are defined as
\begin{align}
E_1^{(1),p} &=
  (\bar p_L \gamma^{\mu_1}\gamma^{\mu_2}\gamma^{\mu_3} T^A b_L) \;
  (\bar D_L \gamma_{\mu_1}\gamma_{\mu_2}\gamma_{\mu_3} T^A p_L) - 16 Q_1^p,\nnb\\
E_2^{(1),p} &=
  (\bar p_L \gamma^{\mu_1}\gamma^{\mu_2}\gamma^{\mu_3} b_L) \;
  (\bar D_L \gamma_{\mu_1}\gamma_{\mu_2}\gamma_{\mu_3} p_L) - 16 Q_2^p,\nnb\\
E_3^{(1)} &=
  (\bar D_L \gamma^{\mu_1}\gamma^{\mu_2}\gamma^{\mu_3}\gamma^{\mu_4}\gamma^{\mu_5} b_L) \;
  {\ts \sum_q}\; (\bar q \gamma_{\mu_1}\gamma_{\mu_2}\gamma_{\mu_3}\gamma_{\mu_4}\gamma_{\mu_5} q) + 64 Q_3 - 20 Q_5,\nnb\\
E_4^{(1)} &=
  (\bar D_L \gamma^{\mu_1}\gamma^{\mu_2}\gamma^{\mu_3}\gamma^{\mu_4}\gamma^{\mu_5} T^A b_L) \; 
  {\ts \sum_q}\; (\bar q \gamma_{\mu_1}\gamma_{\mu_2}\gamma_{\mu_3}\gamma_{\mu_4}\gamma_{\mu_5} T^A q) + 64 Q_4 - 20 Q_6 \, . \label{eq:oneloopevanescent} 
\end{align}
At the two-loop level four more evanescent operators arise, defined as~\cite{Gorbahn:2004my} 
\begin{align} 
E_1^{(2),p} & =
  (\bar p_L \gamma^{\mu_1}\gamma^{\mu_2}\gamma^{\mu_3}\gamma^{\mu_4}\gamma^{\mu_5} T^A b_L) \; 
  (\bar D_L \gamma_{\mu_1}\gamma_{\mu_2}\gamma_{\mu_3}\gamma_{\mu_4}\gamma_{\mu_5} T^A p_L)  - 256 Q_1^p - 20 E^{(1),p}_1 \, , \nnb \\   
E_2^{(2),p} & =
  (\bar p_L \gamma^{\mu_1}\gamma^{\mu_2}\gamma^{\mu_3}\gamma^{\mu_4}\gamma^{\mu_5} b_L) \; 
  (\bar D_L \gamma_{\mu_1}\gamma_{\mu_2}\gamma_{\mu_3}\gamma_{\mu_4}\gamma_{\mu_5} p_L)  - 256 Q_2^p - 20 E^{(1),p}_2 \, , \nnb \\    
E_3^{(2)} & =
  (\bar D_L \gamma^{\mu_1}\gamma^{\mu_2}\gamma^{\mu_3}\gamma^{\mu_4}\gamma^{\mu_5}\gamma^{\mu_6}\gamma^{\mu_7} b_L)
  {\ts \sum_q} (\bar{q} \gamma_{\mu_1}\gamma_{\mu_2}\gamma_{\mu_3}\gamma_{\mu_4}\gamma_{\mu_5}\gamma_{\mu_6}\gamma_{\mu_7} q) + 1280 Q_3 - 336 Q_5 \, , \nnb \\
E_4^{(2)} & = (\bar D_L \gamma^{\mu_1}\gamma^{\mu_2}\gamma^{\mu_3}\gamma^{\mu_4}\gamma^{\mu_5}\gamma^{\mu_6}\gamma^{\mu_7} T^A b_L)
  {\ts \sum_q} (\bar{q} \gamma_{\mu_1}\gamma_{\mu_2}\gamma_{\mu_3}\gamma_{\mu_4}\gamma_{\mu_5}\gamma_{\mu_6}\gamma_{\mu_7} T^A q) \nnb \\
  & \;\;\;\; + 1280 Q_4 - 336 Q_6 \, . 
\label{eq:twoloopevanescent}
\end{align}

\begin{figure}[t]
\begin{center}
 \includegraphics[width=0.3\textwidth]{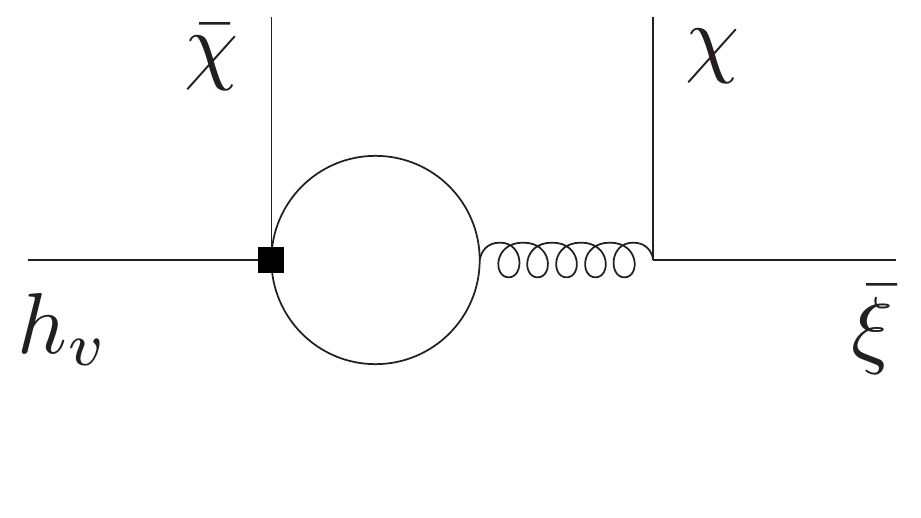}
\end{center}
\vs{-30}
\caption{\label{fig:SCETfields} Fermion lines and associated SCET and 
HQET fields. The black square denotes an insertion of an operator from
the effective weak Hamiltonian.}
\end{figure}

The hard-scattering kernels $T_i^I(u)$ in (\ref{factformula}) are 
determined from matching the QCD matrix elements on the left-hand 
side of the equation to four-quark operators in SCET. Any four-vector 
in SCET can be  decomposed as 
\begin{align}
p^\mu &= (n_{+}p)\frac{n_{-}^\mu}{2} + (n_{-}p)\frac{n_{+}^\mu}{2} 
+ p_\perp^\mu \, , \label{eq:fourvectorscet}
\end{align}
where the two light-like vectors $n_{\mp}^\mu$ which define the collinear 
and anti-collinear directions, satisfy $n_{-}n_{+} =2$.
Eq.~(\ref{eq:fourvectorscet}) also defines the perpendicular 
component. Collinear (anti-collinear) modes have momenta with large 
component $n_+ p = \mathcal{O}(m_b)$ ($n_- p = \mathcal{O}(m_b)$). 

We denote the collinear and anti-collinear SCET fields by $\xi$ and 
$\chi$, respectively, see figure~\ref{fig:SCETfields}.
It turns out that there is only one physical SCET operator in our 
problem. It has the fermion contraction $(\bar\chi\chi)(\bar\xi h_v)$ 
and is given by
\begin{align}
O_1 &= \sum_{q=u,d,s} (\bar \chi_D \,\frac{\nms}{2} (1-\gamma_5)\chi_q) 
\; (\bar \xi_q \, \nps (1-\gamma_5) h_v) \, , \label{eq:scetbasis1}
\end{align}
where $h_v$ is the heavy-quark field in heavy-quark effective 
theory (HQET). 
In contrast, the diagrams relevant to the penguin amplitude $a_4^p$ lead to operators where the fermion lines are contracted in a
different Fierz ordering, $(\bar \xi \chi)(\bar\chi h_v)$, and are therefore of the ``wrong-insertion'' type (see~\cite{Beneke:2009ek}).
The corresponding wrong-insertion SCET operators are conveniently chosen as\footnote{Note that the $\widetilde O_n$ are different from
those for the colour-suppressed tree amplitude in~\cite{Beneke:2009ek}.}
\begin{align}
\widetilde O_n &=
\sum_{q=u,d,s} (\bar \xi_q \,
\gamma_{\perp}^{\alpha}\gamma_{\perp}^{\mu_1}\gamma_{\perp}^{\mu_2}
\ldots \gamma_{\perp}^{\mu_{2 n-2}} \chi_q) \;
(\bar \chi_D (1+\gamma_5)
\gamma_{\perp\alpha}\gamma_{\perp\mu_{2 n-2}}\gamma_{\perp\mu_{2n-3}}
\ldots\gamma_{\perp\mu_1} h_v) \, ,
\label{eq:scetbasis2}
\end{align}
where we will need $n$ up to 4 (strings with seven $\gamma$ matrices in
each bilinear). The operators $\widetilde O_n$ are evanescent
for $n>1$, while $\widetilde O_1$ is Fierz-equivalent to $O_1/2$ in four
dimensions. We therefore add $\widetilde O_1-O_1/2$ as another evanescent
operator. In the equations above we omitted the
Wilson lines necessary to make the SCET operators, which are
non-local on the light-cone \cite{Beneke:2005vv}, gauge-invariant.


\subsection{QCD matrix elements}
\label{sec:QCDmatel}

\begin{figure}[t]
\centerline{\includegraphics[width=0.6\textwidth]
{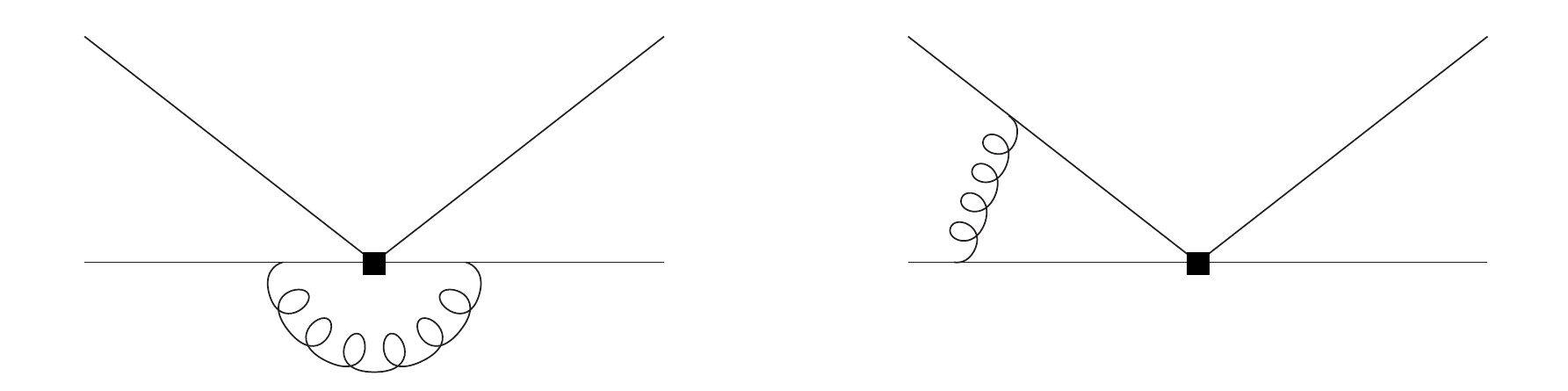}}
\vskip0.0cm
\caption{Examples of factorizable (left panel) and non-factorizable 
(right panel) diagrams.}
\label{fig:facnonfac}
\end{figure}

The renormalized matrix element of a QCD operator $Q_i$ has
the perturbative expansion
\bea
\braket{Q_i} &=& \bigg\{ \widetilde A_{ia}^{(0)} + \frac{\as}{4\pi} 
\left[ \widetilde A_{ia}^{(1)} + Z_{ext}^{(1)} \, \widetilde A_{ia}^{(0)} + 
Z_{ij}^{(1)} \widetilde A_{ja}^{(0)}\right]  \nnb \\
&& \hspace*{0.4cm}+ \,\left(\frac{\as}{4\pi}\right)^2 \left[
\widetilde A_{ia}^{(2)} + Z_{ij}^{(1)} \widetilde A_{ja}^{(1)}
 + Z_{ij}^{(2)} \widetilde A_{ja}^{(0)} + Z_{ext}^{(1)} \,
 \widetilde A_{ia}^{(1)} + Z_{ext}^{(2)} \, \widetilde A_{ia}^{(0)}\right. \nnb 
\\
&&\hspace*{0.4cm} 
\left. + \,Z_{ext}^{(1)} \, Z_{ij}^{(1)} \widetilde A_{ja}^{(0)} + 
Z_{\alpha}^{(1)} \widetilde A_{ia}^{(1)} + 
\, (-i) \, \delta m^{(1)} \, \widetilde A^{\prime (1)}_{ia}\right] + 
{\cal O}(\as^3) \bigg\} \, \braket{\widetilde O_a}^{(0)} \, .
\label{eq:QCDside}
\eea
Throughout the paper, $\alpha_s \equiv \alpha_s(\mu)$ denotes the
five-flavour strong coupling in the $\overline{\rm MS}$ scheme.
Moreover, a superscript on any quantity $G$ labels the order
in $\alpha_s$ according to
\bea
G &=& G^{(0)} + \frac{\alpha_s}{4\pi} \, G^{(1)} + 
\left(\frac{\alpha_s}{4\pi}\right)^2 \, G^{(2)} + {\cal O}(\alpha_s^3) \, .
\label{eq:expansion}
\eea
The $\widetilde A_{ia}^{(\ell)}$ in~(\ref{eq:QCDside}) denote bare $\ell$-loop 
on-shell matrix elements of QCD operators. For $\ell=2$ we only need matrix
elements of the physical operators, but for $\ell < 2$ the tree-level and one-loop matrix
elements of evanescent operators are required in addition
since in terms such as $Z_{ij}^{(1)} \widetilde A_{ja}^{(1)}$ the sum over $j$
includes the evanescent operators.
For the matching procedure it turns out to be convenient to split up 
the amplitudes $\widetilde A_{ia}^{(\ell)}$ further into factorizable and
non-factorizable diagrams, see figure~\ref{fig:facnonfac},
\bea
\widetilde A_{ia}^{(\ell)}&=& \widetilde A_{ia}^{(\ell){\rm f}}
+ \widetilde A_{ia}^{(\ell){\rm nf}} \, .
\eea

The renormalization factors $Z_{ij}$, $Z_{\alpha}$, $\delta m$, 
and $Z_{ext}$ account for operator, coupling, mass, and wave-function
renormalization, respectively. The coupling is renormalized 
in the $\overline{{\rm MS}}$ scheme, whereas the mass and the
fields are renormalized in the on-shell scheme.
In the matrices $Z_{ij}$ the row index runs over $i = \{Q_1^p,Q_2^p,Q_{3-6},Q_{8g}\}$,
while the column index $j$ labels $\{Q_1^p,Q_2^p,Q_{3-6},Q_{8g}
,E_1^{(1),p},E_2^{(1),p},E_3^{(1)},E_4^{(1)}
,E_1^{(2),p},E_2^{(2),p},E_3^{(2)},E_4^{(2)}
\}$. They were computed in~\cite{Gambino:2003zm,Gorbahn:2004my}
but need to be adjusted to our operator basis. We give the  
matrices $Z_{ij}$ explicitly in appendix~\ref{sec:matricesZij}.


\subsection{SCET matrix elements and hard-scattering 
kernels}
\label{sec:SCETmatel}

On the SCET side the matrix elements of $\widetilde O_a$ have a simpler 
structure. Once dimensional regularization is used as infrared regulator the 
on-shell renormalization constants are equal to unity, and the bare matrix 
elements $\widetilde M_{ab}^{(\ell)}$ are non-zero only for $\ell > 1$. At 
two loops, only the diagrams with a massive charm quark-loop insertion into 
the gluon line contribute. Moreover, there is no mass counterterm for a 
HQET quark. One therefore obtains
\bea
\braket{\widetilde O_a} &=& \left\{ \delta_{ab} + 
\frac{\hat{\alpha}_s}{4\pi} \, \widetilde Y_{ab}^{(1)} + 
\left(\frac{\hat{\alpha}_s}{4\pi}\right)^{\!2} \left[ \widetilde M_{ab}^{(2)} +
\widetilde Y_{ab}^{(2)}\right] + {\cal O}(\hat{\alpha}_s^3) \right\} \, \braket{\widetilde O_b}^{(0)} \, .
\label{eq:SCETsidedimreg}
\eea
Here $\hat{\alpha}_s$ denotes the strong coupling in the four-flavour 
theory. Since the non-local SCET operators depend on a variable, the 
renormalization factors $\widetilde Y_{ab}=\widetilde Y_{ab}(u,u')$ are 
functions of two variables and the product in (\ref{eq:SCETsidedimreg}) 
has to be interpreted as a convolution. In the procedure of determining 
the ultraviolet operator renormalization factors $\widetilde Y_{ab}$ one 
has to regulate infrared divergences other than dimensionally. 
While the physical operator $O_1$ is minimally subtracted in the 
$\overline{\rm MS}$ scheme, the evanescent operators are renormalized 
such that their matrix elements with a non-dimensional infrared regulator 
vanish, so that eventually after renormalization and matching 
they can be dropped in the evaluation of a physical decay amplitude. 
Once the $\widetilde Y_{ab}$ are at hand,
one can use~(\ref{eq:SCETsidedimreg}) with dimensional regularization
as infrared and ultraviolet regulator for the on-shell matrix elements
of the $\widetilde O_a$. 

The hard-scattering kernels $\widetilde T_i$ of the operators $Q_i$, 
$i=\{1p,2p,\text{3--6},8g\}$ are then extracted by matching the 
QCD operators onto SCET,
\begin{equation}
\braket{Q_i} = \sum\limits_a \, \widetilde H_{ia}  \, 
\braket{\widetilde O_a}\, ,
\label{eq:WI}
\end{equation}
whose steps are explained in detail in~\cite{Beneke:2009ek} and shall not 
be repeated here. At the end of this procedure we obtain a master formula 
for the expansion of the wrong-insertion hard-scattering kernels, which 
is a generalization of the master formula given in~\cite{Beneke:2009ek}
to the case when the tree-level matching of the $Q_i$ involves evanescent 
SCET operators. For the tree, one-loop and two-loop kernels, we obtain 
\begin{eqnarray}
\frac{1}{2}\,\widetilde T_i^{(0)} &=& \widetilde A^{(0)}_{i1} \, , 
\label{eq:mastertree}\\[0.1cm]
\frac{1}{2}\,\widetilde T_i^{(1)} &=& \widetilde A^{(1){\rm nf}}_{i1}
 + Z_{ij}^{(1)} \, \widetilde A^{(0)}_{j1}
+ \underbrace{\widetilde A^{(1){\rm f}}_{i1} - A^{(1){\rm f}}_{31} \, 
 \widetilde A^{(0)}_{i1}}_{{\cal{O}}(\eps)}
- \underbrace{[\widetilde Y_{11}^{(1)}-Y_{11}^{(1)}]\, 
 \widetilde A^{(0)}_{i1}}_{{\cal{O}}(\eps)} 
- \underbrace{\sum_{b>1} \widetilde A_{ib}^{(0)} \, 
 \widetilde Y_{b1}^{(1)}}_{{\cal{O}}(\eps)} \, ,
\quad\qquad  \label{eq:master1loop} \\[-0.4cm]
\frac{1}{2}\,\widetilde T_i^{(2)} &=& \widetilde A^{(2){\rm nf}}_{i1} + 
Z_{ij}^{(1)} \, \widetilde A^{(1)}_{j1} + Z_{ij}^{(2)} \, 
\widetilde A^{(0)}_{j1}
+ Z_{\alpha}^{(1)} \, \widetilde A^{(1){\rm nf}}_{i1}\nnb \\[0.2cm]
&& + \, (-i) \, \delta m^{(1)} \, \widetilde A^{\prime (1){\rm nf}}_{i1}
+ Z_{ext}^{(1)} \, \big[\widetilde A^{(1){\rm nf}}_{i1}+ Z_{ij}^{(1)} \, 
\widetilde A^{(0)}_{j1}\big]\nnb \\[0.2cm]
&& - \,\frac{1}{2}\,\widetilde T_i^{(1)} \big[  C_{FF}^{(1)} + \widetilde Y_{11}^{(1)}\big] 
- \sum_{b>1} \widetilde H_{ib}^{(1)} \, \widetilde Y_{b1}^{(1)} \nnb\\[0.1cm]
&& + \,[\widetilde A^{(2){\rm f}}_{i1} - A^{(2){\rm f}}_{31} \, 
\widetilde A^{(0)}_{i1}] + \, (-i) \, \delta m^{(1)} \, 
[\widetilde A^{\prime (1){\rm f}}_{i1}
- A^{\prime (1){\rm f}}_{31} \, \widetilde A^{(0)}_{i1}]\nnb\\[0.2cm]
&& + \,(Z_{\alpha}^{(1)}+Z_{ext}^{(1)})\,
[\widetilde A^{(1){\rm f}}_{i1} - A^{(1){\rm f}}_{31} \, 
\widetilde A^{(0)}_{i1}]\nnb \\[0.2cm]
&& - \,[\widetilde M^{(2)}_{11} - M^{(2)}_{11} ] \, 
\widetilde A^{(0)}_{i1} \nnb \\[0.2cm]
&& - \,(C_{FF}^{(1)}-\xi_{45}^{(1)})\, 
[\widetilde Y_{11}^{(1)}-Y_{11}^{(1)}] \, \widetilde A^{(0)}_{i1} - 
[\widetilde Y_{11}^{(2)}-Y_{11}^{(2)}]\,
\widetilde A^{(0)}_{i1}  \nnb \\[0.2cm]
&& - \,\sum_{b>1} \widetilde A_{ib}^{(0)} \, \widetilde M_{b1}^{(2)} 
- \,\sum_{b>1} \widetilde A_{ib}^{(0)} \, \widetilde Y_{b1}^{(2)} \, . \label{eq:master2loop}
\end{eqnarray}
The factor $1/2$ on the left-hand side appears, since quantities 
such as $\widetilde A^{(n)}_{i1}$ are defined as the coefficients 
of $\widetilde{O}_1$, which after renormalization is replaced by 
$O_1/2$. Together with a factor of 4 from extracting the factor 
1/4 in (\ref{factformula}) this implies 
$\widetilde T_i^{(0)} = 2\widetilde A^{(0)}_{i1}$ 
etc.\footnote{Note that this factor 1/2 when ever $\widetilde T_i^{(\ell)}$ 
appears is missing 
in eqs.~(7)--(9) in \cite{Bell:2015koa}.} 

To arrive at (\ref{eq:master2loop}) we traded the four-flavour 
coupling for the 
five-flavour one by means of the $D$-dimensional relation
$\hat\alpha_s = \xi_{45}^{-1} \alpha_s$, where $\xi_{45}=1+\mathcal{O}(\alpha_s)$ is 
given explicitly in~\cite{Beneke:2008ei}.
The matching coefficient $C_{FF}$ can be determined from matching calculations for 
the $b\to u$ transition~\cite{Bonciani:2008wf,Asatrian:2008uk,Beneke:2008ei,Bell:2008ws}
and reads ($L = \ln(\mu^2/m_b^2)$)
\bea
C_{FF} &=& 1 - \frac{\as}{4\pi} \, \frac{C_F}{12} \left(6 L^2+30 L+\pi^2+72\right) + 
{\cal O}(\as^2) \, .
\eea
The terms $A^{(\ell){\rm f}}_{31}$ denote $\ell$-loop factorizable matrix elements of $Q_3$ of the
\emph{right-insertion} type~\cite{Beneke:2009ek}, whose result is proportional to the tree-level matrix element of
the physical SCET operator $O_1$.

All except three terms in~\eqref{eq:mastertree}~--~\eqref{eq:master2loop} 
have an analogue in the wrong-insertion master formula for the tree 
amplitudes, which was discussed at length in~\cite{Beneke:2009ek}. The 
last term in~\eqref{eq:master1loop} and the last line 
in~\eqref{eq:master2loop} are new. They stem from tree-level matrix 
elements of the $Q_i$ proportional to evanescent SCET operators, 
convoluted with SCET matrix elements and renormalization constants that 
describe the mixing of evanescent with physical SCET operators.
Such non-vanishing tree-level SCET evanescent operator contributions can 
appear whenever the operator in $\calH_\text{eff}$ contains a 
fermion bilinear with more than one Dirac matrix as is the case 
for $Q_{5,6}$ and the evanescent QCD operators in 
(\ref{eq:oneloopevanescent}), (\ref{eq:twoloopevanescent}).

%
%
%

\section{NNLO calculation}
\label{sec:calculation}


\subsection{Operator insertions and diagrams}
\label{sec:insertionsdiagrams}

\begin{figure}[t]
\begin{center}
 \includegraphics[width=.9\textwidth]{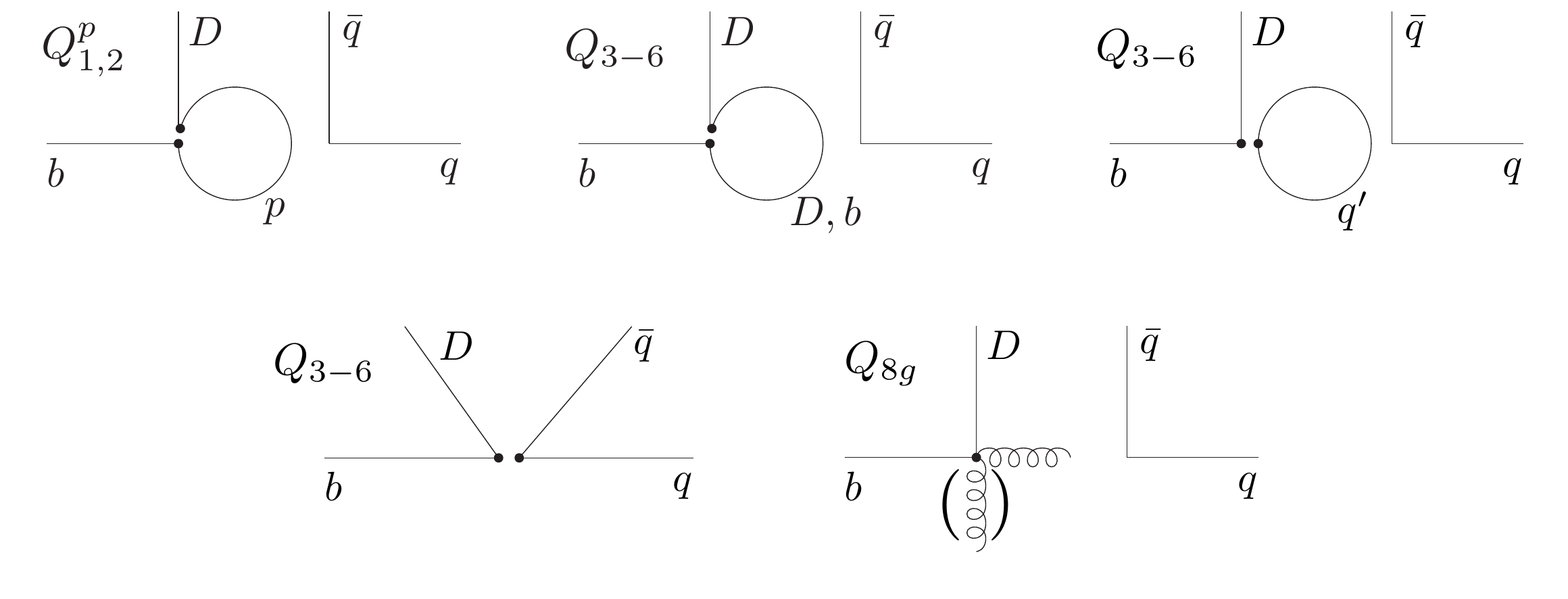}
\end{center}
\vs{-20}
\caption{\label{fig:insertions} Graphical illustration of the various 
operator insertions. The symbols stand for
$p \in \{u,c\}$, $D \in \{d,s\}$, $q \in \{u,d,s\}$ and $q^\prime \in \{u,d,s,c,b\}$. The black dots denote the operator insertion from the effective
weak Hamiltonian. Disconnected parts are understood to be connected by 
gluons. }
\end{figure}

Several insertions of the operators from the effective weak Hamiltonian 
contribute to the computation of the penguin amplitudes at higher orders. 
They are depicted in figure~\ref{fig:insertions}. 
Whereas the current-current operators $Q_{1,2}^p$ can be inserted
in a single manner only (see first panel of figure~\ref{fig:insertions}), there 
exist several ways of inserting the QCD penguin operators $Q_{3-6}$. Besides the ``penguin-type'' contractions
in the second and third panel of figure~\ref{fig:insertions}, also insertions into
the ``tree-type'' diagrams have to be considered, see the lower left panel of figure~\ref{fig:insertions}.
Finally, there is the insertion of $Q_{8g}$ whose contribution at a given order in $\alpha_s$
involves one loop less compared to the other operators. For the counterterm contribution of the
master formulas~\eqref{eq:mastertree}~--~\eqref{eq:master2loop} also insertions of evanescent QCD operators
are required, which are not shown in the figure. Keeping track of all these insertions leads to quite
some bookkeeping during the calculation.

The leading order (LO) and next-to-leading order (NLO) contributions 
to the QCD penguin amplitudes have been known since 
long~\cite{Beneke:1999br,Beneke:2001ev,Beneke:2003zv}. Also the 
calculation of the NNLO  $\calO(\as^2)$ correction involving 
one-loop spectator scattering dates back more than a 
decade~\cite{Beneke:2006mk}. The first NNLO calculation  
of a vertex correction to the leading QCD penguin amplitudes 
was the one-loop $\calO(\as^2)$ insertion of the 
chromomagnetic dipole operator $Q_{8g}$~\cite{Kim:2011jm}.
More recently the current-current operator contribution has been completed~\cite{Bell:2015koa},
and in the present work we compute the remaining insertions from figure~\ref{fig:insertions} at NNLO.

\begin{figure}[t]
\begin{center}
 \includegraphics[width=.95\textwidth]{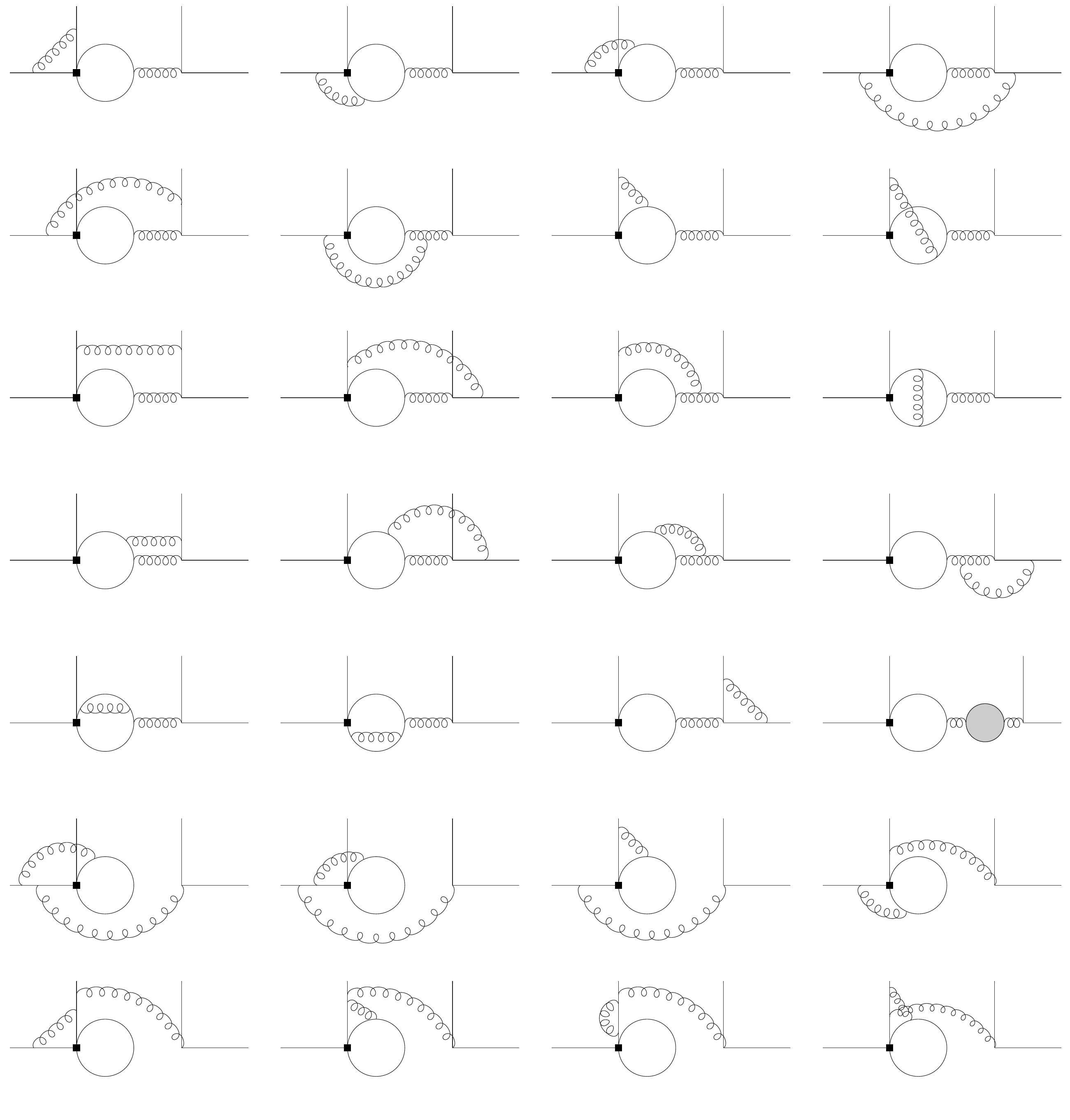}
\end{center}
\vs{-25}
\caption{\label{fig:penguins} Two-loop penguin diagrams I.}
\end{figure}

\begin{figure}[t]
\begin{center}
 \includegraphics[width=.95\textwidth]{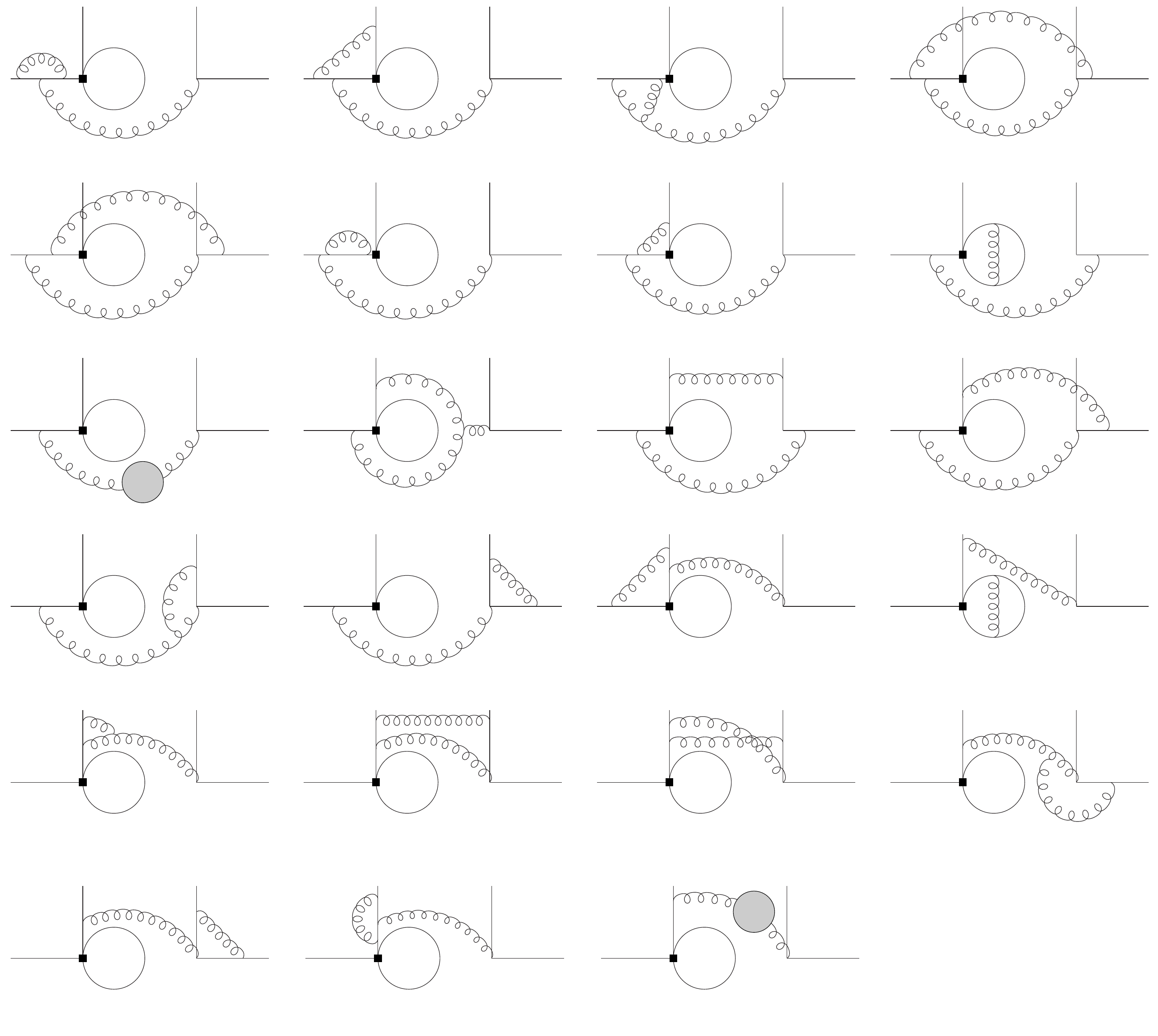}
\end{center}
\vs{-25}
\caption{\label{fig:penguins0} Two-loop penguin diagrams II. The sum of 
these diagrams vanishes for all operator insertions.}
\end{figure}

\begin{figure}[t]
\begin{center}
 \includegraphics[width=.95\textwidth]{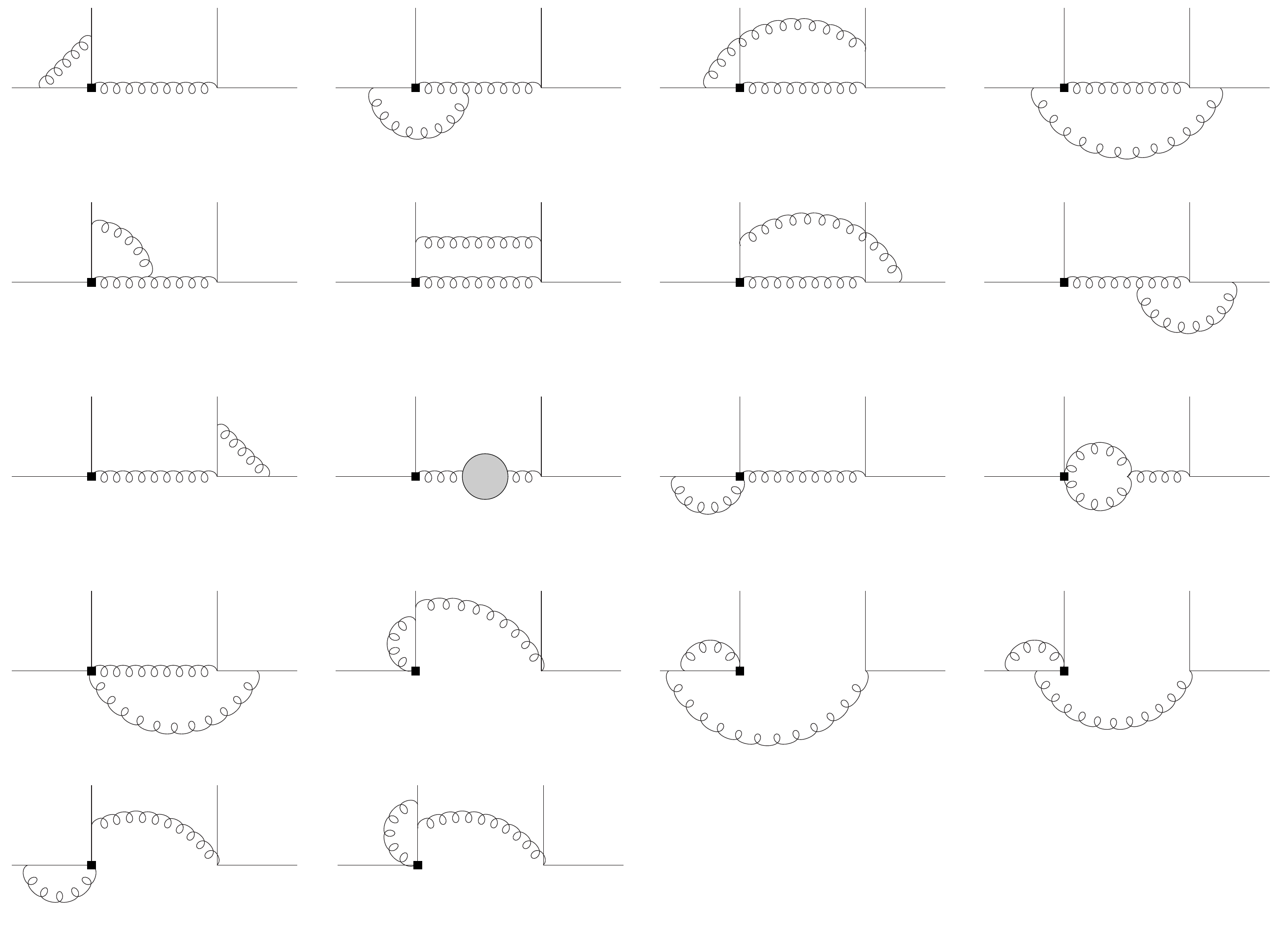}
\end{center}
\vs{-25}
\caption{\label{fig:dipoles} One-loop diagrams of the chromomagnetic
dipole operator $Q_{8g}$.}
\end{figure}

There are in total more than a hundred diagrams that one has to compute 
at NNLO. Those of the ``penguin-type'' contractions are shown in 
figures~\ref{fig:penguins} and~\ref{fig:penguins0}. We found
after explicit calculation that the sum of all diagrams in 
figure~\ref{fig:penguins0} vanishes for all operator insertions. In 
addition, one has to insert the QCD penguin operators $Q_{3-6}$ into the
``tree-type'' diagrams which are shown explicitly in section~5 
of~\cite{Beneke:2000ry}. Finally, the one-loop diagrams for the 
chromomagnetic dipole operator $Q_{8g}$ are depicted in 
figure~\ref{fig:dipoles}.

\subsection{Details of the two-loop calculation}
\label{sec:2loopcalc}

To obtain the vertex kernels $T_i^I(u)$ the quark matrix 
elements $\langle D(uq)\bar{q}(\bar{u}q) q(p)|Q_i|b(p_b)\rangle$ 
must be calculated 
at the two-loop order. Some general kinematic features can be seen 
from the one-loop penguin contraction:
\begin{center}
 \includegraphics[width=0.3\textwidth]{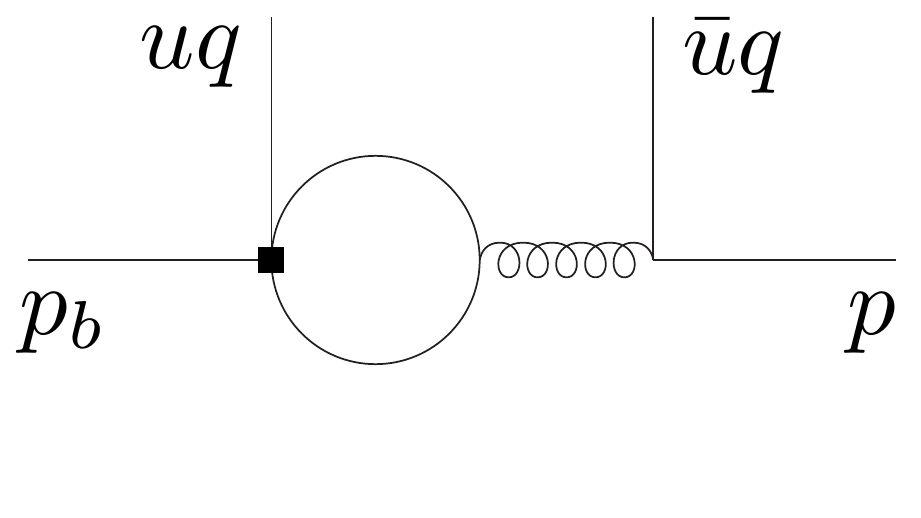}
\end{center}
\vs{-25}

The light quark flavours are taken to be massless and therefore the 
particle in the fermion loop (solid circle) can have mass $m_f=0$ (light 
quarks), $m_f=m_c$ (charm quark) or $m_f=m_b$ (bottom quark). The external 
states are on-shell and satisfy 
$p_b^2=m_b^2$ and $p^2=q^2=0$. The quark that goes into meson $M_2$ 
carries momentum fraction $u \in [0,1]$ of $q$, the anti-quark the 
remaining fraction $\ub \equiv 1-u$. There is a kinematic threshold 
at $\ub =4 m_c^2/m_b^2$ in the charm-loop diagrams.

The problem at hand is a genuine two-scale problem with dimensionless 
quantities $\ub$ and $z_f = (m_f^2-i\eta)/m_b^2$, where the 
infinitesimally small quantity $\eta>0$ determines the sign of the analytic 
continuation. For later convenience
we also define the following additional kinematic variables~\cite{Bell:2014zya}
\begin{align}
r & =\sqrt{1-4 z_c} \, , & s&=\sqrt{1-4 z_c/\ub} \, , \nnb \\[0.1em]
s_1&=\sqrt{1-4z_b/\ub} \, , & s_u & =\sqrt{1-4 z_c/u} \, , \nnb \\[0.1em]
s_{u,1} &=\sqrt{1-4z_b/u} \, , && \nnb \\[0.1em]
v &= \frac{1 + s_1}{2} + \frac{1 - s_1}{2} \sqrt{1 + \frac{8 (1 + s_1) z_c}{(1 - s_1)^2}} \, , &
t &= \frac{1 - s_1}{2} + \frac{1 + s_1}{2} \sqrt{1 + \frac{8 (1 - s_1) z_c}{(1 + s_1)^2}} \, ,  \nnb \\[0.1em]
v_0 &= \frac{1 - i \sqrt{3}}{2} \, r + \frac{1+i \sqrt{3}}{2} = v(u=0) \, , &
t_0 &= \frac{1 + i \sqrt{3}}{2} \, r + \frac{1-i \sqrt{3}}{2} = t(u=0) \, ,  \nnb \\[0.1em]
p & = \frac{1 - \sqrt{u^2 + 4 \ub z_c}}{\ub} \, , & p_u & = \frac{1 - \sqrt{\ub^2 + 4 u z_c}}{u} \, . \label{eq:kinematics}
\end{align}

The methods that we apply in the two-loop calculation have become 
standard in contemporary advanced multi-loop computations. We work in 
dimensional regularization with $D=4-2\eps$, where ultraviolet and infrared 
(soft and collinear) divergences appear as poles in $\ep$. We first apply a 
Passarino~--~Veltman~\cite{Passarino:1978jh} reduction to the tensor structure 
of the amplitude. The Dirac and colour algebra is then performed by
means of in-house routines. Subsequently, the dimensionally 
regularized scalar integrals are reduced to master integrals using the Laporta 
algorithm~\cite{Laporta:1996mq,Laporta:2001dd} based on 
integration-by-parts~(IBP) identities~\cite{Tkachov:1981wb,Chetyrkin:1981qh}. 
To this end we use the package {\tt FIRE}~\cite{Smirnov:2008iw} and an in-house routine.
The master integrals that stem from the insertion of penguin operators into the
``tree-type'' diagrams are known analytically since long~\cite{Bell:2007tv,Bell:2009nk,Huber:2009se,Beneke:2009ek} and evaluate
to harmonic polylogarithms (HPLs)~\cite{Remiddi:1999ew}. Those that come from the reduction of the
diagrams in figures~\ref{fig:penguins} and~\ref{fig:penguins0} were computed analytically
in~\cite{Bell:2014zya} in terms of iterated integrals over generalized weight functions.

In the notation of~\cite{Bell:2014zya} the generalized HPLs are defined as
\begin{align}\label{eq:defHPL}
H_{\vec0_n}(x) & = \frac{1}{n!} \ln^n (x) \, , \nnb \\
H_{a_1 , a_2 , \ldots, a_n }(x) & = \int_0^x dt \; f_{a_1}(t) \, H_{a_2, \ldots, a_n }(t)  \, .
\end{align}
For the weight functions we have $f_{0}(x) = 1/x$, while for any expression 
with $w \neq 0$ we define
\begin{align}\label{eq:genweights}
{f_{w}(x) = \frac{1}{w-x}}\,, \qquad {f_{-w}(x) = \frac{1}{w+x}}\, .
\end{align}
Moreover it turns out to be convenient to define the linear combinations
\begin{align}
f_{w^+}(x) = & f_{w}(x) + f_{-w}(x) = \frac{2w}{w^2-x^2}\,, \nnb \\
f_{w^-}(x) = & f_{w}(x) - f_{-w}(x) = \frac{2x}{w^2-x^2}\,. \label{eq:genpm}
\end{align}
In our calculation, we encounter the following expressions for $w$,
\begin{align}
w_1 = & \,1\, , & w_4 &= 1 + \sqrt{1-r^2}\, , & 
w_6 &= {w_2}_{\big| z_c \to z_b}  \, , \nnb \\
w_2 = & \,r\, , & w_5 &= 1 - \sqrt{1-r^2}\, , & 
w_7 &= {w_4}_{\big| z_c \to z_b} \, , \nnb \\
w_3 = & \,\frac{r^2+1}{2} \, .& & 
\end{align}
Finally there is one new master integral that stems from the
one-loop diagrams in figure~\ref{fig:dipoles}, whose analytic 
result reads
\begin{align}
  & \int\!\frac{d^Dk}{(2\pi)^D} \, \frac{u}{\left[k^2\right] \, \left[(k+u q)^2-m_b^2\right] \, \left[(k+p_b)^2-m_b^2\right]} \\[0.6em]
  &= -(i\ESGamma) \, (m_b^2)^{-1-\ep} \, \Gamma(1-\ep) \Gamma(1+\ep)
     \left(3 H_{w_1^+,w_1^+}(s_1)-3i\pi H_{w_1^+}(s_1)-8\zeta_2+\calO(\ep)\right) \, , \nnb
\end{align}
where $\ESGamma = 1/((4\pi)^{D/2}\,\Gamma(1-\ep))$ and $\zeta_2=\pi^2/6$.


\subsection{Calculation of the counterterms}
\label{sec:counterterms}

The diagrams shown in figures~\ref{fig:penguins}~--~\ref{fig:dipoles} 
and their calculation discussed in the previous subsection refer to 
the term $\widetilde A^{(2){\rm nf}}_{i1}$ in the expression 
for  $\widetilde T_i^{(2)}$ in the master formula (\ref{eq:master2loop}). 
In the following we remark on the remaining terms in 
(\ref{eq:master2loop}) complementing the discussion of the 
corresponding terms in \cite{Beneke:2009ek} for the calculation of 
the colour-allowed and colour-suppressed tree amplitudes.

We begin with the terms on the right-hand side of the expression 
(\ref{eq:master1loop}) for the one-loop kernels $\widetilde T_i^{(1)}$. 
The first term, $Z_{ij}^{(1)} \, \widetilde A^{(0)}_{j1}$, following 
the bare non-factorizable one-loop amplitudes 
$\widetilde A^{(1){\rm nf}}_{i1}$, is the counterterm that 
renormalizes the operator $Q_i$. The implicit sum over $j$ includes 
the evanescent operators in the effective weak Hamiltonian. 
The next term, $\widetilde A^{(1){\rm f}}_{i1} - A^{(1){\rm f}}_{31} \, 
\widetilde A^{(0)}_{i1}$ arises, because the factorizable one-loop 
terms that are not part of the kernels but of the full QCD form 
factors correspond to right insertions of $Q_3$, while the 
fermion lines of the $Q_i$ are contracted in the wrong-insertion 
order for the QCD penguin amplitude. The difference must be 
put into $\widetilde T_i^{(1)}$ to reproduce the full amplitude. 
However, it turns out to be ${\cal{O}}(\eps)$ and can therefore 
be dropped in the limit $\epsilon\to 0$. The last two terms 
in (\ref{eq:master1loop}) appear, because the matrix elements of 
the evanescent SCET operators $\widetilde O_1-O_1/2$, 
$\widetilde O_b$ with $b>1$ with a non-dimensional infrared 
regulator must be renormalized to zero. The non-minimal 
one-loop renormalization factors $\widetilde Y_{11}^{(1)}, 
\widetilde Y_{b1}^{(1)}$ may give finite contributions to the 
kernels. Once again we find that these contributions  
actually vanish as $\epsilon\to 0$. However, this no longer 
holds true at the two-loop order.

The counter- and subtraction terms needed to obtain finite 
two-loop hard-scattering kernels $\widetilde T_i^{(2)}$ in 
(\ref{eq:master2loop}) are much more involved. The first two lines 
of (\ref{eq:master2loop}) represent ultraviolet counterterms 
for the operators $Q_i$, including external field as well as bottom 
and charm mass renormalization. The internal quark masses 
are taken to be the pole masses, hence $\delta m^{(1)}$ is 
the counterterm for the pole mass. There are no internal massive 
quark lines in the $\mathcal{O}(\alpha_s)$ matrix element of 
the chromomagnetic dipole operator $Q_{8g}$. However, the mass 
parameter $\overline{m}_b$ that appears 
in the definition (\ref{eq:Q8gdef}) of the operator is 
understood to be the $\overline{\rm MS}$ mass and the mass 
counterterm is not included in the anomalous dimension 
matrix from \cite{Gorbahn:2004my}. The conversion to the pole mass 
then implies that the same counterterm  $\delta m^{(1)}$ is 
applied multiplicatively to the bare amplitude 
$\widetilde A^{(1){\rm nf}}_{8g1}$. 

\begin{figure}[t]
\begin{center}
 \includegraphics[width=.32\textwidth]{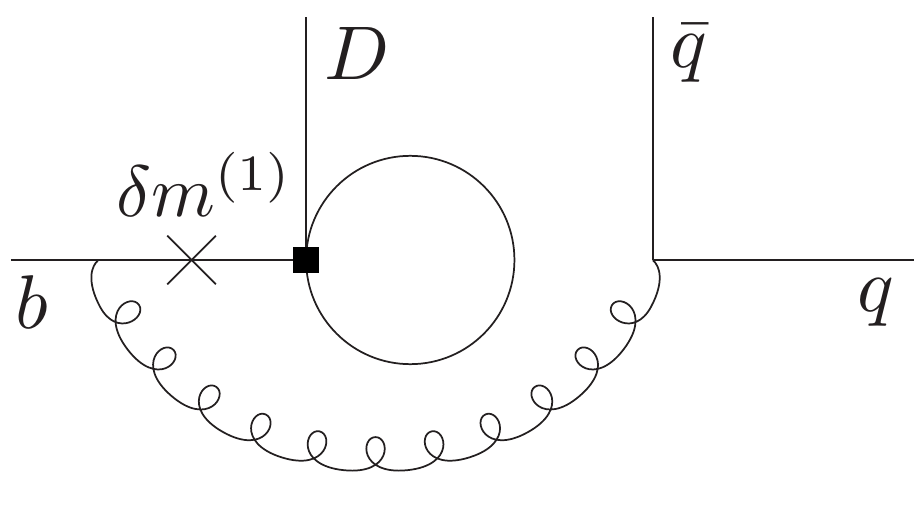}
\end{center}
\vspace*{-0.3cm}
\caption{\label{fig:tadpoles} Non-vanishing tadpole contribution 
to $ (-i) \, \delta m^{(1)} \, \widetilde A^{\prime (1){\rm nf}}_{i1}$, 
$i=3,\ldots, 6$.}
\end{figure}

It is interesting to note that 
$(-i) \, \delta m^{(1)} \, \widetilde A^{\prime (1){\rm nf}}_{i1}$ is 
the only place where fermion-tadpole contractions of the 
four-quark operators survive in the counterterms. The relevant diagram is 
shown in figure~\ref{fig:tadpoles}. For the one-loop tadpole diagrams 
without the mass counterterm insertion there is a cancellation 
between the two diagrams obtained from attaching the gluon to the external
fermion lines to the left and above the four-quark vertex. This cancellation 
does not occur for the mass counterterm tadpole diagram, since 
it only exists for the massive $b$-quark line. 
The non-vanishing tadpole contribution is divergent and required to make  
the final result for $\widetilde T_i^{(2)}$ finite. The corresponding 
tadpole contribution vanishes for the $Q_{1,2}^p$ matrix elements 
due to the $(V-A)\otimes(V-A)$ structure of these operators.
In the bare two-loop matrix elements almost all the tadpole contractions
cancel (see figure~\ref{fig:penguins0}), and only two diagrams give a net 
contribution (see last line of figure~\ref{fig:penguins}). 

The remaining terms on the right-hand side of the two-loop master 
formula (\ref{eq:master2loop}) represent generalizations of similar 
structures in (\ref{eq:master1loop}) and account for i) the use of 
the full QCD rather than SCET $B\to M$ form factors, ii) the 
renormalization of the SCET operators, iii) non-vanishing two-loop 
SCET bare matrix elements due to massive internal charm loops, 
iv) the finite subtractions required to make the renormalized 
matrix elements of evanescent operators vanish.

As was the case in the calculation of the topological tree 
amplitudes, the expression $- \sum_{b>1} \widetilde H_{ib}^{(1)} \, 
\widetilde Y_{b1}^{(1)}$ in the third line of (\ref{eq:master2loop}) 
vanishes in the limit $\epsilon\to 0$. The finite one-loop hard-scattering 
kernels $\widetilde H_{ib}^{(1)}$ are non-zero for 
$b=2,3$, but the mixing of the evanescent operators 
into $\widetilde O_1$, $\widetilde Y_{b1}^{(1)}$ turns out to 
be $\mathcal{O}(\eps)$. We recall that the renormalization factors 
$\widetilde Y_{a1}^{(\ell)}$ are functions of two momentum fractions $u',u$. 
Expressions such as  $- \sum_{b>1} \widetilde H_{ib}^{(1)} \, 
\widetilde Y_{b1}^{(1)}$ therefore represent 
convolutions in $u'$ with the hard-scattering kernels. 
In case of $\widetilde T_i^{(1)} \widetilde Y_{11}^{(1)}$ in the 
third line, since $\widetilde Y_{11}^{(1)}$ contains a 
$1/\eps^2$ singularity, the $D$-dimensional hard-scattering kernel 
$\widetilde T_i^{(1)}$ must be expanded to $\mathcal{O}(\eps^2)$. The 
convolution can then be quite involved. 

It is advantageous to rearrange some of the terms in 
(\ref{eq:master2loop}). To this end, we define the one-loop kernel 
without the terms from the SCET evanescent operator renormalization, 
\begin{equation}
\frac{1}{2}\,\widetilde T_i^{(1)F} \equiv 
\widetilde A^{(1){\rm nf}}_{i1}+ Z_{ij}^{(1)} \, 
 \widetilde A^{(0)}_{j1}
+\widetilde A^{(1){\rm f}}_{i1} - A^{(1){\rm f}}_{31} \, 
 \widetilde A^{(0)}_{i1}\,.
\end{equation}
Notice that $\widetilde T_i^{(1)F}$ must be computed to 
$\mathcal{O}(\epsilon^2)$, since it multiplies the renormalization 
constant $\widetilde Y_{11}^{(1)}$, which contains $1/\epsilon^2$ 
poles. Hence, the term 
$\widetilde A^{(1){\rm f}}_{i1} - A^{(1){\rm f}}_{31}
\widetilde A^{(0)}_{i1}$, which is 
$\mathcal{O}(\epsilon)$ cannot be dropped here. 
We then combine terms in (\ref{eq:master2loop}) as follows:
\begin{eqnarray}
&& \hspace*{-0.6cm} - \frac{1}{2}\,\widetilde T_i^{(1)} 
\widetilde Y_{11}^{(1)}  - 
[\widetilde Y_{11}^{(2)}-Y_{11}^{(2)}]\,
\widetilde A^{(0)}_{i1} - 
\sum_{b>1} \widetilde A_{ib}^{(0)} \, \widetilde Y_{b1}^{(2)}
\nonumber\\
&=&  
- \frac{1}{2}\,\widetilde T_i^{(1)F} \widetilde Y_{11}^{(1)}
+ \underbrace{\left\{[\widetilde Y_{11}^{(1)}-Y_{11}^{(1)}]\, 
\widetilde Y_{11}^{(1)} - 
[\widetilde Y_{11}^{(2)}-Y_{11}^{(2)}]\right\} \widetilde A^{(0)}_{i1}
}_{\equiv \,\hat{\Delta}} 
\nonumber\\
&& + \,\underbrace{\sum_{b>1} \widetilde A_{ib}^{(0)} \, 
 \left(\widetilde Y_{b1}^{(1)}\widetilde Y_{11}^{(1)}-\widetilde Y_{b1}^{(2)}\right)}_{\equiv \,\hat{\mathcal{E}}}\,.
\end{eqnarray}
The combination of renormalization constants $\hat{\Delta}$ was already 
computed in \cite{Beneke:2009ek}. The quantity $\hat{\mathcal{E}}$ 
appears only for the matrix elements of $Q_{5,6}$, which have 
non-vanishing tree-level contributions $\widetilde A_{ib}^{(0)}$ 
proportional to the evanescent SCET operator $\widetilde{O}_2$. 
As discussed in  \cite{Beneke:2009ek}, the advantage of defining 
$\hat{\Delta}$, $\hat{\mathcal{E}}$ is that while the individual terms 
in these expressions depend on the infrared regulator chosen to 
compute the non-minimal evanescent operator renormalization 
constants, $\hat{\Delta}$, $\hat{\mathcal{E}}$ are regulator-independent, 
as indeed the final result must be. 
By applying infrared rearrangements, in which the product of 
one-loop renormalization factors appears as subgraph of the two-loop 
factor, $\hat{\Delta}$, $\hat{\mathcal{E}}$  
can be computed directly without the need of 
ever introducing an explicit infrared regulator. 

With these remarks we provide explicit expressions for 
\begin{eqnarray}
&& \widetilde A^{(2){\rm f}}_{i1} - A^{(2){\rm f}}_{31} \, 
\widetilde A^{(0)}_{i1} 
= \left\{1,\frac{4}{3},32,\frac{128}{3}\right\} \,\frac{8}{27} 
\left(\frac{1}{\eps}+2 L\right) +
\left\{1,\frac{4}{3},32,\frac{128}{3}\right\} \frac{4}{27} n_0 T_f 
\nonumber\\
&&\hspace*{3.55cm} 
+\left\{1,\frac{4}{3},64,\frac{256}{3}\right\} \frac{7}{27}\,,
\qquad
\\[0.2cm]
&& (-i) \, \delta m^{(1)} \, 
[\widetilde A^{\prime (1){\rm f}}_{i1}
- A^{\prime (1){\rm f}}_{31} \, \widetilde A^{(0)}_{i1}]
= \mathcal{O}(\ep)\,,
\\[0.3cm]
&& (Z_{\alpha}^{(1)}+Z_{ext}^{(1)})\,
[\widetilde A^{(1){\rm f}}_{i1} - A^{(1){\rm f}}_{31} \, 
\widetilde A^{(0)}_{i1}] = 
\left\{1,\frac{4}{3},32,\frac{128}{3}\right\} 
\left(-\frac{8}{27} n_f T_f +\frac{26}{9}\right), 
\\[0.2cm]
&& - \,[\widetilde M^{(2)}_{11} - M^{(2)}_{11} ] \, 
\widetilde A^{(0)}_{i1} = 
\left\{1,\frac{4}{3},16,\frac{64}{3}\right\}\frac{4}{27}T_f\,,
\\[0.2cm]
&& - \,(C_{FF}^{(1)}-\xi_{45}^{(1)})\, 
[\widetilde Y_{11}^{(1)}-Y_{11}^{(1)}] \, \widetilde A^{(0)}_{i1} 
= \mathcal{O}(\ep)\,,
\\[0.3cm]
&&- \,\sum_{b>1} \widetilde A_{ib}^{(0)}\,\widetilde M_{b1}^{(2)}
= \left\{0,0,1,\frac{4}{3}\right\}\,\frac{64}{27}T_f\,,
\\[0.2cm] 
&& \hat{\Delta} =  \left\{1,\frac{4}{3},16,\frac{64}{3}\right\}
\left(-\frac{4}{27} (n_0+1) T_f +\frac{17}{27}\right), 
\\[0.2cm]
&& \hat{\mathcal{E}} = \left\{0,0,1,\frac{4}{3}\right\}
\left(-\frac{64}{27} (n_0+1) T_f +\frac{16}{27}\right), 
\end{eqnarray}
which we expanded in $\eps$ up to $\mathcal{O}(\ep^0)$. 
The four entries in curly brackets refer to $i=3,4,5,6$.
$L=\ln\frac{\mu^2}{m_b^2}$, $T_f=1/2$. $n_0=3$ denotes the number of 
massless flavours, and $n_f=5$ the total number of flavours.


\subsection{Hard-scattering kernels}
\label{sec:HSK}

Although most of the individual terms in the master 
formulas~\eqref{eq:mastertree}~--~\eqref{eq:master2loop} have poles in the 
dimensional regulator $\ep$, the total expression for the
hard-scattering kernels (HSK) $\widetilde T_i^{(\ell)}$ must be free of 
poles in $\ep$, which we checked analytically to $\calO(\as^2)$ and for all
$i \in \{1u,2u,1c,2c,\text{3-6},8g\}$. 
 
The tree-level and one-loop HSK are known from \cite{Beneke:1999br}. 
However, the calculation was performed with another operator basis 
for the effective weak Hamiltonian \cite{Buchalla:1995vs}, which is 
less suitable for NNLO calculations than the CMM 
basis~\cite{Chetyrkin:1997gb}. For completeness, we therefore 
begin by summarizing the tree-level and one-loop HSK in the 
CMM basis and in the notation of the present paper. 

At tree-level the 
HSK relevant
for the penguin amplitude read
\begin{align}
\widetilde T^{(0)}_{1u} &= 0 \, , & \widetilde T^{(0)}_{2c} &= 0 \, , & \widetilde T^{(0)}_{5} &= 16/N_c \, , \nnb \\
\widetilde T^{(0)}_{2u} &= 0 \, , & \widetilde T^{(0)}_{3} &= 1/N_c \, , & \widetilde T^{(0)}_{6} &= 16 C_F/N_c \, , \nnb \\
\widetilde T^{(0)}_{1c} &= 0 \, , & \widetilde T^{(0)}_{4} &= C_F/N_c \, , & \widetilde T^{(0)}_{8g} &= 0 \,,
\end{align}
with $N_c=3$ and $C_F = (N_c^2-1)/(2N_c)=4/3$. At $\calO(\as)$ the HSK 
are conveniently expressed in terms of the following functions, 
\begin{align}
 G(z_c,\ub) &=  \frac{2 (12 z_c + 5 \ub - 3 \ub \ln(z_c))}{9\ub} -
  \frac{2 s (2 z_c + \ub)}{3\ub} \ln \left(\frac{s + 1}{s - 1}\right) \nnb \\
            &=  \frac{2 (12 z_c + 5 \ub - 3 \ub \ln(z_c))}{9 \ub} -
  \frac{2 s (2 z_c + \ub)}{3\ub} \left(H_{w_1^+}(s)-i \pi \right) \, , \label{eq:Gc} \\
 G(z_b,\ub) &=  \frac{2 (12 + 5 \ub)}{9\ub} - \frac{2 s_1 (2 + \ub)}{3\ub} \ln \left(\frac{s_1 + 1}{s_1 - 1}\right) \nnb \\
            &=  \frac{2 (12 + 5 \ub)}{9\ub} - \frac{2 s_1 (2 + \ub)}{3\ub} \left(H_{w_1^+}(s_1)-i \pi \right) \, , \label{eq:Gb} \\
 G(0,\ub)   &= -\frac{2}{3} \, \ln (\ub)+\frac{2}{3} i \pi + \frac{10}{9} \, , \label{eq:G0} \\
 t_1(u)     &= -22 -3 i \pi +3 \left(1-\frac{u}{\ub}\right) \ln u \nnb\\
& \quad \, + \bigg[2 \, {\rm Li}_2(u) - \ln^2(u) - \left(1+2i\pi-\frac{2u}{\ub}\right) \ln(u) -(u\to\ub) \bigg]. \label{eq:t1} 
 \end{align}
The variable $s$ is as in~\eqref{eq:kinematics} and also 
$z_f = (m_f^2-i\eta)/m_b^2$ is as before. The penguin function 
$G(z_f,\bar u)$ coincides with the corresponding function defined in 
\cite{Beneke:2001ev}, while $t_1(u)$ is related to the one-loop 
vertex function $g(u)$ in that reference by $t_1(u) = g(u)-22$.
One then has
\begin{align}
\widetilde T^{(1)}_{1u} &= -\frac{C_F}{2 N_c^2} \left[ - \frac{2}{3} \, L + \frac{2}{3} - G(0,\ub) \right] \, ,& \quad
\widetilde T^{(1)}_{2u} &= - 2 N_c \, \widetilde T^{(1)}_{1u} \, , \nnb \\
\widetilde T^{(1)}_{1c} &= -\frac{C_F}{2 N_c^2} \left[ - \frac{2}{3} \, L + \frac{2}{3} - G(z_c,\ub) \right] \, , & \quad
\widetilde T^{(1)}_{2c} &= - 2 N_c \, \widetilde T^{(1)}_{1c} \, , \nnb
\end{align}
\begin{align}
\widetilde T^{(1)}_{3}  &= \frac{C_F}{N_c} \left[- \frac{22}{3} \, L + t_1(u)  + \frac{4}{3} - G(0,\ub) - G(z_b,\ub)\right]
                           + \widetilde T^{(1)}_{8g} \, , \nnb \\
\widetilde T^{(1)}_{4}  &= -\frac{1}{2N_c} \, \widetilde T^{(1)}_{3} -C_F + \frac{C_F}{N_c} \left[ -\frac{2}{3} \, n_f \, L  - n_0 \, G(0,\ub) - G(z_c,\ub) - G(z_b,\ub) \right] \, , \nnb \\
\widetilde T^{(1)}_{5}  &= 16 \, \widetilde T^{(1)}_{3}	+\frac{16 C_F}{3 \, N_c}   + 4 \, \widetilde T^{(1)}_{8g} \, , \nnb \\		   
\widetilde T^{(1)}_{6}  &= -\frac{3}{N_c} \, \widetilde T^{(1)}_{3} + 10 \, \widetilde T^{(1)}_{4} - \frac{2}{N_c} \, \widetilde T^{(1)}_{8g}
                             - \frac{4 C_F}{N_c} \left(\frac{15}{2} N_c+\frac{2}{3 N_c} -n_f \right)\, , \nnb \\
\widetilde T^{(1)}_{8g} &= -\frac{2 C_F}{N_c \ub} \, .
\label{eq:oneloopTs}
\end{align}
Note that the HSK must be calculated to higher orders in $\ep$ when they 
enter a term in the master formula at higher loop order, since they 
usually multiply terms that are divergent in $\ep$. We refrain from giving
terms beyond $\calO(\ep^0)$ here since they are straightforward to derive.

At $\calO(\as^2)$ only the kernel $\widetilde T^{(2)}_{8g}$ of the 
chromomagnetic dipole operator is of a length suitable for printing, 
and given by the expression
\begin{align}
\widetilde T^{(2)}_{8g} & = \frac{8 L}{9 \ub}+\frac{64}{27 \ub} \, L \, \ln (\ub)+\frac{8 \ln (u)}{27 \ub}
-\frac{16 (\ub-4 z_c) (\ub+2 z_c)}{27 \, s \, \ub^3} \, (H_{w_1^+}(s) - i \pi) -\frac{16}{27 \ub} \, \text{Li}_2(\ub) \nnb \\
&-\frac{4 (\ub-4) \left(\ub^3+5 \ub^2-12 \ub+8\right)}{27 s_1 u^2 \ub^3} \, (H_{w_1^+}(s_1) - i \pi)
+\frac{8 \left(49 \ub^2-61 \ub+24\right)}{81 u \ub^2} +\frac{64 z_c}{27 \ub^2}\nnb \\
&-\frac{8 (\ub-2)}{9 u^3 \ub} \, (H_{w_1^+,w_1^+}(s_1) - i \pi\, H_{w_1^+}(s_1)) -\frac{4 \pi ^2 (\ub-3) \left(\ub^2-5\right)}{81 u^3 \ub}
-\frac{4 i \pi  \left(3 \ub^2+\ub+14\right)}{27 u^2 \ub}\nnb \\
&+\frac{4 \left(5 \ub^2-17 \ub+30\right)}{27 u^2 \ub} \, \ln (\ub)
+\frac{4 \left(7 \ub^3-21 \ub^2+29 \ub-6\right)}{27 u^3 \ub} \, \ln ^2(\ub)
-\frac{8}{27 \ub} \, i \pi \ln(u) \nnb \\
&-\frac{8  \left(9 \ub^3-27 \ub^2+35\ub-8\right)}{27 u^3 \ub} \, i \pi \ln (\ub)
-\frac{16 \ln (z_c)}{27 \ub}-\frac{4}{27 \ub} \, \ln ^2(u)
-\frac{8 \ln (\ub) \ln (u)}{27 \ub} \, ,
\label{eq:Q8g2loopkernel}
\end{align}
where we substituted numerical values for the colour and flavour 
factors. The explicit expressions for $H_{w_1^+}(s)$ and 
$H_{w_1^+,w_1^+}(s)$ are 
\begin{eqnarray}
H_{w_1^+}(s) &=& 
\ln\left(\frac{1+s}{1-s}\right) = \ln\left(\frac{s+1}{s-1}\right) + i\pi\,,
\\ 
H_{w_1^+,w_1^+}(s) &=& \frac{1}{2}\, H^2_{w_1^+}(s) \,.
\end{eqnarray}
The kernel $\widetilde T^{(2)}_{8g}$ was previously 
calculated in \cite{Kim:2011jm}. Our expression 
(\ref{eq:Q8g2loopkernel}) confirms this result.\footnote{
The $\overline{\rm MS}$ heavy-quark 
masses are used in  \cite{Kim:2011jm}, whereas we employ pole masses. 
However, since $\widetilde T^{(1)}_{8g}$ does not depend on quark 
masses, the scheme conversion does not affect the one-loop 
kernel (\ref{eq:Q8g2loopkernel}). Note that if one does not 
convert the $\overline{\rm MS}$ mass $\overline{m}_b$ in the 
definition of $Q_{8g}$ into the pole mass, the tree-level HSK 
$\widetilde T^{(1)}_{8g}$ is proportional to 
$\overline{m}_b/m_b$, where the pole mass in the denominator 
arises from on-shell kinematics. In order to arrive at the expression 
in the last line of (\ref{eq:oneloopTs}), one must then express 
one mass definition in terms of the other. To obtain the 
correct result for $\widetilde T^{(2)}_{8g}$, the conversion 
must be done with one-loop accuracy. The one-loop correction 
contributes to $\widetilde T^{(2)}_{8g}$.}
The expressions for the two-loop penguin HSK of the operators 
$Q_{1,2}^p,Q_{3-6}$ 
are long and complicated, although they are available analytically as a 
linear combination of the generalized HPLs introduced in 
section~\ref{sec:2loopcalc}. We provide the analytic expressions for 
all HSK at $\mathcal{O}(\alpha_s^2)$ electronically 
as supplementary material to the present article. 


\subsection{Convolution in Gegenbauer moments}
\label{sec:convolution}

The HSK enter the formula of the QCD penguin amplitudes
via the convolution with the LCDA $\phi_{M}(u)$
of the light meson. The LCDA is expanded into the eigenfunctions of
the one-loop renormalization kernel,
\bea
\phi_{M}(u) &=& 6 \, u \, \ub \, \left[1+\sum_{n=1}^{\infty}a_n^{M} \, 
C^{(3/2)}_n(2u-1)\right] \, ,
\label{gegenbauerexp}
\eea
where $a_n^{M} \equiv a_n^{M}(\mu)$ and $C^{(3/2)}_n(x)$ are the Gegenbauer 
moments and polynomials, respectively. We truncate the Gegenbauer expansion 
(\ref{gegenbauerexp}) at $n=2$, which is sufficient in practice. 

The convolution of those parts of the HSK that come from operator insertions 
into ``tree-type'' diagrams is performed in the same way as 
in~\cite{Bell:2007tv,Bell:2009nk,Beneke:2009ek} and can be done completely 
analytically. The terms that stem from ``penguin-type'' diagrams, however,
require a different and more refined treatment. A method that is applicable 
to the majority of the terms is to trade $u$ for $s=\sqrt{1-4 z_c/\ub}$ as 
integration variable, which results in
$s=r$ and $s=+i\infty$ as integration limits for $s$,
\begin{align}
\int\limits_0^1 \!du \; \widetilde T_{i}(u) \; \phi_{M}(u) =
\int\limits_r^{+i\infty} \!\! ds \; \frac{2 s (r^2-1)}{(1-s^2)^2} \, \widetilde T_{i}(u(s)) \; \phi_{M}(u(s)) \, .
\end{align}
The threshold at $\ub=4z_c$ is mapped to $s=0$. The main advantage of this substitution
is the ability to perform the integration over $s$ in terms of the same iterated integrals
as in section~\ref{sec:2loopcalc}. Subtleties arise, however, when taking the limits $s \to r$
and $s \to +i\infty$ of the integral function. In case of the lower limit individual terms
contain power divergences proportional to $1/(r-s)^n$. They can be isolated via a Taylor expansion
about $s=r$ of the corresponding numerators and disappear in the sum of all terms.
Taking the upper limit requires an argument inversion of the generalized HPLs. This is done
recursively via the following formulas:
\begin{align}
H_{w^+,\vec w}\left(\frac{1}{y}\right) &=  H_{w^+,\vec w}(a) - \int\limits_{y}^{\frac{1}{a}} \!dt \; \frac{2 \, {\textstyle{\frac{1}{w}}}}{\left({\textstyle{\frac{1}{w}}}\right)^2-t^2} \, H_{\vec w}\left(\frac{1}{t}\right) \, , \nnb \\[0.3em]
H_{w^-,\vec w}\left(\frac{1}{y}\right) &=  H_{w^-,\vec w}(a) - \int\limits_{y}^{\frac{1}{a}} \!dt \; \left[\frac{2}{t} + \frac{2 \, t}{\left({\textstyle{\frac{1}{w}}}\right)^2-t^2}\right] \, H_{\vec w}\left(\frac{1}{t}\right) \, , \nnb \\[0.3em]
H_{0,\vec w}\left(\frac{1}{y}\right) &=  H_{0,\vec w}(a) + \int\limits_{y}^{\frac{1}{a}} \!dt \; \frac{1}{t} \, H_{\vec w}\left(\frac{1}{t}\right) \, .
\label{eq:HPLinversion}
\end{align}
The recursion ends at HPLs of weight one, where the explicit form of the argument inversion can be easily computed, e.g.\
\begin{align}
H_{w_2^-}\left(\frac{1}{t}\right) &=  H_{{\textstyle{(\frac{1}{w_2})}}^-}(t) + 2 H_0(t) + 2 H_0(r)+ i \pi \, ,
\end{align}
which holds for arbitrary values of $1/t$ on the positive, 
imaginary axis. 
In this way, logarithmic divergences contained in generalized HPLs as $s \to +i\infty $ are made explicit, as in
\begin{align}
H_{w_2^-,0}(s) &\stackrel{s \to +i\infty}{\longrightarrow}
- 2 H_{0,0}(s) + H_{{\textstyle{(\frac{1}{w_2})}}^-,0}(-i) + H_{w_2^-,0}(i) - \frac{\pi^2}{4} + \calO \left(\frac{1}{s}\right)\, .
\end{align}
The divergences that arise in individual terms as $s \to +i\infty$ have to cancel in the end as well.
This procedure yields generalized HPLs up to weight five, which we evaluate numerically by means of the
program GiNaC~\cite{Bauer:2000cp,Vollinga:2004sn}. The complex number $a$ 
in (\ref{eq:HPLinversion}) is in principle arbitrary. The choice $a=i$ in the argument inversion
turns out to be convenient for the numerical evaluation.

There are a few terms to which the above procedure cannot be applied since 
the dependence on the kinematic variables is more involved, e.g.\ in 
products of HPLs of different $u$-dependent arguments. Fortunately, these 
terms are easy to integrate numerically and/or are free from kinematic 
thresholds. In the latter case we derive Mellin-Barnes (MB)
representations for the integrals and perform the convolution over $u$ analytically. The subsequent integration over the
MB variables can be done numerically to high accuracy using {\tt MB.m}~\cite{Czakon:2005rk}.

In this way, we obtain all terms in the leading QCD penguin amplitudes that involve
powers of $L = \ln(\mu^2/m_b^2)$ completely analytically. In the $L^0$ pieces a few terms are obtained only as
an interpolation in $z_c$. We present all the expressions out of which the penguin amplitudes are built
in the next section.

As an independent check we evaluated the convolution integrals for
a fixed value of the charm-quark mass numerically, based on a
grid of 230 points in $u$ that was designed to capture the singular
behavior of the hard-scattering kernels at the endpoints $u\to
0,1$ and at the charm threshold $u=1-4 z_c$. We then fitted the
integrands to a suitable ansatz in the singular regions, while
we used interpolating functions in the regions where the integrands
are smooth. In this way we compared our results for the convolution
integrals for all considered Gegenbauer moments and 25 different
values of the charm-quark mass, and found agreement between the
two methods at the subpercent level, except for a few outliers
that are due to large numerical cancellations.

%
%
%

\section{Penguin amplitudes with NNLO accuracy}
\label{sec:amplitudes}


\subsection{Analytic expressions}
\label{sec:a4analytic}

In the CMM basis the leading QCD penguin amplitudes $a_4^p$ with $p=u,c$ become, after expanding
through to the second Gegenbauer moment,
\begin{align}
a_4^p = & \; \frac{C_3}{N_c} + \frac{C_F}{N_c} C_4 + \frac{16C_5}{N_c} + \frac{16C_F}{N_c} C_6\nnb\\
& + \asfourpi \frac{C_F}{N_c} \left[ \left( C_3 - \frac{C_4}{2N_c} +16C_5 - \frac{8C_6}{N_c}\right) (-6L+I^{(1)}_t)\right.\nnb\\
& \qquad\qquad\quad + \left(C_{8g} + C_3 - \frac{C_4}{2 \, N_c} + 20 C_5 - \frac{10 \, C_6}{N_c}\right) (-2)\,I^{(1)}_{8g} \nnb\\
& \qquad\qquad\quad +\left( C_2 - \frac{C_1}{2N_c} \right) \left( -\frac{2}{3} \, L + \frac{2}{3} - \delta_{pu} \, I^{(1)}_0 - \delta_{pc} \, I^{(1)}_c\right) \nnb\\
& \qquad\qquad\quad + \left(C_3 - \frac{C_4}{2N_c} +16 C_5 - \frac{8C_6}{N_c} \right)
\left( -\frac{4}{3} \, L + \frac{4}{3} - I^{(1)}_0 - I^{(1)}_b \right) \nnb\\
&\qquad\qquad\quad + \Big( C_4 + 10 C_6 \Big)
\left( -\frac{2}{3} \, n_f \, L  - n_0 \, I^{(1)}_0 - I^{(1)}_c - I^{(1)}_b \right) \nnb\\
&\qquad\qquad\quad \left.
-N_c C_4  + \frac{16}{3} C_5 - 4 \left(10N_c+\frac{2}{3N_c} -n_f \right) C_6\right] \nnb\\
& + \left(\asfourpi\right)^2 \bigg[C_1 \left(\delta_{pu} \, I^{(2)}_{1u} + \delta_{pc} \, I^{(2)}_{1c} \right) + C_2 \left(\delta_{pu} \, I^{(2)}_{2u} + \delta_{pc} \, I^{(2)}_{2c} \right) \nnb \\
& \qquad\qquad\quad + \sum\limits_{i=3}^{6} C_i \, I^{(2)}_{i} + C_{8g} \, I^{(2)}_{8g} \bigg] \, .
\end{align}
As before, we use $L = \ln(\mu^2/m_b^2)$ and $z_f = (m_f^2-i\eta)/m_b^2$. The functions $I^{(1)}_{t,0,c,b,8g}$ denote the convolution 
of the functions in~\eqref{eq:Gc}~--~\eqref{eq:t1} with the 
Gegenbauer expansion  (\ref{gegenbauerexp}) of the light-meson LCDA. 
They are known analytically \cite{Beneke:2001ev},
\begin{align}
I^{(1)}_t & = \int\limits_0^1 \!du \; t_1(u) \; \phi_{M}(u) = -\frac{45}{2}-3 i \pi
+ a_1^M \left(\frac{11}{2}-3 i \pi \right)
-\frac{21}{20} \, a_2^M \; , \\
I^{(1)}_0 & = \int\limits_0^1 \!du \; G(0,\ub) \; \phi_{M}(u) = \frac{5}{3}+\frac{2 i \pi }{3}
+\frac{1}{2} \, a_1^M 
+\frac{1}{5} \, a_2^M  \; ,
\end{align}
\begin{eqnarray}
I^{(1)}_c &=& \int\limits_0^1 \!du \; G(z_c,\ub) \; \phi_{M}(u) \, = \, \frac{r^2+2}{2} \, \left(r^2-1\right)^2 g_3(r)-\frac{r}{3} \left(3 r^4-7 r^2+6\right) g_1(r)-\frac{2}{3} \ln(z_c) \nnb \\[-0.2cm]
&& \hspace*{1.2cm}+\,\frac{1}{3} \left(r^2-2\right) \left(3 r^2-8\right)
\nonumber\\
&& +\, a_1^M \left[\frac{9}{16} \left(3 r^4+2 r^2+3\right) \left(r^2-1\right)^2 g_3(r)-\frac{27}{8} r \left(r^2+1\right) \! \left(r^2-1\right)^2 \! g_1(r)\right. \nnb \\
&&\hspace*{1.2cm} 
\left.+\frac{1}{8} \left(27 r^6-18 r^4-69 r^2+64\right)\right]
\nonumber\\
&&   +\,a_2^M \left[\frac{9}{8} \left(5 r^6+r^2+2\right) \left(r^2-1\right)^2 g_3(r)-\frac{3}{4} r \left(15 r^4+5 r^2+6\right) \left(r^2-1\right)^2 g_1(r) \right. \nnb \\
&&\hspace*{1.2cm}\left. +\frac{1}{20} \left(225 r^8-300 r^6+85 r^4-230 r^2+224\right)\right]\,, 
\end{eqnarray}
\begin{eqnarray}
I^{(1)}_b &=& \int\limits_0^1 \!du \; G(z_b,\ub) \; \phi_{M}(u) = 
\frac{85}{3}-6 \sqrt{3} \pi +\frac{4 \pi ^2}{9}
+ a_1^M \left( -\frac{155}{2}+36 \sqrt{3} \pi -12 \pi ^2\right) \nnb \\
&& + \, a_2^M \left(\frac{7001}{5}-504 \sqrt{3} \pi +136 \pi ^2 \right) 
\, , \\
I^{(1)}_{8g} & =& \int\limits_0^1 \!du \; \frac{\phi_{M}(u)}{\bar u} = 
3\left(1+a_1^M+a_2^M\right)\; ,
\end{eqnarray}
where we used the variable $r=\sqrt{1-4 z_c}$. The functions $g_1(r)$ and $g_3(r)$ can be found in~\eqref{eq:funcgi}.
At $\calO(\as^2)$, all terms involving $L$ were also obtained in a fully analytic manner, while the $L^0$ terms
are analytic only for $I^{(2)}_{1u}$, $I^{(2)}_{2u}$ and $I^{(2)}_{8g}$.
For the $L^0$ term in $I^{(2)}_{1c}$, $I^{(2)}_{2c}$ and $I^{(2)}_{3-6}$ we calculated numerical values for
several hundred points in the interval $0.01 \le z_c \le 1$. We provide the data tables for these functions
electronically in ancillary files, and in this write-up present fits which reproduce the data at the level of $5$~per mille
for all $0.01 \le z_c \le 1$. For the physical region $0.05 < z_c < 0.2$ the agreement is well below the per-mille level.
All functions that contribute at $\calO(\as^2)$ are collected in appendix~\ref{sec:ampfunc}.


\subsection{Numerical results}
\label{sec:a4numeric}

The convoluted kernels and fit coefficient tables allow for a fast 
evaluation of the two-loop penguin amplitude. In the following we 
discuss the size of the new correction and present for the first time 
the numerical value of the complete leading QCD penguin amplitude 
at NNLO. For the sake of comparison with previous partial results, 
in particular \cite{Bell:2015koa}, we adopt the same values for the 
hadronic parameters as in that reference, which in turn are mostly 
the same as adopted in \cite{Beneke:2009ek} for the evaluation of 
the topological tree amplitudes at NNLO in QCD factorization. 
The form-factor contribution to $a_4^p$ depends only on the 
Gegenbauer moments of the light-meson LCDA. The spectator-scattering 
contribution further depends on form factors, the $B$-meson decay 
constant and LCDA, and light-quark masses.

The penguin amplitude coefficients $a_4^{u,c}$ depend on the 
final state mesons through the hadronic parameters. We present 
the result for the $\pi\bar K$ final state:
\begin{eqnarray}
a_4^u(\pi \bar{K})/10^{-2} &=& -2.87 -
[0.09 + 0.09i]_{\rm V_1} + [0.49-1.32i]_{\rm P_1} \nnb \\
&&- [0.32+0.71i]_{\rm P_{2}, \, Q_{1,2}}
+ [0.33 +0.38 i]_{\rm P_{2}, \, Q_{3-6,8g}}
\nonumber \\[0.2cm]
&& +\,\left[ \frac{r_{\rm sp}}{0.434} \right]
  \Big\{ [0.13]_{\rm LO} + [0.14 +0.12i]_{\rm HV} - [0.01-0.05i]_{\rm HP}
  + [0.07]_{\rm tw3} \Big\} \nonumber \\[0.2cm]
&=& (-2.12^{+0.48}_{-0.29}) + (- 1.56^{+0.29}_{-0.15})i\,,
\label{eq:a4unum} \\[1em]
a_4^c(\pi \bar{K})/10^{-2} &=& -2.87 -
[0.09 + 0.09i]_{\rm V_1} + [0.05-0.62i]_{\rm P_1} \nnb \\
&&- [0.77+0.50i]_{\rm P_{2}, \, Q_{1,2}}
+ [0.33 +0.38 i]_{\rm P_{2}, \, Q_{3-6,8g}}
\nonumber \\[0.2cm]
&& +\,\left[ \frac{r_{\rm sp}}{0.434} \right]
  \Big\{ [0.13]_{\rm LO} + [0.14 +0.12i]_{\rm HV} + [0.01+0.03i]_{\rm HP}
  + [0.07]_{\rm tw3} \Big\} \nonumber \\[0.2cm]
&=& (-3.00^{+0.45}_{-0.32}) + (-0.67^{+0.50}_{-0.39})i\,.
\label{eq:a4cnum}
\end{eqnarray}
In both equations the first two lines originate from the form-factor 
term (kernels $T_i^I$ in (\ref{factformula})). The third line is 
the spectator-scattering contribution (kernels $T_i^{II}$ in 
(\ref{factformula})), which was computed to $\mathcal{O}(\alpha_s^2)$ 
in~\cite{Beneke:2006mk}. Spectator scattering has only a small 
effect on the penguin amplitude for $r_{\rm sp}=0.434$ and we do 
not discuss it further.

The second line in (\ref{eq:a4unum}), (\ref{eq:a4cnum}) represents 
the NNLO correction to the form-factor term, which is further split 
into the contribution 
from the current-current operators $Q_{1,2}^p$ and the penguin 
operators $Q_{3-6,8g}$. The first reproduces the result from 
\cite{Bell:2015koa}. The second, 
$[0.33 +0.38 i]_{\rm P_{2}, \, Q_{3-6,8g}}$, is the main result 
of this paper. Note that the penguin operators contribute equally to 
$a_4^u$ and $a_4^c$. Comparing the new correction to the LO (unlabelled) 
and NLO (labelled V$_1$ and 
P$_1$) contribution, we see that the real part constitutes a 
$(10-15)\%$ correction relative to LO and is a sizeable fraction of 
the NLO term. The imaginary part is especially important as it 
is responsible for the existence of a direct CP asymmetry and 
vanishes at LO. In case of $a_4^u(\pi \bar{K})$, the new 
contribution represents a $-27\%$ correction, and for 
$a_4^c(\pi \bar{K})$ it reaches $-54\%$. However, for both, the real 
and imaginary part there is a strong cancellation between the 
NNLO correction from the current-current operators $Q_{1,2}^p$  
and from the penguin operators $Q_{3-6,8g}$, resulting in a much 
reduced overall NNLO correction. 

\begin{figure}[t]
\begin{center}
 \includegraphics[width=.70\textwidth]{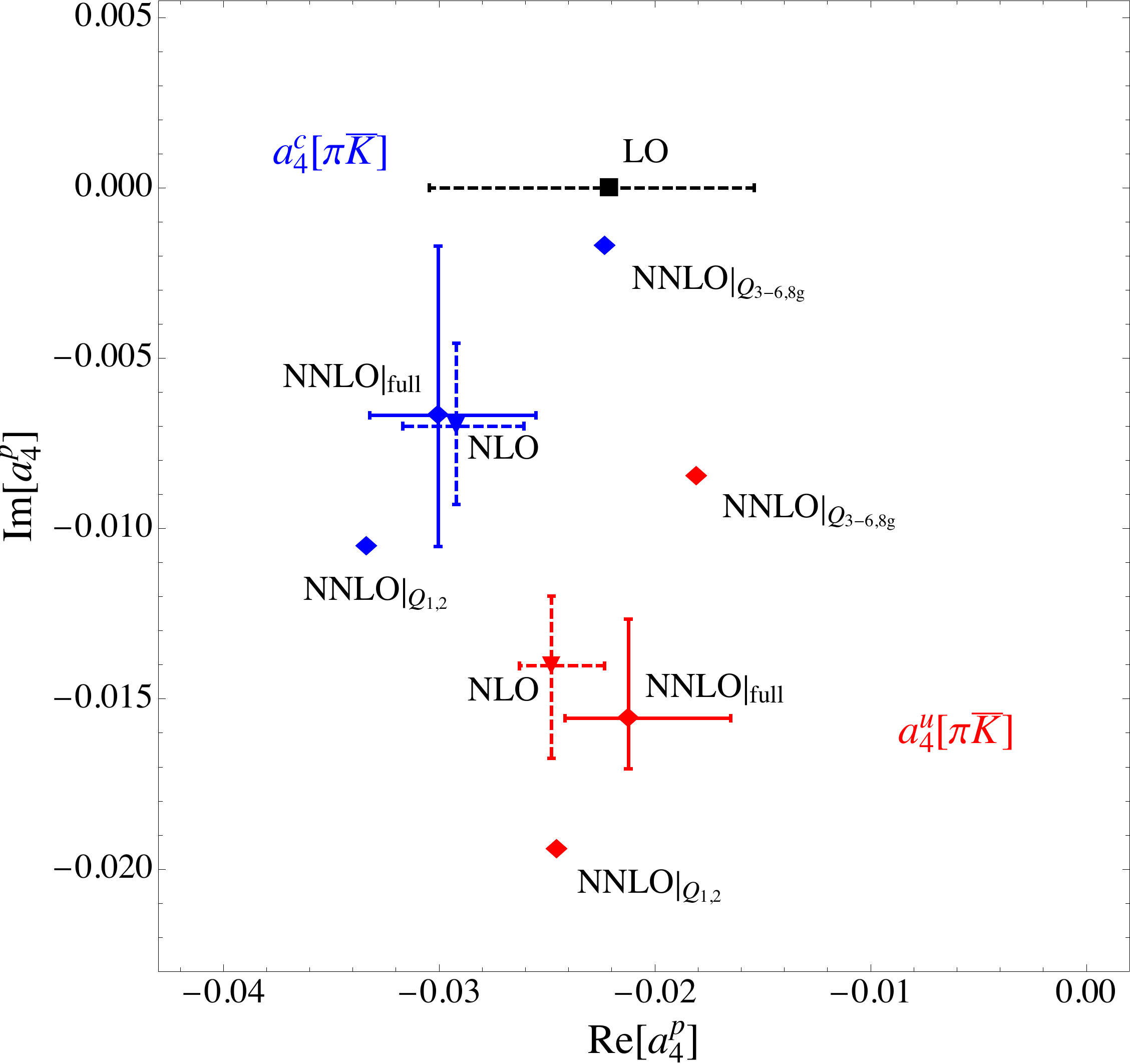}
\end{center}
\caption{\label{fig:anatomya4ua4c} 
Anatomy of QCD corrections to $a_4^u$ 
and $a_4^c$. The points to the lower right (red in colour) refer to 
$a_4^u$, those to the upper left (blue in colour) to $a_4^c$. 
The LO point is equal for both. See text for further explanation.}
\end{figure}

The numerical situation is illustrated in figure~\ref{fig:anatomya4ua4c} 
in the complex $a_4^p$ plane, which shows the LO, NLO and NNLO 
approximation to the QCD penguin amplitude, including the 
spectator-scattering term. At NNLO, we show in addition the 
previous result (labelled NNLO$|_{Q_{1,2}}$) from 
\cite{Bell:2015koa} and for comparison the result when the 
current-current operator rather than the penguin operator 
contributions to the NNLO form-factor term are excluded 
(labelled NNLO$|_{Q_{3-6,8g}}$). That the full NNLO result 
lies between the two previous points illustrates the above 
mentioned cancellation. Note that the LO and NLO numerical 
values shown in the figure do not correspond precisely to 
the values of the LO and NLO terms in (\ref{eq:a4unum}), 
(\ref{eq:a4cnum}). The reason is that in the NNLO result we 
employ Wilson coefficients $C_i$ in the CMM basis evolved to 
the bottom-quark mass scale with next-to-next-to-leading logarithmic 
accuracy, while in the NLO (LO) result we use only 
next-to-leading logarithmic (leading logarithmic) accuracy. 
Eqs.~(\ref{eq:a4unum}), (\ref{eq:a4cnum}), on the other hand, represents 
a split-up of the NNLO expression into terms of different 
orders in $\alpha_s$, but always employing 
next-to-next-to-leading logarithmic Wilson coefficients. As a 
consequence the full NNLO result is even closer to the NLO 
result than the cancellations in~(\ref{eq:a4unum}), (\ref{eq:a4cnum}) 
suggest.\footnote{Note also that the LO and NLO points shown in 
figure~\ref{fig:anatomya4ua4c} are slightly different from those 
shown in \cite{Bell:2019qya}, since in that reference the operator 
basis of \cite{Buchalla:1995vs} was used at LO and NLO.} 

\begin{figure}[t]
\begin{center}
\includegraphics[width=0.48\textwidth]{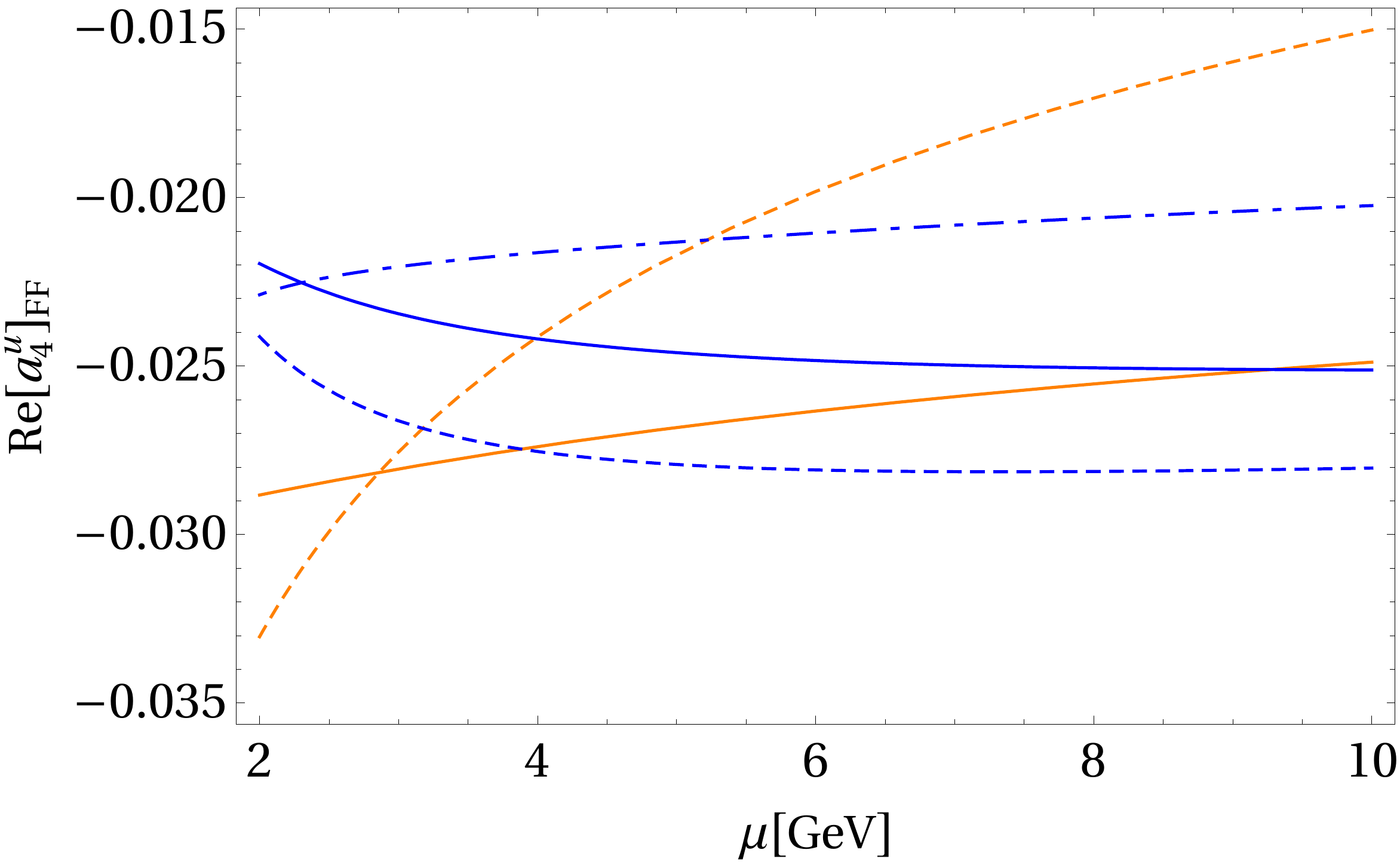} \, 
\includegraphics[width=0.48\textwidth]{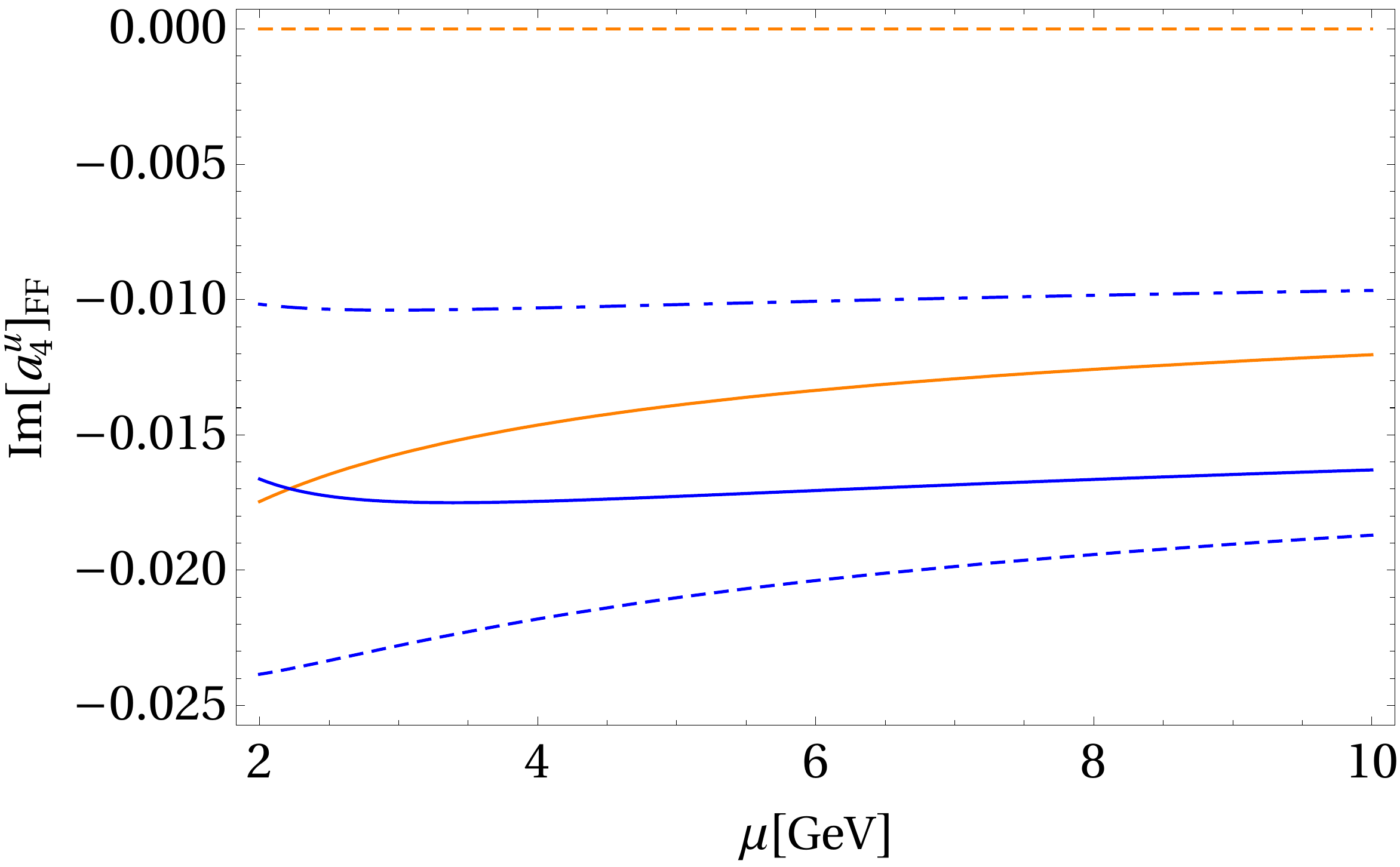}
\vskip0.5cm
\includegraphics[width=0.48\textwidth]{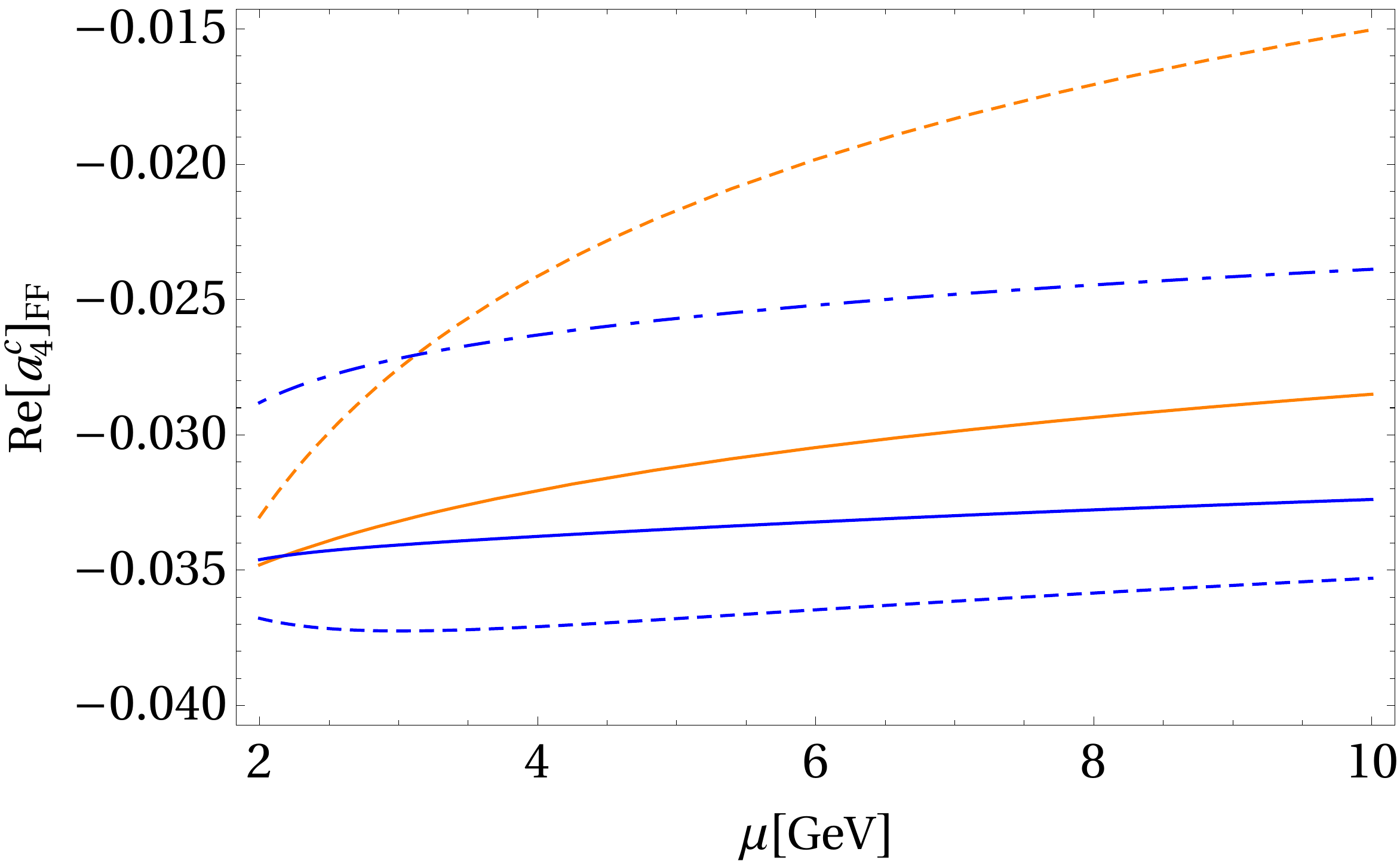} \, 
\includegraphics[width=0.48\textwidth]{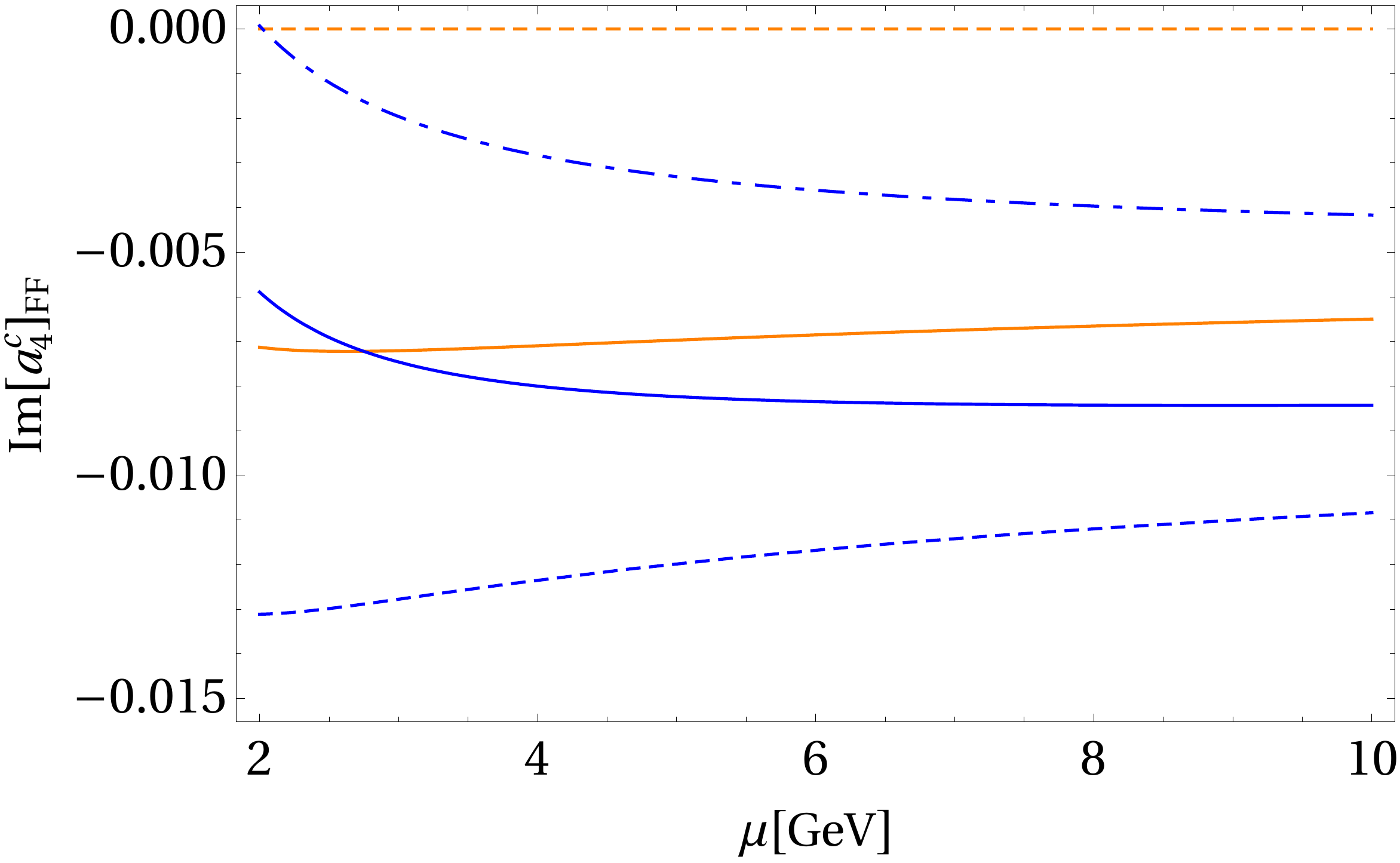}
\end{center}
\vskip-0.2cm
\caption{\label{fig:scalea4p} Scale dependence of $a_4^u(\pi \bar{K})$ 
(upper row) and $a_4^c(\pi \bar{K})$ (lower row). The spectator-scattering 
contribution is not included here. Real part to the left, 
imaginary part to the right. Line coding: 
full NNLO (solid, dark-grey/blue),  
NNLO$|_{Q_{3-6,8g}}$  (dash-dotted, dark-grey/blue),  
NNLO$|_{Q_{1,2}}$  (dashed, dark-grey/blue), 
NLO (solid, light-grey/orange), LO (dashed, light-grey/orange).}
\end{figure}

The natural renormalization scale of the form factor term is 
the bottom-quark mass scale. 
In figure~\ref{fig:scalea4p} we display the 
residual renormalization scale dependence of the form-factor 
contribution to $a_4^p$. The various lines refer to the 
approximations of figure~\ref{fig:anatomya4ua4c} as explained in 
the caption, with 
the solid, dark-grey (blue) line representing the full NNLO 
result. At this order the scale dependence is negligible, especially 
for $\mu> 4\,$GeV. The last line in (\ref{eq:a4unum}) and 
(\ref{eq:a4cnum}) provides an estimate of the theoretical 
uncertainty of the entire QCD penguin amplitude coefficient 
$a_4^p$, which is also shown in figure~\ref{fig:anatomya4ua4c}. 
Figure~\ref{fig:anatomya4ua4c} also displays the uncertainty 
at LO and NLO for comparison. Somewhat surprisingly, the uncertainty is larger 
at NNLO than at NLO. Most of this effect is caused by 
the NNLO $\mathcal{O}(\alpha_s^2)$ correction to 
spectator scattering, and hence was present already in 
\cite{Beneke:2006mk}. The total uncertainty arises to a large part 
from the uncertainty in hadronic input parameters, which cannot be 
reduced by perturbative calculations. The dominant hadronic 
uncertainties are due to the second Gegenbauer moment of the 
light-meson LCDAs and, despite the smallness 
of the spectator-scattering contribution, the inverse moment 
$\lambda_B$ of the $B$-meson LCDA. We note, however, that 
the two-loop correction to the form-factor term calculated in the 
present paper roughly doubles
the sensitivity to the first two Gegenbauer moments of the 
light-meson LCDA. Similarly, the sensitivity of 
$\mbox{Im} \,[a_4^c]_{\rm FF}$ to the value of the charm-quark 
mass is larger in the NNLO approximation than at NLO.

%
%
%

\section{Conclusion}
\label{sec:conclusion}

In this paper we computed the remaining missing piece to the leading 
QCD penguin amplitude at NNLO in QCD factorization from the two-loop 
matrix elements of the penguin operators $Q_{3-6}$ in the 
effective weak Hamiltonian. The new contribution is sizeable and 
cancels in part the previously known current-current operator 
contribution. With this result, direct CP asymmetries in charmless 
$B$ decays, which are largely determined by the interference of 
the QCD penguin amplitude with the topological tree amplitudes, 
can for the first time be calculated including the first 
QCD correction. The Belle~II experiment is expected to measure 
many charmless $B$ decays with unprecedented precision in the 
near future, motivating a reconsideration of the analysis of 
complete sets of charmless final states with pseudoscalar and/or 
vector mesons in the framework of QCD factorization. Such analyses 
can now be performed with NNLO perturbative calculations.

%
%
%

\subsubsection*{Acknowledgements}
We thank Stefan Weinzierl for useful correspondence on GiNaC 
and the authors of~\cite{Kim:2011jm} for correspondence on 
their result.
The work of GB and TH was supported by DFG Forschergruppe 
FOR 1873 ``Quark Flavour Physics and Effective Field Theories''. 
The work of MB is supported by the DFG 
Sonderforschungsbereich/Transregio 110  ``Symmetries and the 
Emergence of Structure in QCD''. The work of XL is supported by the 
National Natural Science Foundation of China under Grant Nos.~11675061 
and 11435003, and by the Fundamental Research Funds for the Central 
Universities under Grant No.~CCNU18TS029.

%
%
%

\begin{appendix}

\section{Renormalization constants}
\label{sec:matricesZij}

Here we list the matrices $Z_{ij}$ needed for operator renormalization.
The row index runs over $i = \{Q_1^p,Q_2^p,Q_{3-6},Q_{8g}\}$,
while the column index $j$ labels 
\begin{equation}
\{\underbrace{Q_1^p,Q_2^p}_{C},
\underbrace{Q_{3-6}}_{P},\underbrace{Q_{8g}}_{M}
,\underbrace{E_1^{(1),p},E_2^{(1),p}, E_3^{(1)},E_4^{(1)}}_{E^{(1)}}
,\underbrace{E_1^{(2),p},E_2^{(2),p},E_3^{(2)},E_4^{(2)}}_{E^{(2)}}\}\,.
\end{equation}
The matrices $Z_{ij}$ were computed in~\cite{Gambino:2003zm,Gorbahn:2004my},
but need to be adjusted to our operator basis since our definition of $Q_{8g}$ differs from
that in~\cite{Gambino:2003zm,Gorbahn:2004my}. 
We split up the matrices into several blocks,
\begin{equation}
Z^{(1)} = \frac{1}{\ep} \left( \begin{array}{ccccc}
(Z^{(1)}_{CC})_{2\times 2} &
(Z^{(1)}_{CP})_{2\times 4} &
(Z^{(1)}_{CM})_{2\times 1} &
(Z^{(1)}_{CE^{(1)}})_{2\times 4} &
(Z^{(1)}_{CE^{(2)}})_{2\times 4} \\[2mm]
(Z^{(1)}_{PC})_{4\times 2} &
(Z^{(1)}_{PP})_{4\times 4} &
(Z^{(1)}_{PM})_{4\times 1} &
(Z^{(1)}_{PE^{(1)}})_{4\times 4} &
(Z^{(1)}_{PE^{(2)}})_{4\times 4}\\[2mm]
0 &
0 &
(Z^{(1)}_{MM})_{1\times 1} &
0 &
0
\end{array} \right)\, ,
\end{equation}
\begin{equation}
Z^{(2)} = \sum\limits_{k=1}^{2} \frac{1}{\ep^k} \left( \begin{array}{ccccc}
(Z^{(2),k}_{CC})_{2\times 2} &
(Z^{(2),k}_{CP})_{2\times 4} &
(Z^{(2),k}_{CM})_{2\times 1} &
(Z^{(2),k}_{CE^{(1)}})_{2\times 4} &
(Z^{(2),k}_{CE^{(2)}})_{2\times 4} \\[2mm]
(Z^{(2),k}_{PC})_{4\times 2} &
(Z^{(2),k}_{PP})_{4\times 4} &
(Z^{(2),k}_{PM})_{4\times 1} &
(Z^{(2),k}_{PE^{(1)}})_{4\times 4} &
(Z^{(2),k}_{PE^{(2)}})_{4\times 4}\\[2mm]
0 &
0 &
(Z^{(2),k}_{MM})_{1\times 1} &
0 & 
0
\end{array} \right)\, ,
\end{equation}
which in our operator basis, for $n_f=5$ quark flavours and $T_f=1/2$
read
\begin{align}
Z^{(1)}_{CC} & =
\left( \begin{array}{cc}
-2 & \frac{4}{3} \\[0.1cm]
6 & 0
\end{array} \right) \,, &
Z^{(1)}_{CP} & =
\left( \begin{array}{cccc}
0 & -\frac{1}{9} & 0 & 0 \\[0.1cm]
0 &  \frac{2}{3} & 0 & 0
\end{array} \right) \,, &
Z^{(1)}_{CM} & =
\left( \begin{array}{c}
\frac{167}{648} \\[0.1cm]
\frac{ 19}{ 27}
\end{array} \right) \,, \nnb\\[0.3cm]
Z^{(1)}_{CE^{(1)}} & =
\left( \begin{array}{cccc}
\frac{5}{12} & \frac{2}{9} & 0 & 0 \\[0.1cm]
1 &           0 & 0 & 0
\end{array} \right) \,, &
Z^{(1)}_{CE^{(2)}} & =
\left( \begin{array}{cccc}
0 & 0 & 0 & 0 \\[0.1cm]
0 & 0 & 0 & 0
\end{array} \right)\,,
\end{align}
\begin{align}
Z^{(2),1}_{CC} & =
\left( \begin{array}{cc}
\frac{317}{36} & -\frac{515}{54} \\[0.1cm]
\frac{349}{12} & 3
\end{array} \right) \,, &
Z^{(2),1}_{CP} & =
\left( \begin{array}{cccc}
-\frac{353}{243} & -\frac{1567}{972} & \frac{67}{486} & -\frac{35}{648} \\[0.1cm]
-\frac{104}{81} & \frac{338}{81} & \frac{14}{81} & \frac{35}{108} 
\end{array} \right) \,, & 
Z^{(2),1}_{CM} & =
\left( \begin{array}{c}
-\frac{9625}{8748} \\[0.1cm]
\frac{5749}{5832}
\end{array} \right)\,, \nnb\\[0.3cm]
Z^{(2),1}_{CE^{(1)}} & =
\left( \begin{array}{cccc}
\frac{4493}{864} & -\frac{49}{648} & 0 & 0 \\[0.1cm]
\frac{1031}{144} & \frac{8}{9} & 0 & 0 
\end{array} \right) \,, &
Z^{(2),1}_{CE^{(2)}} & =
\left( \begin{array}{cccc}
\frac{1}{384} & -\frac{35}{864} & 0 & 0 \\[0.1cm]
-\frac{35}{192} & -\frac{7}{72} & 0 & 0
\end{array} \right)\,, 
\end{align}
\begin{align}
Z^{(2),2}_{CC} & =
\left( \begin{array}{cc}
\frac{41}{3} & -\frac{58}{9} \\[0.1cm]
-29 & 4
\end{array} \right) \,, &
Z^{(2),2}_{CP} & =
\left( \begin{array}{cccc}
\frac{10}{81} & \frac{209}{162} & -\frac{1}{81} & -\frac{5}{216} \\[0.1cm]
-\frac{20}{27} & -\frac{128}{27} & \frac{2}{27} & \frac{5}{36}
\end{array} \right) \,, &
Z^{(2),2}_{CM} & =
\left( \begin{array}{c}
-\frac{28189}{17496} \\[0.1cm]
-\frac{7249}{1458}
\end{array} \right)\,, \nnb\\[0.3cm]
Z^{(2),2}_{CE^{(1)}} & =
\left( \begin{array}{cccc}
-\frac{125}{36} & -\frac{73}{54} & 0 & 0 \\[0.1cm]
-\frac{73}{12} & 0 & 0 & 0
\end{array} \right) \,, &
Z^{(2),2}_{CE^{(2)}} & =
\left( \begin{array}{cccc}
\frac{19}{96} & \frac{5}{108} & 0 & 0 \\[0.1cm]
\frac{5}{24} & \frac{1}{9} & 0 & 0
\end{array} \right)\,,
\end{align}
\begin{align}
Z^{(1)}_{PC} & =
\left( \begin{array}{cc}
0 & 0 \\[0.1cm]
0 & 0 \\[0.1cm]
0 & 0 \\[0.1cm]
0 & 0 \\[0.1cm]
\end{array} \right) \,, &
Z^{(1)}_{PP} & =
\left( \begin{array}{cccc}
0 & -\frac{26}{3} & 0 & 1 \\[0.1cm]
-\frac{20}{9} & -\frac{50}{9} & \frac{2}{9} & \frac{5}{12} \\[0.1cm]
0 & -\frac{128}{3} & 0 & 10 \\[0.1cm]
-\frac{128}{9} & \frac{28}{9} & \frac{20}{9} & -\frac{1}{3}
\end{array} \right) \,, &
Z^{(1)}_{PM} & =
\left( \begin{array}{c}
\frac{92}{27} \\[0.1cm]
-\frac{1409}{648} \\[0.1cm]
\frac{3407}{27} \\[0.1cm]
-\frac{1081}{81}
\end{array} \right)\,, \nnb\\[0.3cm]
Z^{(1)}_{PE^{(1)}} & =
\left( \begin{array}{cccc}
0 & 0 & 0 & 0 \\[0.1cm]
0 & 0 & 0 & 0 \\[0.1cm]
0 & 0 & 0 & 1 \\[0.1cm]
0 & 0 & \frac{2}{9} & \frac{5}{12}
\end{array} \right) \,, &
Z^{(1)}_{PE^{(2)}} & =
\left( \begin{array}{cccc}
0 & 0 & 0 & 0 \\[0.1cm]
0 & 0 & 0 & 0 \\[0.1cm]
0 & 0 & 0 & 0 \\[0.1cm]
0 & 0 & 0 & 0
\end{array} \right)\,,
\end{align}
\begin{align}
Z^{(2),1}_{PC} & =
\left( \begin{array}{cc}
0 & 0 \\[0.1cm]
0 & 0 \\[0.1cm]
0 & 0 \\[0.1cm]
0 & 0 \\[0.1cm]
\end{array} \right) \,, &
Z^{(2),1}_{PP} & =
\left( \begin{array}{cccc}
-\frac{1117}{81} & -\frac{31469}{324} & \frac{100}{81} & \frac{3373}{432} \\[0.1cm]
-\frac{4079}{486} & -\frac{59399}{972} & \frac{269}{1944} & \frac{12899}{2592} \\[0.1cm]
-\frac{83080}{81} & -\frac{159926}{81} & \frac{8839}{81} & \frac{14573}{54} \\[0.1cm]
\frac{70100}{243} & -\frac{231956}{243} & -\frac{11501}{243} & \frac{78089}{648}
\end{array} \right) \,, &
Z^{(2),1}_{PM} & =
\left( \begin{array}{c}
\frac{35113}{729} \\[0.1cm]
-\frac{1356773}{34992} \\[0.1cm]
\frac{3116449}{1458} \\[0.1cm]
-\frac{20383751}{17496} 
\end{array} \right)\,, \nnb
\end{align}
\begin{align}
Z^{(2),1}_{PE^{(1)}} & =
\left( \begin{array}{cccc}
0 & 0 & -\frac{7}{72} & -\frac{35}{192} \\[0.1cm]
0 & 0 & -\frac{35}{864} & \frac{1}{384} \\[0.1cm]
0 & 0 & \frac{23}{18} & \frac{449}{36} \\[0.1cm]
0 & 0 & \frac{179}{162} & \frac{463}{108}
\end{array} \right) \,, &
Z^{(2),1}_{PE^{(2)}} & =
\left( \begin{array}{cccc}
0 & 0 & 0 & 0 \\[0.1cm]
0 & 0 & 0 & 0 \\[0.1cm]
0 & 0 & -\frac{7}{72} & -\frac{35}{192} \\[0.1cm]
0 & 0 & -\frac{35}{864} & \frac{1}{384} 
\end{array} \right)\,,
\end{align}
\begin{align}
Z^{(2),2}_{PC} & =
\left( \begin{array}{cc}
0 & 0 \\[0.1cm]
0 & 0 \\[0.1cm]
0 & 0 \\[0.1cm]
0 & 0 \\[0.1cm]
\end{array} \right) \,, &
Z^{(2),2}_{PP} & =
\left( \begin{array}{cccc}
\frac{68}{27} & \frac{1589}{27} & \frac{4}{27} & -\frac{209}{36} \\[0.1cm]
\frac{950}{81} & \frac{6847}{162} & -\frac{163}{162} & -\frac{305}{108} \\[0.1cm]
-\frac{640}{27} & \frac{8036}{27} & \frac{172}{27} & -\frac{440}{9} \\[0.1cm]
\frac{4328}{81} & -\frac{556}{81} & -\frac{692}{81} & \frac{323}{54}
\end{array} \right) \,, &
Z^{(2),2}_{PM} & =
\left( \begin{array}{c}
-\frac{65311}{2916} \\[0.1cm]
\frac{106981}{4374} \\[0.1cm]
-\frac{649774}{729} \\[0.1cm]
\frac{1044751}{4374}
\end{array} \right)\,, \nnb
\end{align}
\begin{align}
Z^{(2),2}_{PE^{(1)}} & =
\left( \begin{array}{cccc}
0 & 0 & \frac{1}{9} & \frac{5}{24} \\[0.1cm]
0 & 0 & \frac{5}{108} & \frac{19}{96} \\[0.1cm]
0 & 0 & -\frac{4}{9} & \frac{5}{12} \\[0.1cm]
0 & 0 & -\frac{211}{54} & -\frac{29}{24}
\end{array} \right) \,, &
Z^{(2),2}_{PE^{(2)}} & =
\left( \begin{array}{cccc}
0 & 0 & 0 & 0 \\[0.1cm]
0 & 0 & 0 & 0 \\[0.1cm]
0 & 0 & \frac{1}{9} & \frac{5}{24} \\[0.1cm]
0 & 0 & \frac{5}{108} & \frac{19}{96}
\end{array} \right)\,,
\end{align}
\begin{align}
(Z^{(1)}_{MM})_{1\times 1} &=\frac{14}{3} \,, &
(Z^{(2),1}_{MM})_{1\times 1}&=\frac{4063}{108} \,, &
(Z^{(2),2}_{MM})_{1\times 1}&=-\frac{337}{12}\,.
\end{align}


\section{Amplitude functions}
\label{sec:ampfunc}

For presenting the expressions of the two-loop penguin amplitudes we use the following abbreviations,
\be
\dps L = \ln\left(\frac{\mu^2}{m_b^2}\right) \, , \quad \dps r=\sqrt{1-4 z_c} \, , \quad \dps \psi^{(1)}(z) = \frac{d^2}{dz^2} \ln \Gamma(z) \, .
\ee
Moreover, we define the following combinations of functions
\begin{align}
g_0(r) = & \, H_{w_1^-}(r)+2 \ln(2) = - \ln(z_c) \, , \nnb \\[0.2em]
g_1(r) = & \, H_{w_1^+}(r)-i \pi = \ln\left(\frac{1+r}{1-r}\right)-i \pi = \ln\left(\frac{r+1}{r-1}\right)\, , \nnb \\[0.2em]
g_2(r) = & \, \psi^{(1)}\!\!\left(\frac{1}{6}\right) - 2 \pi^2 \, , \nnb \\[0.2em]
g_3(r) = & \, H_{w_1^+,w_1^+}(r)-i \pi \, g_1(r) + \frac{\pi ^2}{2}
       = \frac{1}{2} \, \ln^2\left(\frac{1+r}{1-r}\right)-i \pi \, g_1(r) + \frac{\pi ^2}{2} \, , \nnb \\[0.2em]
g_4(r) = & \, H_{w_1^-,w_1^+,w_1^+}(r)-i \pi \, H_{w_1^-,w_1^+}(r)+\frac{7\zeta(3)}{2}-\frac{\pi ^2}{2} \, g_0(r) + i \pi \, g_3(r) \nnb \\[0.2em]
       = & \, 2 \, \text{Li}_3\left(\frac{1-r}{2}\right) + 2 \, \text{Li}_3\left(\frac{1+r}{2}\right) +\ln\left(\frac{1-r}{2}\right) \ln \left(\frac{1+r}{2}\right) \ln (z_c)\nnb \\[0.2em]
        & \, -\left[\text{Li}_2\left(\frac{1+r}{2}\right)-\text{Li}_2\left(\frac{1-r}{2}\right)\right] \left[\ln \left(\frac{1+r}{1-r}\right)-i \pi \right]+\pi ^2 \ln \left(\frac{1+r}{1-r}\right) \nnb \\[0.2em]
	& \, +i \pi  \ln \left(\frac{1+r}{2}\right) \ln
   \left(\frac{1+r}{1-r}\right)-\frac{1}{6} \ln ^3(z_c)+\frac{\pi ^2}{3}  \ln (z_c)-\frac{i \pi ^3}{2}\, . \label{eq:funcgi}
\end{align}

While all terms involving $L$ were obtained in a fully analytic manner, 
the $L^0$ terms are available in analytic form only for $I^{(2)}_{1u}$, 
$I^{(2)}_{2u}$ and $I^{(2)}_{8g}$. For the $L^0$ term in $I^{(2)}_{1c}$, 
$I^{(2)}_{2c}$ and $I^{(2)}_{3-6}$ we calculated numerical values for 
318 points in the interval $0.01 \le z_c \le 1$. We provide the data 
tables for these functions electronically in ancillary files. In the 
following write-up we present fits which reproduce the data at the level 
of $5$~per mille for all $0.01 \le z_c \le 1$. For the physical region 
$0.05 < z_c < 0.2$ the agreement is well below the per-mille level. 
We construct this fit by making the following ansatz  
for the zeroth, first and second Gegenbauer moment, respectively:
\begin{align}
{\cal F}^{(i,0)}(z_c) = & \left(a^{(i,0)}_1 \, z_c+a^{(i,0)}_2\right) \, z_c^2 \, g_4(r) +  \left(a^{(i,0)}_3 \, z_c+a^{(i,0)}_4\right) \, z_c^2 \, g_3(r) \nnb \\[0.2em]
   &+\left(a^{(i,0)}_5 \, z_c^2+a^{(i,0)}_6 \, z_c+a^{(i,0)}_7\right) \, r \, g_1(r)+\left(a^{(i,0)}_8 \, z_c^2+a^{(i,0)}_9 \, z_c+a^{(i,0)}_{10}\right) g_0(r) \nnb \\[0.2em]
   &+\left(a^{(i,0)}_{11} \, z_c^2+a^{(i,0)}_{12} \, z_c+a^{(i,0)}_{13}\right) +i \pi  \left[r \left(b^{(i,0)}_1 \, z_c^2+b^{(i,0)}_2 \, z_c+b^{(i,0)}_3\right) g_1(r) \right.\nnb \\[0.2em]
   & \left.+b^{(i,0)}_4 \, g_0(r)+\left(b^{(i,0)}_5 \, z_c^2+b^{(i,0)}_6 \, z_c+b^{(i,0)}_7\right)\right] \; , \\[0.8em]
{\cal F}^{(i,1)}(z_c) = & \left(a^{(i,1)}_1 \, z_c^2 + a^{(i,1)}_2 \, z_c+a^{(i,1)}_3\right) \, z_c^2 \, g_4(r) +  \left(a^{(i,1)}_4 \, z_c^2 +a^{(i,1)}_5 \, z_c+a^{(i,1)}_6\right) \, z_c^2 \, g_3(r) \nnb \\[0.2em]
   &+\left(a^{(i,1)}_7 \, z_c^3+a^{(i,1)}_8 \, z_c^2+a^{(i,1)}_9 \, z_c+a^{(i,1)}_{10}\right) \, r \, g_1(r)+a^{(i,1)}_{11}\, g_0(r) \nnb \\[0.2em]
   &+\left(a^{(i,1)}_{12} \, z_c^3+a^{(i,1)}_{13} \, z_c^2+a^{(i,1)}_{14} \, z_c+a^{(i,1)}_{15}\right) \nnb \\[0.2em]
   &+i \pi  \left[r \left(b^{(i,1)}_1 \, z_c^3+b^{(i,1)}_2 \, z_c^2+b^{(i,1)}_3 \, z_c+b^{(i,1)}_4\right) g_1(r) \right.\nnb \\[0.2em]
   & \left.+b^{(i,1)}_5 \, g_0(r)+\left(b^{(i,1)}_6 \, z_c^3+b^{(i,1)}_7 \, z_c^2+b^{(i,1)}_8 \, z_c+b^{(i,1)}_9\right)\right] \; , \\[0.8em]
{\cal F}^{(i,2)}(z_c) = & \left(a^{(i,2)}_1 \, z_c^3 + a^{(i,2)}_2 \, z_c^2 + a^{(i,2)}_3 \, z_c+a^{(i,2)}_4\right) \, z_c^2 \, g_4(r) \nnb \\[0.2em]
   &+  \left(a^{(i,2)}_5 \, z_c^3 +a^{(i,2)}_6 \, z_c^2 +a^{(i,2)}_7 \, z_c+a^{(i,2)}_8\right) \, z_c^2 \, g_3(r) \nnb \\[0.2em]
   &+\left(a^{(i,2)}_9 \, z_c^4+a^{(i,2)}_{10} \, z_c^3+a^{(i,2)}_{11} \, z_c^2+a^{(i,2)}_{12} \, z_c+a^{(i,2)}_{13}\right) \, r \, g_1(r)+a^{(i,2)}_{14}\, g_0(r) \nnb \\[0.2em]
   &+\left(a^{(i,2)}_{15} \, z_c^4+a^{(i,2)}_{16} \, z_c^3+a^{(i,2)}_{17} \, z_c^2+a^{(i,2)}_{18} \, z_c+a^{(i,2)}_{19}\right) \nnb \\[0.2em]
   &+i \pi  \left[r \left(b^{(i,2)}_1 \, z_c^4+b^{(i,2)}_2 \, z_c^3+b^{(i,2)}_3 \, z_c^2+b^{(i,2)}_4 \, z_c+b^{(i,2)}_5\right) g_1(r) \right.\nnb \\[0.2em]
   & \left.+b^{(i,2)}_6 \, g_0(r)+\left(b^{(i,2)}_7 \, z_c^4+b^{(i,2)}_8 \, z_c^3+b^{(i,2)}_9 \, z_c^2+b^{(i,2)}_{10} \, z_c+b^{(i,2)}_{11}\right)\right] \; .
\end{align}
The first and second superscript index denotes the operator and Gegenbauer 
moment, respectively. The fitted coefficients $a_n^{(i,j)}$ and 
$b_m^{(i,j)}$ are given numerically in 
tables~\ref{tab:coeffsF1}~--~\ref{tab:coeffsF6} below.

The amplitude functions $I^{(2)}_{1u}, I^{(2)}_{2u},I^{(2)}_{1c},
I^{(2)}_{2c}$  in sections \ref{sec:b1} -- \ref{sec:b4} were computed in 
\cite{Bell:2015koa}. We provide here the analytic expression not 
given in the letter publication.

\subsection{The amplitude function $I^{(2)}_{1u}$ of $Q_{1}^u$}
\label{sec:b1}

\begin{align}
I^{(2)}_{1u} = \; \; \; \; \, & \frac{280}{729} \, L^2 + L \left[\frac{32}{81} (4 z_c-3) \, z_c^2 \, g_3(r)
     +\frac{8}{243} \, r \left(24 z_c^2+2 z_c+1\right)  \, g_1(r) -\frac{8}{243} g_0(r) \right. \nnb \\[0.2em]
   &\left.  -\frac{64}{81} \, z_c^2-\frac{128}{243} \, z_c-\frac{40 \pi^2}{2187}+\frac{20 \pi }{27 \sqrt{3}} 
     +\frac{6113}{2187}+\frac{586}{729} \, i \pi \right] +\frac{32}{81} (4 z_c-3) \, z_c^2 \, g_4(r) \nnb \\[0.2em]
   & -\frac{112}{81} \, z_c^2 \, g_3(r) -\frac{88 \pi}{3645 \sqrt{3}} \, g_2(r) -\frac{2}{135} \, g_2(r) +\frac{4}{81} \, r \left(32 z_c^2+2 z_c+1\right) g_1(r) \nnb \\[0.2em]
   & -\frac{4}{81} \, g_0(r)+\frac{8 \zeta(3)}{9}-\frac{64}{27} \, z_c^2-\frac{32}{27} \, z_c-\frac{2194 \pi ^2}{6561}+\frac{346 \pi }{243 \sqrt{3}}+\frac{47093}{13122}
     +i\pi \left(\frac{4432}{2187} \right. \nnb \\[0.2em]
   & \left.+\frac{8}{243} \, r \left(24 z_c^2+2 z_c+1\right) g_1(r)-\frac{8}{243} \, g_0(r)-\frac{64}{81} \, z_c^2-\frac{128}{243} \, z_c-\frac{82 \pi ^2}{729}+ \frac{8 \pi }{9\sqrt{3}}\right) \nnb \\[0.2em]
   + a_1^M & \left\{ L \left[-\frac{32}{9} \left(6 z_c^2-4 z_c+1\right) z_c^2 \, g_3(r)-\frac{16}{3} \, r \, (2 z_c-1) \, z_c^2 \, g_1(r)+\frac{32}{3} \, z_c^3-\frac{56}{9} \, z_c^2\right.\right.\nnb \\[0.2em]
   & \left. -\frac{16}{27} \, z_c +\frac{40 \pi^2}{81}-\frac{40 \pi }{9\sqrt{3}} +\frac{142}{27}+\frac{2}{81} \, i \pi \right]
    -\frac{32}{9} \left(6 z_c^2-4 z_c+1\right) z_c^2 \, g_4(r) -\frac{2}{81} g_0(r) \nnb \\[0.2em]
   & + \frac{16}{27} \left(9 z_c^2-7\right) z_c^2 \, g_3(r) +\frac{104 \pi}{135 \sqrt{3}} \, g_2(r)+\frac{4}{45} \, g_2(r)+ \frac{88}{3} \, z_c^3-\frac{530}{27} \, z_c^2-\frac{164}{81} \, z_c \nnb \\[0.2em]
   & -\frac{2}{81} \, r \left(756 z_c^3-474 z_c^2-2 z_c-1\right) g_1(r) -\frac{380 \zeta(3)}{27}-\frac{230 \pi^2}{729}-\frac{179 \pi }{27 \sqrt{3}}+\frac{29777}{2916} \nnb \\[0.2em]
   & \left.  +i \pi  \left(-\frac{16}{3} \, r \, (2 z_c-1) \, z_c^2 \, g_1(r)+\frac{32}{3} \, z_c^3-\frac{56}{9} \, z_c^2-\frac{16}{27} \, z_c
   +\frac{82 \pi^2}{81}-\frac{16 \pi }{3 \sqrt{3}}+\frac{50}{81}\right)\right\} \nnb \\[0.2em]
   + a_2^M & \left\{ L \left[\frac{64}{9} \left(40 z_c^3-30 z_c^2+8 z_c-1\right) z_c^2 \, g_3(r) + \frac{32}{27} \, r \left(120 z_c^2-70 z_c+13\right) z_c^2 \, g_1(r)\right.\right. \nnb \\[0.2em]
   & \left. -\frac{1280}{9} \, z_c^4+\frac{2560}{27} \, z_c^3-\frac{1712}{81} \, z_c^2-\frac{16}{27} \, z_c-\frac{1360 \pi^2}{243}+\frac{560 \pi }{9 \sqrt{3}}-\frac{9053}{162} \right]
    -\frac{4}{405} \, g_0(r)\nnb \\[0.2em]
   & +\frac{64}{9} \left(40 z_c^3-30 z_c^2+8 z_c-1\right) z_c^2 \, g_4(r) -\frac{32}{27} \left(88 z_c^3-45 z_c^2+7\right) z_c^2 \, g_3(r) \nnb \\[0.2em]
   & -\frac{1184 \pi}{135\sqrt{3}} \, g_2(r)-\frac{56}{45} \, g_2(r)+\frac{4}{405} \, r \! \left(23520 z_c^4-15780 z_c^3+3786 z_c^2+2 z_c+1\right) \! g_1(r) \nnb \\[0.2em]
   & -\frac{10112}{27} \, z_c^4+272 \, z_c^3-\frac{85924}{1215} \, z_c^2-\frac{1012}{405} z_c+\frac{14408 \zeta (3)}{81}+\frac{19132 \pi ^2}{3645} +\frac{2824 \pi }{27 \sqrt{3}} \nnb \\[0.2em]
   & -\frac{1092649}{7290}+i \pi  \left(\frac{32}{27} \, r \left(120 z_c^2-70 z_c+13\right) z_c^2 \, g_1(r)-\frac{1280}{9} \, z_c^4+\frac{2560}{27} \, z_c^3 \right.\nnb \\[0.2em]
   &\left.\left. -\frac{1712}{81} \, z_c^2-\frac{16}{27} \, z_c-\frac{76 \pi^2}{9}+\frac{224 \pi }{3 \sqrt{3}}-\frac{125209}{2430}\right)\right\} \, .
\end{align}
\subsection{The amplitude function $I^{(2)}_{2u}$ of $Q_{2}^u$}
\begin{align}
I^{(2)}_{2u} = & -6\, I^{(2)}_{1u} + \frac{4}{3} \, L^2+ L \left[16+\frac{8 }{3} \, i \pi \right] -\frac{8 \pi ^2}{9} +\frac{455}{27}+8 i \pi \nnb \\[0.2em]
   & + a_1^M \left\{ 14 \, L+\frac{179}{6}+6 \, i \pi \right\} + a_2^M \left\{ \frac{64}{5} \, L+\frac{2227}{75}+\frac{24}{5} \, i \pi\right\} \, .
\end{align}
\subsection{The amplitude function $I^{(2)}_{1c}$ of $Q_{1}^c$}
\begin{align}
I^{(2)}_{1c} = \; \; \; \; \, & \frac{280}{729} \, L^2 + L \left[ -\frac{272}{27} (4 z_c-3) \, z_c^2 \, g_3(r)
   -\frac{68}{81} \, r \left(24 z_c^2+2 z_c+1\right) \, g_1(r)+\frac{68}{81} \, g_0(r) \right. \nnb \\[0.2em]
   &\left. +\frac{544}{27} \, z_c^2+\frac{1088}{81} \, z_c-\frac{40 \pi ^2}{2187}+\frac{20 \pi }{27 \sqrt{3}}+\frac{6113}{2187}-\frac{50}{729} \, i \pi \right]
   + {\cal F}^{(1,0)}(z_c) \nnb \\[0.2em]
   + a_1^M & \left\{L \left[\frac{2960}{27} \left(6 z_c^2-4 z_c+1\right) z_c^2 \, g_3(r) + \frac{1480}{9} \, r \, (2 z_c-1) \, z_c^2 \, g_1(r)-\frac{2960}{9} \, z_c^3 \right.\right.\nnb \\[0.2em]
   & \left. \left. +\frac{5180}{27} \, z_c^2+\frac{1480}{81} \, z_c+\frac{40 \pi ^2}{81}-\frac{40 \pi}{9 \sqrt{3}} +\frac{142}{27} +\frac{2}{81} \, i \pi \right] + {\cal F}^{(1,1)}(z_c) \right\} \nnb \\[0.2em]
   + a_2^M & \left\{ L \left[ -\frac{6496}{27} \left(40 z_c^3-30 z_c^2+8 z_c-1\right) z_c^2 \, g_3(r) 
   -\frac{1360 \pi ^2}{243}+\frac{560 \pi }{9 \sqrt{3}}-\frac{9053}{162} \right.\right. \nnb \\[0.2em]
   &   \nnb  -\frac{3248}{81} \, r \, \left(120 z_c^2-70 z_c+13\right) \, z_c^2 \, g_1(r) +\frac{129920}{27} \, z_c^4-\frac{259840}{81} \, z_c^3 \\[0.2em]
   & \left.\left. +\frac{173768}{243} \, z_c^2+\frac{1624}{81} \, z_c \right] + {\cal F}^{(1,2)}(z_c) \right\} \, .
\end{align}
\subsection{The amplitude function $I^{(2)}_{2c}$ of $Q_{2}^c$}
\label{sec:b4}

\begin{align}
I^{(2)}_{2c} = \; \; \; \; \, & -\frac{236}{243} \, L^2 + L \left[  \frac{256}{9} (4 z_c-3) z_c^2 \, g_3(r)
     +\frac{64}{27} \, r \, \left(24 z_c^2+2 z_c+1\right) g_1(r)-\frac{64}{27} \, g_0(r) \right. \nnb \\[0.2em]
   & \left. -\frac{512}{9} \, z_c^2-\frac{1024}{27} \, z_c+\frac{80 \pi ^2}{729}-\frac{40 \pi }{9 \sqrt{3}} -\frac{562}{729}+\frac{100}{243} \, i \pi \right] + {\cal F}^{(2,0)}(z_c) \nnb \\[0.2em]
   + a_1^M & \left\{ L \left[-\frac{3328}{9} \left(6 z_c^2-4 z_c+1\right) z_c^2 \, g_3(r) -\frac{1664}{3} \, r \, (2 z_c-1) z_c^2 \, g_1(r)+\frac{3328}{3} \, z_c^3
    \right.\right.\nnb \\[0.2em]
   & \left.\left. -\frac{5824}{9} \, z_c^2-\frac{1664}{27} \, z_c-\frac{80 \pi ^2}{27}+\frac{80 \pi }{3\sqrt{3}} -\frac{158}{9} -\frac{4}{27} \, i \pi \right] + {\cal F}^{(2,1)}(z_c) \right\} \nnb \\[0.2em]
   + a_2^M & \left\{ L \left[ \frac{7808}{9} \left(40 z_c^3-30 z_c^2+8 z_c-1\right) z_c^2 \, g_3(r)+\frac{2720 \pi ^2}{81}-\frac{1120 \pi }{3 \sqrt{3}}+\frac{46993}{135} \right.\right. \nnb \\[0.2em]
   &  +\frac{3904}{27} \, r \, \left(120 z_c^2-70 z_c+13\right) z_c^2 \, g_1(r)-\frac{156160}{9} \, z_c^4+\frac{312320}{27} \, z_c^3 \nnb \\[0.2em]
   &  \left.\left. -\frac{208864}{81} \, z_c^2-\frac{1952}{27} \, z_c \right] + {\cal F}^{(2,2)}(z_c) \right\} \, .
\end{align}
\subsection{The amplitude function $I^{(2)}_{3}$ of $Q_{3}$}
\begin{align}
I^{(2)}_{3} = \; \; \; \; \, &  -\frac{3280}{243} \, L^2 + L \left[   -\frac{128}{27} (4 z_c-3) z_c^2 \, g_3(r)
   -\frac{32}{81} \, r \, \left(24 z_c^2+2 z_c+1\right) g_1(r) +\frac{544 \pi }{9 \sqrt{3}} \right. \nnb \\[0.2em]
   & \left. +\frac{32}{81} \, g_0(r)+\frac{256}{27} \, z_c^2+\frac{512}{81} \, z_c-\frac{1088 \pi ^2}{729} -\frac{134216}{729} -\frac{3232}{243} \, i \pi \right] 
     + {\cal F}^{(3,0)}(z_c)  \nnb \\[0.2em]
   + a_1^M & \left\{ L \left[\frac{128}{3} \left(6 z_c^2-4 z_c+1\right) z_c^2 \, g_3(r) + 64 \, r \, (2 z_c-1) z_c^2 \, g_1(r) -128\, z_c^3 +\frac{224}{3} \, z_c^2 \right.\right.\nnb \\[0.2em]
   & \left. \left. +\frac{64}{9} \, z_c+\frac{1600 \pi ^2}{27}-\frac{1600 \pi }{3 \sqrt{3}} +\frac{31976}{81}-\frac{448}{27} \, i \pi \right] + {\cal F}^{(3,1)}(z_c) \right\} \nnb \\[0.2em]
   + a_2^M & \left\{ L \left[-\frac{256}{3} \left(40 z_c^3-30 z_c^2+8 z_c-1\right) z_c^2 \, g_3(r)-\frac{64192 \pi ^2}{81}+\frac{26432 \pi }{3 \sqrt{3}}-\frac{663298}{81}
     \right.\right. \nnb \\[0.2em]
   &  -\frac{128}{9} \, r \, \left(120 z_c^2-70 z_c+13\right) z_c^2 \, g_1(r)+\frac{5120}{3} \, z_c^4-\frac{10240}{9} \, z_c^3 \nnb \\[0.2em]
   &  \left.\left. +\frac{6848}{27} \, z_c^2+\frac{64}{9} z_c \right] + {\cal F}^{(3,2)}(z_c)\right\} \, .
\end{align}
\subsection{The amplitude function $I^{(2)}_{4}$ of $Q_{4}$}
\begin{align}
I^{(2)}_{4} = \; \; \; \; \, & -\frac{2008}{729} \, L^2 + L \left[ \frac{2608}{81} (4 z_c-3) z_c^2 \, g_3(r)+\frac{652}{243} \, r \left(24 z_c^2+2 z_c+1\right) g_1(r)+\frac{1432\pi }{27 \sqrt{3}}
 \right. \nnb \\[0.2em]
   & \left. -\frac{652}{243} \, g_0(r)-\frac{5216}{81} \, z_c^2-\frac{10432}{243} \, z_c-\frac{2864 \pi ^2}{2187} -\frac{220898}{2187}-\frac{2986}{729} \, i \pi \right]
   + {\cal F}^{(4,0)}(z_c) \nnb \\[0.2em]
   + a_1^M & \left\{ L \left[ -\frac{3632}{9} \left(6 z_c^2-4 z_c+1\right) z_c^2 \, g_3(r) -\frac{1816}{3} \, r \, (2 z_c-1) z_c^2 \, g_1(r)
   +\frac{3632}{3} \, z_c^3 \right.\right.\nnb \\[0.2em]
   & \left. \left. -\frac{6356}{9} \, z_c^2-\frac{1816}{27} \, z_c +\frac{4144 \pi ^2}{81}-\frac{4144 \pi}{9 \sqrt{3}}+\frac{75470}{243}+\frac{326}{81} \, i \pi \right] 
   + {\cal F}^{(4,1)}(z_c) \right\} \nnb \\[0.2em]
   + a_2^M & \left\{ L \left[\frac{8416}{9} \left(40 z_c^3-30 z_c^2+8 z_c-1\right) z_c^2 \, g_3(r)
   -\frac{165376 \pi ^2}{243}+\frac{68096 \pi }{9 \sqrt{3}}-\frac{17037031}{2430} \right.\right. \nnb \\[0.2em]
   & +\frac{4208}{27} \, r \, \left(120 z_c^2-70 z_c+13\right) z_c^2 \, g_1(r) -\frac{168320}{9} \, z_c^4+\frac{336640}{27} \, z_c^3 \nnb \\[0.2em]
   & \left.\left. -\frac{225128}{81} \, z_c^2-\frac{2104}{27} \, z_c \right] + {\cal F}^{(4,2)}(z_c)\right\} \, .
\end{align}
\subsection{The amplitude function $I^{(2)}_{5}$ of $Q_{5}$}
\begin{align}
I^{(2)}_{5} = \; \; \; \; \, & -\frac{46000}{243} \, L^2 + L \left[ \frac{5504}{27} (3-4 z_c) z_c^2 \, g_3(r)-\frac{1376}{81} \, r \left(24 z_c^2+2 z_c+1\right) g_1(r)  \right. \nnb \\[0.2em]
   & +\frac{1376}{81} \, g_0(r)+\frac{11008}{27} \, z_c^2+\frac{22016}{81} \, z_c-\frac{12224 \pi ^2}{729}+\frac{6112 \pi }{9 \sqrt{3}} -\frac{1811552}{729} \nnb \\[0.2em]
   &\left. -\frac{43936}{243} \, i \pi \right] + {\cal F}^{(5,0)}(z_c) \nnb \\[0.2em]
   + a_1^M & \left\{ L \left[ \frac{5504}{3} \left(6 z_c^2-4 z_c+1\right) z_c^2 \, g_3(r) +2752 \, r \, (2 z_c-1) z_c^2 \, g_1(r)
   -5504 \, z_c^3 \right.\right.\nnb \\[0.2em]
   & \left.\left. +\frac{9632}{3} \, z_c^2+\frac{2752}{9} \, z_c+\frac{20416 \pi ^2}{27}-\frac{20416 \pi }{3 \sqrt{3}}+\frac{433592}{81}-\frac{7168}{27} \, i \pi \right] 
   + {\cal F}^{(5,1)}(z_c) \right\} \nnb \\[0.2em]
   + a_2^M & \left\{ L \left[ -\frac{11008}{3} \left(40 z_c^3-30 z_c^2+8 z_c-1\right) z_c^2 \, g_3(r)
   -\frac{850816 \pi ^2}{81}+\frac{350336 \pi }{3 \sqrt{3}} \right.\right. \nnb \\[0.2em]
   &  -\frac{8778976}{81}-\frac{5504}{9} \, r \, \left(120 z_c^2-70 z_c+13\right) z_c^2 \, g_1(r) +\frac{220160}{3} \, z_c^4-\frac{440320}{9} \, z_c^3 \nnb \\[0.2em]
   & \left.\left. +\frac{294464}{27} \, z_c^2+\frac{2752}{9} \, z_c \right] + {\cal F}^{(5,2)}(z_c)\right\} \, .
\end{align}
\subsection{The amplitude function $I^{(2)}_{6}$ of $Q_{6}$}
\begin{align}
I^{(2)}_{6} = \; \; \; \; \, & \frac{8552}{729} \, L^2 + L \left[ \frac{22144}{81} (4 z_c-3) z_c^2 \, g_3(r)+\frac{5536}{243} \, r \, \left(24 z_c^2+2 z_c+1\right)
   g_1(r) \right. \nnb \\[0.2em]
   & -\frac{5536}{243} \, g_0(r)-\frac{44288}{81} \, z_c^2-\frac{88576}{243} \, z_c
   -\frac{15008 \pi ^2}{2187}+\frac{7504 \pi }{27 \sqrt{3}} -\frac{645452}{2187} \nnb \\[0.2em]
   & \left. +\frac{248}{729} \, i \pi \right]+ {\cal F}^{(6,0)}(z_c) \nnb \\[0.2em]
   + a_1^M & \left\{ L \left[ -\frac{16192}{3} \, r \, (2 z_c-1) z_c^2 \, g_1(r)-\frac{32384}{9} \left(6 z_c^2-4 z_c+1\right) z_c^2 \, g_3(r)
   +\frac{32384}{3} \, z_c^3 \right.\right.\nnb \\[0.2em]
   & \left.\left. -\frac{56672}{9} \, z_c^2-\frac{16192}{27} \, z_c +\frac{26272 \pi^2}{81}-\frac{26272 \pi }{9 \sqrt{3}} +\frac{490100}{243}+\frac{5576}{81} \, i \pi \right]
   \! + {\cal F}^{(6,1)}(z_c) \!\right\} \nnb \\[0.2em]
   + a_2^M & \left\{ L \left[ \frac{76288}{9} \left(40 z_c^3-30 z_c^2+8 z_c-1\right) z_c^2 \, g_3(r)-\frac{1108672 \pi ^2}{243}+\frac{456512 \pi }{9\sqrt{3}} \right.\right. \nnb \\[0.2em]
   &  -\frac{56903078}{1215}+\frac{38144}{27} \, r \left(120 z_c^2-70 z_c+13\right) z_c^2 \, g_1(r)-\frac{1525760}{9} \, z_c^4+\frac{3051520}{27} \, z_c^3 \nnb \\[0.2em]
   & \left.\left. -\frac{2040704}{81} \, z_c^2-\frac{19072}{27} \, z_c \right] + {\cal F}^{(6,2)}(z_c)\right\} \, .
\end{align}
\subsection{The amplitude function $I^{(2)}_{8g}$ of $Q_{8g}$}
\begin{align}
I^{(2)}_{8g} = \; \; \; \; \, & - 8 \, L-\frac{64}{3} \, z_c^2 \,  g_3(r) +\frac{16 \pi }{135 \sqrt{3}} \, g_2(r)
     +\frac{16}{9} \, r \, (10 z_c-1) \, g_1(r) + \frac{16}{9} \, g_0(r) \nnb \\[0.2em]
   &  -\frac{400 \zeta (3)}{27} -\frac{416}{9} \, z_c +\frac{56 \pi ^2}{81}+\frac{188 \pi }{9 \sqrt{3}} -\frac{1964}{27}
      +i \pi  \left(\frac{28}{3}-\frac{64 \pi^2}{27}\right) \nnb \\[0.2em]
   + a_1^M & \left\{-\frac{472}{27} \, L +\frac{64}{3} \, (8 z_c-9) \, z_c^2 \, g_3(r)+\frac{16 \pi }{45 \sqrt{3}} \, g_2(r)
   +\frac{16}{9} \, r \left(48 z_c^2+34 z_c-1\right)  \, g_1(r) \right.\nnb \\[0.2em]
   &+ \frac{16}{9}  \, g_0(r)+\frac{1232\zeta (3)}{9} -\frac{256}{3} \, z_c^2-\frac{1760}{9} \, z_c +\frac{184 \pi ^2}{27}+\frac{236 \pi }{3\sqrt{3}}-\frac{36484}{81} \nnb \\[0.2em]
   &\left.  +i \pi  \left(\frac{208\pi^2}{9}-\frac{6844}{27}\right)\right\} \nnb \\[0.2em]
   + a_2^M & \left\{-\frac{616}{27} \, L -\frac{128}{3} \left(45 z_c^2-40 z_c+18\right) z_c^2 \, g_3(r) -\frac{128 \pi}{45 \sqrt{3}} \, g_2(r) + \frac{16}{9} g_0(r)+\frac{2618}{81} \right. \nnb \\[0.2em]
   & -\frac{16}{9} \, r \left(540 z_c^3-390 z_c^2-70 z_c+1\right) \, g_1(r) -\frac{4960 \zeta (3)}{9} + 960 \, z_c^3 -\frac{2320}{3} \, z_c^2 \nnb \\[0.2em]
   &\left. -\frac{4256}{9} \, z_c +\frac{752 \pi ^2}{27}+\frac{596 \pi }{3\sqrt{3}} +i \pi \left(\frac{24272}{27}-\frac{848 \pi^2}{9}\right)\right\} \, .
\end{align}


\begin{table}[!p]
\begin{center}
\scalebox{0.8}{\begin{tabular}{||l|c||l|c||l|c||} \hline\hline
%
 \rule{0pt}{18pt}  $a^{(1,0)}_1$ & $2.10150698  $ & $a^{(1,0)}_8$    & $0.1218737677$ & $b^{(1,0)}_1$ & $21.43576977   $ \\[0.2em] \hline
 \rule{0pt}{18pt}  $a^{(1,0)}_2$ & $-16.0193528 $ & $a^{(1,0)}_9$    & $1.251355621 $ & $b^{(1,0)}_2$ & $2.117746614   $ \\[0.2em] \hline
 \rule{0pt}{18pt}  $a^{(1,0)}_3$ & $-482.457262 $ & $a^{(1,0)}_{10}$ & $2.490976541 $ & $b^{(1,0)}_3$ & $0.0707330472  $ \\[0.2em] \hline
 \rule{0pt}{18pt}  $a^{(1,0)}_4$ & $0.5294506079$ & $a^{(1,0)}_{11}$ & $-907.3052851$ & $b^{(1,0)}_4$ & $-0.06402063106$ \\[0.2em] \hline
 \rule{0pt}{18pt}  $a^{(1,0)}_5$ & $333.7983801 $ & $a^{(1,0)}_{12}$ & $56.80265498 $ & $b^{(1,0)}_5$ & $-41.8211093   $ \\[0.2em] \hline
 \rule{0pt}{18pt}  $a^{(1,0)}_6$ & $-13.62934048$ & $a^{(1,0)}_{13}$ & $2.265572489 $ & $b^{(1,0)}_6$ & $-8.580816057  $ \\[0.2em] \hline
 \rule{0pt}{18pt}  $a^{(1,0)}_7$ & $-2.446234588$ &                  &                & $b^{(1,0)}_7$ & $-0.02097859805$ \\[0.2em] \hline\hline
\end{tabular}}

\vspace*{10pt}

\scalebox{0.8}{\begin{tabular}{||l|c||l|c||l|c||} \hline\hline
%
 \rule{0pt}{18pt}  $a^{(1,1)}_1$ & $3543.885047 $  & $a^{(1,1)}_{10}$ & $-1.081300063$ & $b^{(1,1)}_1$ & $-10740.46597   $ \\[0.2em] \hline
 \rule{0pt}{18pt}  $a^{(1,1)}_2$ & $6937.213584 $  & $a^{(1,1)}_{11}$ & $1.02297185  $ & $b^{(1,1)}_2$ & $-2396.113061   $ \\[0.2em] \hline
 \rule{0pt}{18pt}  $a^{(1,1)}_3$ & $601.8671257 $  & $a^{(1,1)}_{12}$ & $5727.932916 $ & $b^{(1,1)}_3$ & $-8.547573142   $ \\[0.2em] \hline
 \rule{0pt}{18pt}  $a^{(1,1)}_4$ & $17096.14195 $  & $a^{(1,1)}_{13}$ & $-5513.652015$ & $b^{(1,1)}_4$ & $0.003377968074 $ \\[0.2em] \hline
 \rule{0pt}{18pt}  $a^{(1,1)}_5$ & $-5755.173571$  & $a^{(1,1)}_{14}$ & $109.7280251 $ & $b^{(1,1)}_5$ & $-0.008316256123$ \\[0.2em] \hline
 \rule{0pt}{18pt}  $a^{(1,1)}_6$ & $-1712.046812$  & $a^{(1,1)}_{15}$ & $4.641502534 $ & $b^{(1,1)}_6$ & $23252.85106    $ \\[0.2em] \hline
 \rule{0pt}{18pt}  $a^{(1,1)}_7$ & $2296.042054 $  &                  &                & $b^{(1,1)}_7$ & $6618.495416    $ \\[0.2em] \hline
 \rule{0pt}{18pt}  $a^{(1,1)}_8$ & $3636.785527 $  &                  &                & $b^{(1,1)}_8$ & $48.2883333     $ \\[0.2em] \hline
 \rule{0pt}{18pt}  $a^{(1,1)}_9$ & $-7.215453488$  &                  &                & $b^{(1,1)}_9$ & $-0.1184666327  $ \\[0.2em] \hline\hline
\end{tabular}}

\vspace*{10pt}

\scalebox{0.8}{\begin{tabular}{||l|c||l|c||l|c||} \hline\hline
%
 \rule{0pt}{18pt}  $a^{(1,2)}_1$    & $-149598.8345$  & $a^{(1,2)}_{12}$ & $-46.27940526$ & $b^{(1,2)}_1$    & $522039.7536  $ \\[0.2em] \hline
 \rule{0pt}{18pt}  $a^{(1,2)}_2$    & $-399918.2771$  & $a^{(1,2)}_{13}$ & $-0.903465316$ & $b^{(1,2)}_2$    & $218323.5282  $ \\[0.2em] \hline
 \rule{0pt}{18pt}  $a^{(1,2)}_3$    & $-75528.64587$  & $a^{(1,2)}_{14}$ & $0.8761145517$ & $b^{(1,2)}_3$    & $8625.134364  $ \\[0.2em] \hline
 \rule{0pt}{18pt}  $a^{(1,2)}_4$    & $-1708.141004$  & $a^{(1,2)}_{15}$ & $648342.4958 $ & $b^{(1,2)}_4$    & $2.539544345  $ \\[0.2em] \hline
 \rule{0pt}{18pt}  $a^{(1,2)}_5$    & $-472315.4181$  & $a^{(1,2)}_{16}$ & $589322.219  $ & $b^{(1,2)}_5$    & $0.1074203745 $ \\[0.2em] \hline
 \rule{0pt}{18pt}  $a^{(1,2)}_6$    & $315953.0032 $  & $a^{(1,2)}_{17}$ & $40100.82211 $ & $b^{(1,2)}_6$    & $-0.086103876 $ \\[0.2em] \hline
 \rule{0pt}{18pt}  $a^{(1,2)}_7$    & $201779.1562 $  & $a^{(1,2)}_{18}$ & $291.8809093 $ & $b^{(1,2)}_7$    & $-1118878.898 $ \\[0.2em] \hline
 \rule{0pt}{18pt}  $a^{(1,2)}_8$    & $9206.592797 $  & $a^{(1,2)}_{19}$ & $3.117205128 $ & $b^{(1,2)}_8$    & $-555832.9763 $ \\[0.2em] \hline
 \rule{0pt}{18pt}  $a^{(1,2)}_9$    & $-479649.846 $  &                  &                & $b^{(1,2)}_9$    & $-27420.79856 $ \\[0.2em] \hline
 \rule{0pt}{18pt}  $a^{(1,2)}_{10}$ & $-367021.3802$  &                  &                & $b^{(1,2)}_{10}$ & $-42.6189988  $ \\[0.2em] \hline
 \rule{0pt}{18pt}  $a^{(1,2)}_{11}$ & $-22103.98708$  &                  &                & $b^{(1,2)}_{11}$ & $-0.5043636115$ \\[0.2em] \hline\hline
\end{tabular}}
\caption{Coefficients of the fit functions ${\cal F}^{(1,j)}$ for $j = 0, \, 1, \, 2$.
 \label{tab:coeffsF1}}
\end{center}
\end{table}


\begin{table}[!p]
\begin{center}
\scalebox{0.8}{\begin{tabular}{||l|c||l|c||l|c||} \hline\hline
%
 \rule{0pt}{18pt}  $a^{(2,0)}_1$ & $-0.7180032569$ & $a^{(2,0)}_8$    & $-2.372611446$ & $b^{(2,0)}_1$ & $-154.0632213 $ \\[0.2em] \hline
 \rule{0pt}{18pt}  $a^{(2,0)}_2$ & $107.9658183  $ & $a^{(2,0)}_9$    & $-9.410812047$ & $b^{(2,0)}_2$ & $-13.69976447 $ \\[0.2em] \hline
 \rule{0pt}{18pt}  $a^{(2,0)}_3$ & $2071.099721  $ & $a^{(2,0)}_{10}$ & $-6.600312423$ & $b^{(2,0)}_3$ & $-0.4345976781$ \\[0.2em] \hline
 \rule{0pt}{18pt}  $a^{(2,0)}_4$ & $18.94368907  $ & $a^{(2,0)}_{11}$ & $3951.521604 $ & $b^{(2,0)}_4$ & $0.3556285549 $ \\[0.2em] \hline
 \rule{0pt}{18pt}  $a^{(2,0)}_5$ & $-1461.355037 $ & $a^{(2,0)}_{12}$ & $-240.9025729$ & $b^{(2,0)}_5$ & $307.7876363  $ \\[0.2em] \hline
 \rule{0pt}{18pt}  $a^{(2,0)}_6$ & $71.06867778  $ & $a^{(2,0)}_{13}$ & $-6.459822118$ & $b^{(2,0)}_6$ & $55.58969633  $ \\[0.2em] \hline
 \rule{0pt}{18pt}  $a^{(2,0)}_7$ & $6.485773797  $ &                  &                & $b^{(2,0)}_7$ & $0.1562463812 $ \\[0.2em] \hline\hline
\end{tabular}}

\vspace*{10pt}

\scalebox{0.8}{\begin{tabular}{||l|c||l|c||l|c||} \hline\hline
%
 \rule{0pt}{18pt}  $a^{(2,1)}_1$ & $-40016.16773$  & $a^{(2,1)}_{10}$ & $-2.686555839$ & $b^{(2,1)}_1$ & $120168.5609  $ \\[0.2em] \hline
 \rule{0pt}{18pt}  $a^{(2,1)}_2$ & $-76727.44329$  & $a^{(2,1)}_{11}$ & $2.509391671 $ & $b^{(2,1)}_2$ & $25101.75297  $ \\[0.2em] \hline
 \rule{0pt}{18pt}  $a^{(2,1)}_3$ & $-5917.913804$  & $a^{(2,1)}_{12}$ & $159124.1702 $ & $b^{(2,1)}_3$ & $12.41933868  $ \\[0.2em] \hline
 \rule{0pt}{18pt}  $a^{(2,1)}_4$ & $-119981.7486$  & $a^{(2,1)}_{13}$ & $97090.5571  $ & $b^{(2,1)}_4$ & $0.5073244648 $ \\[0.2em] \hline
 \rule{0pt}{18pt}  $a^{(2,1)}_5$ & $102219.4405 $  & $a^{(2,1)}_{14}$ & $193.2593089 $ & $b^{(2,1)}_5$ & $-0.4587752903$ \\[0.2em] \hline
 \rule{0pt}{18pt}  $a^{(2,1)}_6$ & $24558.67505 $  & $a^{(2,1)}_{15}$ & $0.4663987469$ & $b^{(2,1)}_6$ & $-260345.2975 $ \\[0.2em] \hline
 \rule{0pt}{18pt}  $a^{(2,1)}_7$ & $-119561.4974$  &		      &                & $b^{(2,1)}_7$ & $-70206.09881 $ \\[0.2em] \hline
 \rule{0pt}{18pt}  $a^{(2,1)}_8$ & $-55052.26705$  &		      &                & $b^{(2,1)}_8$ & $-217.3938691 $ \\[0.2em] \hline
 \rule{0pt}{18pt}  $a^{(2,1)}_9$ & $-60.90388979$  &		      &                & $b^{(2,1)}_9$ & $-2.471675606 $ \\[0.2em] \hline\hline
\end{tabular}}

\vspace*{10pt}

\scalebox{0.8}{\begin{tabular}{||l|c||l|c||l|c||} \hline\hline
%
 \rule{0pt}{18pt}  $a^{(2,2)}_1$    & $496051.4944 $  & $a^{(2,2)}_{12}$ & $-279.0286084$ & $b^{(2,2)}_1$    & $-658182.6253  $ \\[0.2em] \hline
 \rule{0pt}{18pt}  $a^{(2,2)}_2$    & $-393454.7326$  & $a^{(2,2)}_{13}$ & $-2.617750056$ & $b^{(2,2)}_2$    & $1325343.833   $ \\[0.2em] \hline
 \rule{0pt}{18pt}  $a^{(2,2)}_3$    & $-692700.5439$  & $a^{(2,2)}_{14}$ & $2.953838046 $ & $b^{(2,2)}_3$    & $158502.4719   $ \\[0.2em] \hline
 \rule{0pt}{18pt}  $a^{(2,2)}_4$    & $-34658.03658$  & $a^{(2,2)}_{15}$ & $-5280092.911$ & $b^{(2,2)}_4$    & $160.2188813   $ \\[0.2em] \hline
 \rule{0pt}{18pt}  $a^{(2,2)}_5$    & $579238.242  $  & $a^{(2,2)}_{16}$ & $-396123.0058$ & $b^{(2,2)}_5$    & $0.2212114815  $ \\[0.2em] \hline
 \rule{0pt}{18pt}  $a^{(2,2)}_6$    & $-3625832.486$  & $a^{(2,2)}_{17}$ & $466690.2142 $ & $b^{(2,2)}_6$    & $-0.04081938371$ \\[0.2em] \hline
 \rule{0pt}{18pt}  $a^{(2,2)}_7$    & $820298.8926 $  & $a^{(2,2)}_{18}$ & $847.6370807 $ & $b^{(2,2)}_7$    & $1564391.01    $ \\[0.2em] \hline
 \rule{0pt}{18pt}  $a^{(2,2)}_8$    & $140739.1835 $  & $a^{(2,2)}_{19}$ & $1.896328744 $ & $b^{(2,2)}_8$    & $-2936443.344  $ \\[0.2em] \hline
 \rule{0pt}{18pt}  $a^{(2,2)}_9$    & $2908868.962 $  & 		 &		  & $b^{(2,2)}_9$    & $-467072.3934  $ \\[0.2em] \hline
 \rule{0pt}{18pt}  $a^{(2,2)}_{10}$ & $-547120.1402$  & 		 &		  & $b^{(2,2)}_{10}$ & $-1315.901432  $ \\[0.2em] \hline
 \rule{0pt}{18pt}  $a^{(2,2)}_{11}$ & $-300366.2871$  & 		 &		  & $b^{(2,2)}_{11}$ & $-2.621246631  $ \\[0.2em] \hline\hline
\end{tabular}}
\caption{Coefficients of the fit functions ${\cal F}^{(2,j)}$ for $j = 0, \, 1, \, 2$.
 \label{tab:coeffsF2}}
\end{center}
\end{table}


\begin{table}[!p]
\begin{center}
\scalebox{0.8}{\begin{tabular}{||l|c||l|c||l|c||} \hline\hline
%
 \rule{0pt}{18pt}  $a^{(3,0)}_1$ & $242.9200651 $ & $a^{(3,0)}_8$    & $-4.079828122$ & $b^{(3,0)}_1$ & $-527.1937569 $ \\[0.2em] \hline
 \rule{0pt}{18pt}  $a^{(3,0)}_2$ & $232.154595	$ & $a^{(3,0)}_9$    & $-7.59265584 $ & $b^{(3,0)}_2$ & $-12.12644895 $ \\[0.2em] \hline
 \rule{0pt}{18pt}  $a^{(3,0)}_3$ & $293.4504276 $ & $a^{(3,0)}_{10}$ & $5.515604792 $ & $b^{(3,0)}_3$ & $-0.6304605763$ \\[0.2em] \hline
 \rule{0pt}{18pt}  $a^{(3,0)}_4$ & $-540.9677681$ & $a^{(3,0)}_{11}$ & $-1307.746094$ & $b^{(3,0)}_4$ & $0.6389032131 $ \\[0.2em] \hline
 \rule{0pt}{18pt}  $a^{(3,0)}_5$ & $783.6486442 $ & $a^{(3,0)}_{12}$ & $-122.6707222$ & $b^{(3,0)}_5$ & $1176.077605  $ \\[0.2em] \hline
 \rule{0pt}{18pt}  $a^{(3,0)}_6$ & $59.108871	$ & $a^{(3,0)}_{13}$ & $-129.7675486$ & $b^{(3,0)}_6$ & $61.77461189  $ \\[0.2em] \hline
 \rule{0pt}{18pt}  $a^{(3,0)}_7$ & $-5.3462818	$ &		     &  	      & $b^{(3,0)}_7$ & $-62.44498622 $ \\[0.2em] \hline\hline
\end{tabular}}

\vspace*{10pt}

\scalebox{0.8}{\begin{tabular}{||l|c||l|c||l|c||} \hline\hline
%
 \rule{0pt}{18pt}  $a^{(3,1)}_1$ & $6960.340592 $  & $a^{(3,1)}_{10}$ & $-0.8463584446$ & $b^{(3,1)}_1$ & $-21490.63623 $ \\[0.2em] \hline
 \rule{0pt}{18pt}  $a^{(3,1)}_2$ & $14133.33176 $  & $a^{(3,1)}_{11}$ & $0.2063978945 $ & $b^{(3,1)}_2$ & $-4828.183102 $ \\[0.2em] \hline
 \rule{0pt}{18pt}  $a^{(3,1)}_3$ & $1098.574764 $  & $a^{(3,1)}_{12}$ & $-59971.08153 $ & $b^{(3,1)}_3$ & $19.40708321  $ \\[0.2em] \hline
 \rule{0pt}{18pt}  $a^{(3,1)}_4$ & $13805.32877 $  & $a^{(3,1)}_{13}$ & $-26927.10186 $ & $b^{(3,1)}_4$ & $-0.6635841086$ \\[0.2em] \hline
 \rule{0pt}{18pt}  $a^{(3,1)}_5$ & $-25314.80715$  & $a^{(3,1)}_{14}$ & $-660.0530122 $ & $b^{(3,1)}_5$ & $0.7779679968 $ \\[0.2em] \hline
 \rule{0pt}{18pt}  $a^{(3,1)}_6$ & $-6237.393152$  & $a^{(3,1)}_{15}$ & $223.1912871  $ & $b^{(3,1)}_6$ & $46461.44669  $ \\[0.2em] \hline
 \rule{0pt}{18pt}  $a^{(3,1)}_7$ & $35176.987	$  &		      & 		& $b^{(3,1)}_7$ & $13431.28847  $ \\[0.2em] \hline
 \rule{0pt}{18pt}  $a^{(3,1)}_8$ & $13959.60126 $  &		      & 		& $b^{(3,1)}_8$ & $-24.95846256 $ \\[0.2em] \hline
 \rule{0pt}{18pt}  $a^{(3,1)}_9$ & $166.5508225 $  &		      & 		& $b^{(3,1)}_9$ & $-40.57571823 $ \\[0.2em] \hline\hline
\end{tabular}}

\vspace*{10pt}

\scalebox{0.8}{\begin{tabular}{||l|c||l|c||l|c||} \hline\hline
%
 \rule{0pt}{18pt}  $a^{(3,2)}_1$    & $9302.02958  $  & $a^{(3,2)}_{12}$ & $144.2121861 $ & $b^{(3,2)}_1$    & $-58362.1387   $ \\[0.2em] \hline
 \rule{0pt}{18pt}  $a^{(3,2)}_2$    & $61766.22269 $  & $a^{(3,2)}_{13}$ & $-2.152740533$ & $b^{(3,2)}_2$    & $-53721.22587  $ \\[0.2em] \hline
 \rule{0pt}{18pt}  $a^{(3,2)}_3$    & $23354.79513 $  & $a^{(3,2)}_{14}$ & $2.150927987 $ & $b^{(3,2)}_3$    & $-4814.731674  $ \\[0.2em] \hline
 \rule{0pt}{18pt}  $a^{(3,2)}_4$    & $1126.099102 $  & $a^{(3,2)}_{15}$ & $-70243.11632$ & $b^{(3,2)}_4$    & $-12.12410291  $ \\[0.2em] \hline
 \rule{0pt}{18pt}  $a^{(3,2)}_5$    & $21780.82993 $  & $a^{(3,2)}_{16}$ & $-110643.8816$ & $b^{(3,2)}_5$    & $0.006626199425$ \\[0.2em] \hline
 \rule{0pt}{18pt}  $a^{(3,2)}_6$    & $7347.139006 $  & $a^{(3,2)}_{17}$ & $-15985.79026$ & $b^{(3,2)}_6$    & $0.001329267819$ \\[0.2em] \hline
 \rule{0pt}{18pt}  $a^{(3,2)}_7$    & $-47835.15093$  & $a^{(3,2)}_{18}$ & $-486.6512829$ & $b^{(3,2)}_7$    & $121375.293    $ \\[0.2em] \hline
 \rule{0pt}{18pt}  $a^{(3,2)}_8$    & $-4787.783531$  & $a^{(3,2)}_{19}$ & $-124.8484997$ & $b^{(3,2)}_8$    & $128986.1226   $ \\[0.2em] \hline
 \rule{0pt}{18pt}  $a^{(3,2)}_9$    & $42892.27432 $  & 		 &		  & $b^{(3,2)}_9$    & $14005.89198   $ \\[0.2em] \hline
 \rule{0pt}{18pt}  $a^{(3,2)}_{10}$ & $76725.29524 $  & 		 &		  & $b^{(3,2)}_{10}$ & $75.10217647   $ \\[0.2em] \hline
 \rule{0pt}{18pt}  $a^{(3,2)}_{11}$ & $9489.604785 $  & 		 &		  & $b^{(3,2)}_{11}$ & $-39.32174887  $ \\[0.2em] \hline\hline
\end{tabular}}
\caption{Coefficients of the fit functions ${\cal F}^{(3,j)}$ for $j = 0, \, 1, \, 2$.
 \label{tab:coeffsF3}}
\end{center}
\end{table}


\begin{table}[!p]
\begin{center}
\scalebox{0.8}{\begin{tabular}{||l|c||l|c||l|c||} \hline\hline
%
 \rule{0pt}{18pt}  $a^{(4,0)}_1$ & $-74.33807565$ & $a^{(4,0)}_8$    & $-1.072113517$ & $b^{(4,0)}_1$ & $-67.3157309  $ \\[0.2em] \hline
 \rule{0pt}{18pt}  $a^{(4,0)}_2$ & $83.51877142 $ & $a^{(4,0)}_9$    & $-7.54309398 $ & $b^{(4,0)}_2$ & $-12.00333253 $ \\[0.2em] \hline
 \rule{0pt}{18pt}  $a^{(4,0)}_3$ & $2211.863641 $ & $a^{(4,0)}_{10}$ & $-7.50854142 $ & $b^{(4,0)}_3$ & $-0.9261306312$ \\[0.2em] \hline
 \rule{0pt}{18pt}  $a^{(4,0)}_4$ & $201.8040524 $ & $a^{(4,0)}_{11}$ & $4793.697814 $ & $b^{(4,0)}_4$ & $0.8610154185 $ \\[0.2em] \hline
 \rule{0pt}{18pt}  $a^{(4,0)}_5$ & $-1864.287186$ & $a^{(4,0)}_{12}$ & $-203.7776759$ & $b^{(4,0)}_5$ & $97.43004472  $ \\[0.2em] \hline
 \rule{0pt}{18pt}  $a^{(4,0)}_6$ & $59.48051719 $ & $a^{(4,0)}_{13}$ & $12.56399994 $ & $b^{(4,0)}_6$ & $51.54366079  $ \\[0.2em] \hline
 \rule{0pt}{18pt}  $a^{(4,0)}_7$ & $7.318121262 $ &		     &                & $b^{(4,0)}_7$ & $-1.797683781 $ \\[0.2em] \hline\hline
\end{tabular}}

\vspace*{10pt}

\scalebox{0.8}{\begin{tabular}{||l|c||l|c||l|c||} \hline\hline
%
 \rule{0pt}{18pt}  $a^{(4,1)}_1$ & $-33263.52427$  & $a^{(4,1)}_{10}$ & $0.5974304872 $ & $b^{(4,1)}_1$ & $101701.3176  $ \\[0.2em] \hline
 \rule{0pt}{18pt}  $a^{(4,1)}_2$ & $-65394.55662$  & $a^{(4,1)}_{11}$ & $-0.4581028326$ & $b^{(4,1)}_2$ & $21596.82836  $ \\[0.2em] \hline
 \rule{0pt}{18pt}  $a^{(4,1)}_3$ & $-5136.278821$  & $a^{(4,1)}_{12}$ & $91174.96835  $ & $b^{(4,1)}_3$ & $24.77527346  $ \\[0.2em] \hline
 \rule{0pt}{18pt}  $a^{(4,1)}_4$ & $-113640.3868$  & $a^{(4,1)}_{13}$ & $77042.06009  $ & $b^{(4,1)}_4$ & $0.4171623618 $ \\[0.2em] \hline
 \rule{0pt}{18pt}  $a^{(4,1)}_5$ & $78403.14527 $  & $a^{(4,1)}_{14}$ & $48.39104275  $ & $b^{(4,1)}_5$ & $-0.3939475425$ \\[0.2em] \hline
 \rule{0pt}{18pt}  $a^{(4,1)}_6$ & $20305.3333	$  & $a^{(4,1)}_{15}$ & $-74.45848329 $ & $b^{(4,1)}_6$ & $-220034.4015 $ \\[0.2em] \hline
 \rule{0pt}{18pt}  $a^{(4,1)}_7$ & $-82313.42428$  &		      & 		& $b^{(4,1)}_7$ & $-60326.64433 $ \\[0.2em] \hline
 \rule{0pt}{18pt}  $a^{(4,1)}_8$ & $-44842.44435$  &		      & 		& $b^{(4,1)}_8$ & $-232.9360438 $ \\[0.2em] \hline
 \rule{0pt}{18pt}  $a^{(4,1)}_9$ & $-48.95679755$  &		      & 		& $b^{(4,1)}_9$ & $4.842367923  $ \\[0.2em] \hline\hline
\end{tabular}}

\vspace*{10pt}

\scalebox{0.8}{\begin{tabular}{||l|c||l|c||l|c||} \hline\hline
%
 \rule{0pt}{18pt}  $a^{(4,2)}_1$    & $763815.9759 $  & $a^{(4,2)}_{12}$ & $-40.97365643 $ & $b^{(4,2)}_1$    & $-2144980.643 $ \\[0.2em] \hline
 \rule{0pt}{18pt}  $a^{(4,2)}_2$    & $1180737.052 $  & $a^{(4,2)}_{13}$ & $0.6946727184 $ & $b^{(4,2)}_2$    & $-51197.38647 $ \\[0.2em] \hline
 \rule{0pt}{18pt}  $a^{(4,2)}_3$    & $-112227.0528$  & $a^{(4,2)}_{14}$ & $-0.4169896858$ & $b^{(4,2)}_3$    & $58241.72209  $ \\[0.2em] \hline
 \rule{0pt}{18pt}  $a^{(4,2)}_4$    & $-13536.98632$  & $a^{(4,2)}_{15}$ & $-5469382.558 $ & $b^{(4,2)}_4$    & $86.72943137  $ \\[0.2em] \hline
 \rule{0pt}{18pt}  $a^{(4,2)}_5$    & $1787287.401 $  & $a^{(4,2)}_{16}$ & $-2454934.018 $ & $b^{(4,2)}_5$    & $-0.1265238057$ \\[0.2em] \hline
 \rule{0pt}{18pt}  $a^{(4,2)}_6$    & $-3099998.914$  & $a^{(4,2)}_{17}$ & $105075.7583  $ & $b^{(4,2)}_6$    & $0.1633022691 $ \\[0.2em] \hline
 \rule{0pt}{18pt}  $a^{(4,2)}_7$    & $-309064.9062$  & $a^{(4,2)}_{18}$ & $-232.8189073 $ & $b^{(4,2)}_7$    & $4671869.217  $ \\[0.2em] \hline
 \rule{0pt}{18pt}  $a^{(4,2)}_8$    & $46980.66442 $  & $a^{(4,2)}_{19}$ & $11.53750538  $ & $b^{(4,2)}_8$    & $367092.4634  $ \\[0.2em] \hline
 \rule{0pt}{18pt}  $a^{(4,2)}_9$    & $3372467.234 $  & 		 &		   & $b^{(4,2)}_9$    & $-163439.2968 $ \\[0.2em] \hline
 \rule{0pt}{18pt}  $a^{(4,2)}_{10}$ & $1073881.732 $  & 		 &		   & $b^{(4,2)}_{10}$ & $-629.2131294 $ \\[0.2em] \hline
 \rule{0pt}{18pt}  $a^{(4,2)}_{11}$ & $-86878.63443$  & 		 &		   & $b^{(4,2)}_{11}$ & $4.61591432   $ \\[0.2em] \hline\hline
\end{tabular}}
\caption{Coefficients of the fit functions ${\cal F}^{(4,j)}$ for $j = 0, \, 1, \, 2$.
 \label{tab:coeffsF4}}
\end{center}
\end{table}


\begin{table}[!p]
\begin{center}
\scalebox{0.8}{\begin{tabular}{||l|c||l|c||l|c||} \hline\hline
%
 \rule{0pt}{18pt}  $a^{(5,0)}_1$ & $3905.581329 $ & $a^{(5,0)}_8$    & $-72.43889949$ & $b^{(5,0)}_1$ & $-8478.618241$ \\[0.2em] \hline
 \rule{0pt}{18pt}  $a^{(5,0)}_2$ & $3736.990603 $ & $a^{(5,0)}_9$    & $-129.5483802$ & $b^{(5,0)}_2$ & $-196.6104915$ \\[0.2em] \hline
 \rule{0pt}{18pt}  $a^{(5,0)}_3$ & $2264.778673 $ & $a^{(5,0)}_{10}$ & $109.3433799 $ & $b^{(5,0)}_3$ & $-10.10763078$ \\[0.2em] \hline
 \rule{0pt}{18pt}  $a^{(5,0)}_4$ & $-9306.551804$ & $a^{(5,0)}_{11}$ & $-27312.86911$ & $b^{(5,0)}_4$ & $10.09635221 $ \\[0.2em] \hline
 \rule{0pt}{18pt}  $a^{(5,0)}_5$ & $15122.34121 $ & $a^{(5,0)}_{12}$ & $-2055.349456$ & $b^{(5,0)}_5$ & $18913.79983 $ \\[0.2em] \hline
 \rule{0pt}{18pt}  $a^{(5,0)}_6$ & $1010.941469 $ & $a^{(5,0)}_{13}$ & $-3725.212494$ & $b^{(5,0)}_6$ & $998.0693091 $ \\[0.2em] \hline
 \rule{0pt}{18pt}  $a^{(5,0)}_7$ & $-106.0075415$ &		     &  	      & $b^{(5,0)}_7$ & $-1133.206617$ \\[0.2em] \hline\hline
\end{tabular}}

\vspace*{10pt}

\scalebox{0.8}{\begin{tabular}{||l|c||l|c||l|c||} \hline\hline
%
 \rule{0pt}{18pt}  $a^{(5,1)}_1$ & $37058.95433 $  & $a^{(5,1)}_{10}$ & $-57.26884721$ & $b^{(5,1)}_1$ & $-122495.0958$ \\[0.2em] \hline
 \rule{0pt}{18pt}  $a^{(5,1)}_2$ & $86434.94551 $  & $a^{(5,1)}_{11}$ & $44.96166644 $ & $b^{(5,1)}_2$ & $-34379.8989 $ \\[0.2em] \hline
 \rule{0pt}{18pt}  $a^{(5,1)}_3$ & $8328.079906 $  & $a^{(5,1)}_{12}$ & $-193387.553 $ & $b^{(5,1)}_3$ & $158.4485277 $ \\[0.2em] \hline
 \rule{0pt}{18pt}  $a^{(5,1)}_4$ & $138006.1508 $  & $a^{(5,1)}_{13}$ & $-182385.6292$ & $b^{(5,1)}_4$ & $-8.510613511$ \\[0.2em] \hline
 \rule{0pt}{18pt}  $a^{(5,1)}_5$ & $-129916.3419$  & $a^{(5,1)}_{14}$ & $-8329.465612$ & $b^{(5,1)}_5$ & $10.40978139 $ \\[0.2em] \hline
 \rule{0pt}{18pt}  $a^{(5,1)}_6$ & $-45882.17487$  & $a^{(5,1)}_{15}$ & $3246.207284 $ & $b^{(5,1)}_6$ & $263518.7876 $ \\[0.2em] \hline
 \rule{0pt}{18pt}  $a^{(5,1)}_7$ & $140460.8266 $  &		      & 	       & $b^{(5,1)}_7$ & $93109.20931 $ \\[0.2em] \hline
 \rule{0pt}{18pt}  $a^{(5,1)}_8$ & $95287.2913	$  &		      & 	       & $b^{(5,1)}_8$ & $-124.8071873$ \\[0.2em] \hline
 \rule{0pt}{18pt}  $a^{(5,1)}_9$ & $2496.299161 $  &		      & 	       & $b^{(5,1)}_9$ & $-786.1795826$ \\[0.2em] \hline\hline
\end{tabular}}

\vspace*{10pt}

\scalebox{0.8}{\begin{tabular}{||l|c||l|c||l|c||} \hline\hline
%
 \rule{0pt}{18pt}  $a^{(5,2)}_1$    & $-522757.1396$  & $a^{(5,2)}_{12}$ & $881.4771555 $ & $b^{(5,2)}_1$    & $4915032.722 $ \\[0.2em] \hline
 \rule{0pt}{18pt}  $a^{(5,2)}_2$    & $-6523145.412$  & $a^{(5,2)}_{13}$ & $-73.62385747$ & $b^{(5,2)}_2$    & $7113454.316 $ \\[0.2em] \hline
 \rule{0pt}{18pt}  $a^{(5,2)}_3$    & $-3241047.26 $  & $a^{(5,2)}_{14}$ & $74.48846037 $ & $b^{(5,2)}_3$    & $641388.2145 $ \\[0.2em] \hline
 \rule{0pt}{18pt}  $a^{(5,2)}_4$    & $-138353.0261$  & $a^{(5,2)}_{15}$ & $-13653184.76$ & $b^{(5,2)}_4$    & $490.3117761 $ \\[0.2em] \hline
 \rule{0pt}{18pt}  $a^{(5,2)}_5$    & $-5917007.059$  & $a^{(5,2)}_{16}$ & $4120332.652 $ & $b^{(5,2)}_5$    & $3.530025843 $ \\[0.2em] \hline
 \rule{0pt}{18pt}  $a^{(5,2)}_6$    & $-8734679.062$  & $a^{(5,2)}_{17}$ & $1893384.372 $ & $b^{(5,2)}_6$    & $-2.307663415$ \\[0.2em] \hline
 \rule{0pt}{18pt}  $a^{(5,2)}_7$    & $4850904.549 $  & $a^{(5,2)}_{18}$ & $-1346.449475$ & $b^{(5,2)}_7$    & $-10091443.38$ \\[0.2em] \hline
 \rule{0pt}{18pt}  $a^{(5,2)}_8$    & $573351.9957 $  & $a^{(5,2)}_{19}$ & $-2606.043971$ & $b^{(5,2)}_8$    & $-16691093.61$ \\[0.2em] \hline
 \rule{0pt}{18pt}  $a^{(5,2)}_9$    & $5216650.702 $  & 		 &		  & $b^{(5,2)}_9$    & $-1910503.367$ \\[0.2em] \hline
 \rule{0pt}{18pt}  $a^{(5,2)}_{10}$ & $-5568616.336$  & 		 &		  & $b^{(5,2)}_{10}$ & $-4756.538617$ \\[0.2em] \hline
 \rule{0pt}{18pt}  $a^{(5,2)}_{11}$ & $-1231649.287$  & 		 &		  & $b^{(5,2)}_{11}$ & $-716.9721919$ \\[0.2em] \hline\hline
\end{tabular}}
\caption{Coefficients of the fit functions ${\cal F}^{(5,j)}$ for $j = 0, \, 1, \, 2$.
 \label{tab:coeffsF5}}
\end{center}
\end{table}


\begin{table}[!p]
\begin{center}
\scalebox{0.8}{\begin{tabular}{||l|c||l|c||l|c||} \hline\hline
%
 \rule{0pt}{18pt}  $a^{(6,0)}_1$ & $-1011.098065$ & $a^{(6,0)}_8$    & $-9.196475916$ & $b^{(6,0)}_1$ & $-135.9357965$ \\[0.2em] \hline
 \rule{0pt}{18pt}  $a^{(6,0)}_2$ & $612.2418916 $ & $a^{(6,0)}_9$    & $-69.88704353$ & $b^{(6,0)}_2$ & $-112.0425213$ \\[0.2em] \hline
 \rule{0pt}{18pt}  $a^{(6,0)}_3$ & $20800.85166 $ & $a^{(6,0)}_{10}$ & $-74.52135156$ & $b^{(6,0)}_3$ & $-8.580246776$ \\[0.2em] \hline
 \rule{0pt}{18pt}  $a^{(6,0)}_4$ & $2483.329368 $ & $a^{(6,0)}_{11}$ & $47022.34261 $ & $b^{(6,0)}_4$ & $7.856428099 $ \\[0.2em] \hline
 \rule{0pt}{18pt}  $a^{(6,0)}_5$ & $-18580.22374$ & $a^{(6,0)}_{12}$ & $-1797.122131$ & $b^{(6,0)}_5$ & $-234.1481254$ \\[0.2em] \hline
 \rule{0pt}{18pt}  $a^{(6,0)}_6$ & $513.5129609 $ & $a^{(6,0)}_{13}$ & $425.0463944 $ & $b^{(6,0)}_6$ & $466.9137641 $ \\[0.2em] \hline
 \rule{0pt}{18pt}  $a^{(6,0)}_7$ & $72.63158237 $ &		     &  	      & $b^{(6,0)}_7$ & $150.3890827 $ \\[0.2em] \hline\hline
\end{tabular}}

\vspace*{10pt}

\scalebox{0.8}{\begin{tabular}{||l|c||l|c||l|c||} \hline\hline
%
 \rule{0pt}{18pt}  $a^{(6,1)}_1$ & $-370556.6243$  & $a^{(6,1)}_{10}$ & $-7.250818623$ & $b^{(6,1)}_1$ & $1130735.088 $ \\[0.2em] \hline
 \rule{0pt}{18pt}  $a^{(6,1)}_2$ & $-726286.536 $  & $a^{(6,1)}_{11}$ & $8.422466929 $ & $b^{(6,1)}_2$ & $238659.3868 $ \\[0.2em] \hline
 \rule{0pt}{18pt}  $a^{(6,1)}_3$ & $-56315.15861$  & $a^{(6,1)}_{12}$ & $1290555.261 $ & $b^{(6,1)}_3$ & $164.9852581 $ \\[0.2em] \hline
 \rule{0pt}{18pt}  $a^{(6,1)}_4$ & $-1183952.802$  & $a^{(6,1)}_{13}$ & $901297.0421 $ & $b^{(6,1)}_4$ & $5.713012677 $ \\[0.2em] \hline
 \rule{0pt}{18pt}  $a^{(6,1)}_5$ & $923055.4177 $  & $a^{(6,1)}_{14}$ & $2551.928407 $ & $b^{(6,1)}_5$ & $-5.566487772$ \\[0.2em] \hline
 \rule{0pt}{18pt}  $a^{(6,1)}_6$ & $232335.3238 $  & $a^{(6,1)}_{15}$ & $-776.9367694$ & $b^{(6,1)}_6$ & $-2446748.927$ \\[0.2em] \hline
 \rule{0pt}{18pt}  $a^{(6,1)}_7$ & $-1033904.509$  &		      & 	       & $b^{(6,1)}_7$ & $-667444.1274$ \\[0.2em] \hline
 \rule{0pt}{18pt}  $a^{(6,1)}_8$ & $-516141.4549$  &		      & 	       & $b^{(6,1)}_8$ & $-2190.014875$ \\[0.2em] \hline
 \rule{0pt}{18pt}  $a^{(6,1)}_9$ & $-867.3699483$  &		      & 	       & $b^{(6,1)}_9$ & $169.6784678 $ \\[0.2em] \hline\hline
\end{tabular}}

\vspace*{10pt}

\scalebox{0.8}{\begin{tabular}{||l|c||l|c||l|c||} \hline\hline
%
 \rule{0pt}{18pt}  $a^{(6,2)}_1$    & $7349040.411 $  & $a^{(6,2)}_{12}$ & $-1438.447187$ & $b^{(6,2)}_1$    & $-18954487.93$ \\[0.2em] \hline
 \rule{0pt}{18pt}  $a^{(6,2)}_2$    & $8615904.5   $  & $a^{(6,2)}_{13}$ & $-3.077387638$ & $b^{(6,2)}_2$    & $2863788.748 $ \\[0.2em] \hline
 \rule{0pt}{18pt}  $a^{(6,2)}_3$    & $-2651744.946$  & $a^{(6,2)}_{14}$ & $6.228365923 $ & $b^{(6,2)}_3$    & $886573.1136 $ \\[0.2em] \hline
 \rule{0pt}{18pt}  $a^{(6,2)}_4$    & $-201649.9631$  & $a^{(6,2)}_{15}$ & $-59838907.02$ & $b^{(6,2)}_4$    & $1164.541363 $ \\[0.2em] \hline
 \rule{0pt}{18pt}  $a^{(6,2)}_5$    & $15229984.0  $  & $a^{(6,2)}_{16}$ & $-21990089.43$ & $b^{(6,2)}_5$    & $0.1548051941$ \\[0.2em] \hline
 \rule{0pt}{18pt}  $a^{(6,2)}_6$    & $-34683397.53$  & $a^{(6,2)}_{17}$ & $1964802.631 $ & $b^{(6,2)}_6$    & $0.6613035272$ \\[0.2em] \hline
 \rule{0pt}{18pt}  $a^{(6,2)}_7$    & $-708824.9359$  & $a^{(6,2)}_{18}$ & $1967.09485  $ & $b^{(6,2)}_7$    & $41583495.74 $ \\[0.2em] \hline
 \rule{0pt}{18pt}  $a^{(6,2)}_8$    & $747462.6945 $  & $a^{(6,2)}_{19}$ & $587.116867  $ & $b^{(6,2)}_8$    & $-4272494.994$ \\[0.2em] \hline
 \rule{0pt}{18pt}  $a^{(6,2)}_9$    & $35564210.45 $  & 		 &		  & $b^{(6,2)}_9$    & $-2537807.82 $ \\[0.2em] \hline
 \rule{0pt}{18pt}  $a^{(6,2)}_{10}$ & $7835697.023 $  & 		 &		  & $b^{(6,2)}_{10}$ & $-8849.79627 $ \\[0.2em] \hline
 \rule{0pt}{18pt}  $a^{(6,2)}_{11}$ & $-1456909.929$  & 		 &		  & $b^{(6,2)}_{11}$ & $118.7297335 $ \\[0.2em] \hline\hline
\end{tabular}}
\caption{Coefficients of the fit functions ${\cal F}^{(6,j)}$ for $j = 0, \, 1, \, 2$.
 \label{tab:coeffsF6}}
\end{center}
\end{table}

\end{appendix}

\newpage

%
%

\renewcommand{\refname}{R\lowercase{eferences}}

\addcontentsline{toc}{section}{References}

\bibliographystyle{JHEP} 

\small

\bibliography{references}

\end{document}